\documentclass[twocolumn]{aastex63}

\usepackage{gensymb,amsmath}

\usepackage{tabularx}

\newcommand\clearrow{\global\let\rowmac\relax}
\usepackage{multirow}
\usepackage{threeparttable}
\usepackage[utf8]{inputenc}
\usepackage{makecell}
\begin{document}

\received{}
\revised{}
\accepted{}

\submitjournal{ApJ}

%
\shorttitle{Isolated Massive Star Formation in G28.20-0.05}
\shortauthors{Law et al.}
%

\newcommand{\am}{NH$_{3}$}
\newcommand{\water}{H$_2$O}
\newcommand{\form}{H$_2$CO}
\newcommand{\co}{$^{12}{\rm CO}(2-1$}
\newcommand{\coo}{${{\rm C}^{18}{\rm O}}(2-1)$}
\newcommand{\halpha}{${\rm H}30\alpha$}
\newcommand{\choh}{${{\rm CH}_{3}{\rm OH}}$}
\newcommand{\sio}{${\rm SiO}(5-4$}
\newcommand{\hto}{${\rm H}_{2}{\rm O}$}
\newcommand{\oh}{${\rm OH}$}

\newcommand{\herschel}{{\it Herschel}~}
\newcommand{\spitzer}{{\it Spitzer}~}
\newcommand{\gaia}{{\it GAIA}~}
\newcommand{\apogee}{{\it APOGEE}~}
\newcommand{\apex}{{\it APEX}~}
\newcommand{\alma}{{\it ALMA}~}
\newcommand{\planck}{{\it PLANCK}~}
\newcommand{\twomass}{{\it 2MASS}~}
\newcommand{\wise}{{\it WISE}~}
\newcommand{\jcmt}{{\it JCMT}~}
\newcommand{\vla}{{\it VLA}~}
\newcommand{\atca}{{\it ATCA}~}
\newcommand{\sofia}{{\it SOFIA}~}

\newcommand{\degs}{$^{\circ}$}
\newcommand{\arcsecond}{$^{\prime\prime}$}
\newcommand{\msun}{M$_{\odot}$}
\newcommand{\mdust}{M$_{\textrm{\footnotesize Dust}}$}
\newcommand{\ra}{$\alpha$(J2000)}
\newcommand{\dec}{$\delta$(J2000)}
\newcommand{\h}{$^{\mathrm{h}}$}
\newcommand{\m}{$^{\mathrm{m}}$}
\newcommand{\s}{$^{\mathrm{s}}$}

\newcommand{\kms}{km~s$^{-1}$}
\newcommand{\kmsp}{km s$^{-1}$ pc$^{-1}$}
\newcommand{\ergs}{ergs s$^{-1}$}
\newcommand{\ergy}{ergs yr$^{-1}$}
\newcommand{\jyb}{Jy b$^{-1}$}
\newcommand{\jybe}{$~\rm mJy\:beam^{-1}$}
\newcommand{\cmq}{cm$^{-3}$}
\newcommand{\cms}{cm$^{-2}$}
\newcommand{\msuny}{M$_{\odot}$ yr$^{-1}$}
\newcommand{\mdot}{$\dot\textrm{M}$}

\title{Isolated Massive Star Formation in G28.20-0.05}

\author[0000-0003-1964-970X]{Chi-Yan Law}
\email{chiyan.law@chalmers.se}
\affiliation{Department of Space, Earth \& Environment, Chalmers University of Technology, SE-412 96 Gothenburg, Sweden}
\affiliation{European Southern Observatory, Karl-Schwarzschild-Strasse 2, D-85748 Garching, Germany}
\author[0000-0002-3389-9142]{Jonathan C. Tan}
\affiliation{Department of Space, Earth \& Environment, Chalmers University of Technology, SE-412 96 Gothenburg, Sweden}
\affiliation{Department of Astronomy, University of Virginia, Charlottesville, VA 22904-4325, USA}
\author[0000-0003-1602-6849]{Prasanta Gorai}
\affiliation{Department of Space, Earth \& Environment, Chalmers University of Technology, SE-412 96 Gothenburg, Sweden}
\author[0000-0001-7511-0034]{Yichen Zhang}
\affiliation{The Institute of Physical and Chemical Research (RIKEN), 2-1, Hirosawa, Wako-shi, Saitama 351-0198, Japan}
\author[0000-0003-4040-4934]{Rub\'en Fedriani}
\affiliation{Department of Space, Earth \& Environment, Chalmers University of Technology, SE-412 96 Gothenburg, Sweden}
\author[0000-0002-2149-2660]{Daniel Tafoya}
\affiliation{Department of Space, Earth and Environment, Chalmers University of Technology, Onsala Space Obsevatory, 439 92 Onsala, Sweden}
\author[0000-0002-6907-0926]{Kei E. I. Tanaka}
\affiliation{Center for Astrophysics and Space Astronomy, University of Colorado Boulder, Boulder, CO 80309, USA}
\affiliation{ALMA Project, National Astronomical Observatory of Japan, Mitaka, Tokyo 181-8588, Japan}
\author[0000-0001-5551-9502]{Giuliana Cosentino}
\affiliation{Department of Space, Earth \& Environment, Chalmers University of Technology, SE-412 96 Gothenburg, Sweden}
\author[0000-0001-8227-2816]{Yao-Lun Yang}
\affiliation{The Institute of Physical and Chemical Research (RIKEN), 2-1, Hirosawa, Wako-shi, Saitama 351-0198, Japan}
\affiliation{Department of Astronomy, University of Virginia, Charlottesville, VA 22904-4325, USA}
\author[0000-0002-5065-9175]{Diego Mardones}
\affiliation{Departamento de Astronom\'ia, Universidad de Chile, Las Condes, Santiago, Chile}
\author[0000-0003-3315-5626]{Maria T. Beltr\'an}
\affiliation{INAF, Osservatorio Astrofisico di Arcetri, Largo E. Fermi 5, I-50125
Firenze, Italy}
\author[0000-0003-1649-7958]{Guido Garay}
\affiliation{Departamento de Astronom\'ia, Universidad de Chile, Las Condes, Santiago, Chile}



\begin{abstract}
We report high-resolution 1.3~mm continuum and molecular line observations of the massive protostar G28.20-0.05 with ALMA. The continuum image reveals a ring-like structure with 2,000~au radius, similar to morphology seen in archival 1.3~cm VLA observations. Based on its spectral index and associated H$30\alpha$ emission, this structure mainly traces ionised gas. However, there is evidence for $\sim30$~M$_{\odot}$ of dusty gas near the main mm continuum peak on one side of the ring, as well as in adjacent regions within 3,000~au. A virial analysis on scales of $\sim$2,000~au from hot core line emission yields a dynamical mass of $\sim80\:M_\odot$. A strong velocity gradient in the H$30\alpha$ emission is evidence for a rotating, ionized disk wind, which drives a larger-scale molecular outflow. An infrared SED analysis indicates a current protostellar mass of $m_*\sim40\:M_\odot$ forming from a core with initial mass $M_c\sim300\:M_\odot$ in a clump with mass surface density of $\Sigma_{\rm cl}\sim 0.8\:{\rm g\:cm}^{-2}$. Thus the SED and other properties of the system can be understood in the context of core accretion models. Structure-finding analysis on the larger-scale continuum image indicates G28.20-0.05 is forming in a relatively isolated environment, with no other concentrated sources, i.e., protostellar cores, above $\sim 1\:M_\odot$ found from $\sim$0.1 to 0.4~pc around the source. This implies that a massive star can form in relative isolation and the dearth of other protostellar companions within the $\sim1$~pc environs is a strong constraint on massive star formation theories that predict the presence of a surrounding protocluster.

\end{abstract}

\section{Introduction}\label{sec:intro}
Massive ($>8\:M_\odot$) stars impact many areas of astrophysics. However, the mechanism of their formation is still under debate. Two main scenarios are (i) Core Accretion \citep[e.g., the Turbulent Core Accretion model of][]{2003ApJ...585..850M} and (ii) Competitive Accretion \citep[e.g.,][]{2001MNRAS.324..573B,2010ApJ...709...27W} (see, e.g., \citealt{2014prpl.conf..149T} for a review). The former is a scaled-up version of the standard model of low-mass star formation \citep[][]{1987ARA&A..25...23S}, although with the internal pressure of the massive pre-stellar core being dominated by turbulence and/or magnetic fields, rather than thermal pressure. Such conditions make it likely that the collapse will be more disordered than in the low-mass case, perhaps including significant accretion via overdense filaments and other sub-structures, e.g., as seen in magnetohydrodynamical (MHD) simulations of such structures \citep[e.g.,][]{2012MNRAS.422..347S,2013ApJ...766...97M,2021MNRAS.502.1104H}. A characteristic feature of core accretion models is a more direct linkage of the pre-stellar core mass function (CMF) and the stellar initial mass function (IMF), although perhaps mediated by effects of a varying core-to-star formation efficiency and binary or small-$N$ multiple formation by disk fragmentation within a core. 

In Competitive Accretion, stars chaotically gain their mass via the global collapse of a cluster-forming clump without passing through the massive pre-stellar core phase. In the context of the Competitive Accretion model, there is no correlation between the CMF and the IMF as the accretion involves ambient gas materials of the cloud.


Identifying relatively isolated massive protostars provides a direct way to constrain massive star formation. These types of sources, i.e., with limited surrounding fragmentation and star formation, indicate that collapse from a massive core has occurred in a relatively monolithic manner. For instance, \citet[][]{2017A&A...600L..10C} studied 35 sources with \alma and found that most of them show limited fragmentation, with at most 3 cores per clump. \citet{2019A&A...622A..99L} also found low levels of fragmentation in the massive cores of the NGC-6334 region. 
On the other hand, \citet[][]{2017MNRAS.468.3694C} studied the massive star-forming region G11.92-0.61 finding that the three massive protostars in the region are surrounded by at least 16 lower mass protostellar sources within a region about 0.3~pc in radius. 

Protostars forming via core accretion, especially in relatively uncrowded environments, are more likely to involve an ordered transition from the infall envelope to a Keplerian disk, as has been claimed in G339.88-1.26 by \citet[][]{2019ApJ...873...73Z}. They are also more likely to exhibit relatively ordered outflows, i.e., launched orthogonally to the accretion disk and maintaining their orientation for relatively long periods. 


Additional observational studies of isolated massive protostars are important to test theoretical models, as they are relatively simple systems that can have high discriminatory power between the different formation scenarios.
In this work, we analyze 1.3~mm (band 6) continuum and line data obtained by \alma observations of the massive protostar G28.20-0.05. This source has been characterized as being a high luminosity ($\sim 1.4 - 1.6\times10^{5}\:L_{\odot}$) \citep{2014ApJ...786...38H,2015MNRAS.453..645M} shell-like hypercompact HII region and a hot molecular core \citep{2003A&A...410..597W,2004ApJ...605..285S,2008ApJ...686L..21Q} at a near kinematic distance of $d=5.7^{+0.5}_{-0.8}\:{\rm kpc}$ \citep{2003ApJ...587..701F}, based on a systemic velocity of $v_{\rm sys}=95.6\pm 0.5\:{\rm km\:s}^{-1}$ \citep{2008ApJ...686L..21Q}, which is consistent with our observations of hot core line tracers in the source (see \S3.3). 
We note that some previous studies adopted the far kinematic distance of $9.1\:{\rm kpc}$ \citep{1994ApJS...91..659K,2020MNRAS.492..895D}, however, as discussed later in \S3.2, we are able to make a new astrometric confirmation of the near distance and so adopt $d=5.7\:$kpc throughout this work.

Based on SMA 1.3~mm continuum emission that is assumed to trace dusty gas within a radius of 0.48~pc, G28.20-0.05 has been estimated to have a gas mass within this region of $33\:M_{\odot}$, and thus a mass surface density of $9.52\times 10^{-3}\:{\rm g\:cm}^{-2}$ \citep{2014ApJ...786...38H}. However, such an estimate is quite uncertain due to assumptions about dust temperatures and may also be subject to missing flux. 
Previous studies (e.g., \citealt{2005ApJ...631..399S}) have suggested the presence of two components: (i) an infalling equatorial torus of molecular gas containing a central ionized region; and (ii) an extended molecular shell, which is associated wide-angle outflow or wind. Furthermore, \citet{2009ApJ...703.1308K} presented SMA observations and inferred from a velocity gradient perpendicular to the outflow direction that warm molecular gas (e.g., as traced by SO$_{2}$) is undergoing bulk rotation. 
\citet{2011A&A...530A..53K} detected a large and wide-angle $\rm{^{12}CO(2-1)}$ outflow based on the \jcmt observations. 
\citet{2008ApJ...686L..21Q} presented a chemical study of the source with the SMA to measure the kinetic temperature and column density of the source. Based on multiple K-components of $\rm{CH_3CN}$ 
the authors measured a rotational temperature of about $300\:$K.


This paper is organized as follows. In \S\ref{sec:observations}, we summarise the ALMA observations and the reduction procedures. Here we also summarize Hubble Telescope (HST) NIR observations of the source.
We study the continuum and molecular line properties of the protostar in \S\ref{sec:g28-cont}, including a discussion of overall morphology, kinematics and dynamics. We measure and model the spectral energy distribution (SED) of the protostar with multi-wavelength data in \S\ref{sec:SED}. In \S\ref{sec:fragmentation}, we discuss the fragmentation and multiplicity properties of the source. Finally, a summary is presented in \S\ref{sec:discussion}.


\section{Observations}\label{sec:observations}

\subsection{ALMA observations \& data reduction}\label{sec:reduction}

G28.20-0.05 was observed with \alma in Band 6 via a Cycle 3 project (PI: Y. Zhang; 2015.1.01454.S) with Compact (C36-2, C) and Intermediate (C36-5, I)\footnote{\url{https://almascience.eso.org/documents-and-tools/cycle3/alma-technical-handbook (Table 7.1)}} array configurations and via a Cycle 4 project (PI: J. Tan; 2016.1.00125.S) with an Extended (C40-9, E)\footnote{\url{https://arc.iram.fr/documents/cycle4/ \\ ALMACycle4TechnicalHandbook-Final.pdf} (Table 7.1)} configuration (see Table~\ref{table:observ_sum_config}).
In each case, a single pointing observation was made with a primary beam size (half power beam width) of $26.9\arcsec$\footnote{\url{https://almascience.nrao.edu/about-alma/alma-basics}}. All the observations have the same spectral set-up, covering frequencies from 216.7 to 234.9~GHz. Information about the spectral band-passes used in the observations is given in Table~\ref{table:observ_sum_spectral}. The total integration times were 12.3, 16.2 and 160 minutes in these configurations, respectively. For the Cycle 3 compact configuration observation, J1751+0939 (1.74~Jy) was used for band-pass and flux calibration and J1830+0619 (0.31~Jy) was used for phase calibration. For the intermediate configuration observation, J1924-2914 (4.03~Jy) was used for band-pass and flux calibration and J1851+0035 (0.24~Jy) was used for phase calibration. For the Cycle 4 observations, J1924-2914 (8.48~Jy) was used for band-pass and flux calibration and J1834-0301 (0.26~Jy) was used for phase calibration. 

All data reduction and imaging was performed with CASA software \citep[][]{2007ASPC..376..127M} using version 5.6.0-60. After pipeline calibration, we performed the continuum and line imaging separately. For the continuum imaging, we image and self-calibrate each spectral window based on line-free channels with task {\it tclean} using `briggs' weighting with a robust parameter of 0.5. We identify line-free channels as follows. We first inspect the raw spectrum of each spectral window and identify representative line-free channels. We then define a threshold by the root mean square (rms; $\sigma_{\mathrm{rms}}=\sqrt{\sum_{i} I_{i}^{2} / N}$) of those channels. Any channels that are within four times the rms are counted as line-free channels. We then use these line-free channels to form the individual spectral window continuum images, as well as the total continuum image. We perform four iterations of phase-only calibrations with solution intervals of 30s, 10s and `int' with manual masking. Finally, an iteration of amplitude calibration with manual masking, which we also apply to the line data.

The resulting synthesized beams of each configuration are summarized in Table~\ref{table:observ_sum_config}. We self-calibrated each configuration before combining them using the CASA function {\it concat} to obtain the final combined continuum images. During combination, we weight each configuration based on their average time intervals (Table~\ref{table:observ_sum_config} Column 6)\footnote{\url{https://casaguides.nrao.edu/index.php \\/DataWeightsAndCombination}}. The final weightings are 1, 1, and 0.33 for the C, I, and E configurations, respectively. The combined (C+I+E) continuum was then cleaned interactively with manual masking with multi-scale deconvolver at scales of 0, 10, 50, 150 pixels. The resulting synthesized beam of the final combined continuum image is $0.060$\arcsec $\times 0.036$\arcsec. A summary of the observations and configuration set up is given in Table~\ref{table:observ_sum_config}. We note that the maximum recoverable scale (MRS) ranges from $0.34^{\prime\prime}$ to $11^{\prime\prime}$, while the resolution ranges from about $0.026^{\prime\prime}$ to $0.813^{\prime\prime}$.  

The rms noise level in a given continuum image (before primary beam correction) is done by sampling 5,000 regions, each with an area equal to that of the synthesized beam. We then fit a Gaussian to this distribution of fluxes and estimate the $1\sigma$ noise level from the standard deviation of this Gaussian. The $1\sigma$ noise levels in the C, C+I and C+I+E configuration images are $1.29$, $0.358$, and $0.170~~\rm mJy\:beam^{-1}$, respectively.

For molecular line imaging, we first subtract the baseline in the uv plane using {\it uvcontsub} in CASA. The continuum emission is obtained by subtracting line-free channels using the {\it uvcontsub} function. We apply manual masking during line imaging. We use {\it tclean} to image the emission lines, again with `briggs' weighting and robust factor of 0.5 and multi-scale deconvolver at scales of 0, 10, 50, 150 pixels. Unless otherwise mentioned specifically, science measurements are performed on the primary beam corrected images.


\begin{deluxetable*}{lcccccccc}
\tabletypesize{\footnotesize}
\tablecaption{Summary of ALMA observations of G28.20-0.05\label{table:observ_sum_config}} 
\tablewidth{18pt}
\tablehead{
\colhead{Start of Obs. (epoch)}  & \colhead{Obs. time}  & \colhead{Config.} & \colhead{Antennas used} & \colhead{Baseline Lengths} & \colhead{Averaged Interval} & \colhead{MRS} & \colhead{Beam size}\\
\colhead{}  & \colhead{(min.)}  & \colhead{} & \colhead{} & \colhead{(m)} & \colhead{(sec.)} & \colhead{($^{\prime\prime}$)} & \colhead{($^{\prime\prime}$)}}
\startdata
\hline
2016/4/24 & & & & & &\\
07:36:04.4  & 12.3 & C (C36-2) & 41 & 14.7-377 & 6.05 & 11.0 & $0.735\times0.813$\\
(2016.31) & & & & & & &\\
\hline
2016/9/11 & & & & & &\\
02:45:50.2  & 16.2 & I (C36-5)& 37 & 47.9-1400.0 & 6.05 & 3.40 & $0.201\times0.217$\\
(2016.78) & & & & & & &\\
\hline
2017/9/30 & & & & & &\\
01:42:37.4  & 45.3 &  & 41 &  & &\\
(2017.75) & & & & & & &\\
2017/11/1 & & & & & &\\
00:08:10.3   & 41.8 & E (C40-9)& 49 &347-9740 & 2.02 &  0.340 &  $0.0260\times 0.0480$\\
(2017.84) & & & & & & &\\
2017/11/5 & & & & & &\\
22:49:20.9   & 72.5 & & 47 & & & &\\
(2017.85) & & & & & & &\\
\enddata
\end{deluxetable*}

\begin{table*}[]
\centering
\caption{Summary of set-ups of ALMA spectral windows}
\label{table:observ_sum_spectral}
\hskip 2.0cm
\begin{tabular}{ccccc}
\hline
Spectral Window & Molecular Line                     & Frequency range (MHz) & Channel Spacing  (kHz) & rms ($~\rm mJy\:beam^{-1}$)\\ \hline\hline
Spw0            & CH$_3$OH($\rm{4_{2,3}-5_{1,4}}$)   & 232928.10 - 234928.10 & 15625.00 &  0.600            \\ 
Spw1            & H(30)$\alpha$                      & 231587.86 - 232056.61 & 488.28 &   0.810             \\
Spw2            & $^{12}$CO(2-1)                     & 230297.25 - 230765.99 & 488.28 & 0.310          \\ 
Spw3            & H$_{2}$CO ($\rm{9_{1,8}-9_{1,9}}$) & 218714.73 - 218656.14 & 122.07 &  0.717            \\
Spw4            & CH$_3$OH($\rm{4_{-2,3}-3_{-1,2}}$) & 21839.79 - 218365.55  & 122.07 &   0.589            \\
Spw5            & C$^{18}$O(2-1)                    & 219514.81 - 219485.58 & 122.07 &   0.460            \\ 
Spw6            & CH$_{3}$CN                                & 220278.12 - 220248.88 & 122.07 &   0.890         \\
Spw7            & SiO(5-4)                           & 217147.97 - 217031.03 & 488.28 &   0.279            \\ 
Spw8            & SO$_{2}$                           & 216685.46 - 216451.08 & 488.28 &   0.710           \\ \hline
\\
\end{tabular}%
\end{table*}

\subsection{HST Observations}\label{sec:hst}

G28.20-0.05 was observed with the Hubble Space Telescope (HST) on the 8th August 2016 (epoch 2016.61) with the Wide Field Camera 3 (WFC3) in the near-infrared (NIR) channel (Project ID: 14494, PI: J. C. Tan). Two broad filters, F110W and F160W, covering the J ($1.1\,\mu$m) and H ($1.6\,\mu$m) bands and two narrow band filters, F128N and F164N, targeting the Pa$\beta$ ($1.28\,\mu$m) and [FeII] ($1.64\,\mu$m) lines were used. The diffraction limited spatial resolution for the images are $0\farcs13$, and $0\farcs16$ for the J and H bands, respectively. The pixel scale is $0\farcs13$. The field of view is $2\arcmin\times2\arcmin$ and the image is centred at (RA, Dec) = (18:42:58.48, -04:13:57.8). The integration times were 202.93\,s in the F110W and F160W filters and 399.23\,s in the F128N and F164N. The reduced images were downloaded from the Hubble Legacy Archive\footnote{\url{https://hla.stsci.edu/hla_welcome.html}} and custom python scripts were used to analyse them. We used the python package DrizzlePac\footnote{\url{https://drizzlepac.readthedocs.io/en/deployment/index.html}} to align the HST image to {\it Gaia}-determined astrometry \citep[][]{2021A&A...649A...1G}. This resulted in an astrometric accuracy of 9\,mas, which is consistent with their documentation\footnote{\url{https://hst-docs.stsci.edu/drizzpac/chapter-4-astrometric-information-in-the-header/4-5-absolute-astrometry}}.


\subsection{SOFIA Observations}\label{sec:sofia}

G28.20-0.05 was observed with the Stratospheric Observatory for Infrared Astronomy ({\it SOFIA}) Faint Object infraRed CAmera \citep[FORCAST,][]{2018JAI.....740005H} at $7.7, 19.7, 31.5$ and $37.1\:{\rm \mu m}$ on the 17th February 2022 (epoch 2022.05) as part of the SOFIA Massive (SOMA) Star Formation Survey (Project ID 09\_0085, PI: J. C. Tan). The corresponding beam FWHM for all four bands is $3\farcs8$. The full imaging data from this observation will be presented elsewhere as part of the SOMA survey. Here we use these data to measure background subtracted fluxes of the protostar to help constrain the spectral energy distribution (SED). To obtain these fluxes, standard analysis methods following those of the SOMA survey papers \citep[][]{2017ApJ...843...33D,2019ApJ...874...16L,2020ApJ...904...75L,2022arXiv220511422F} were used. The photometric calibration error is estimated to be in the range of $\sim3\%$-$7\%$. The astrometric precision is about $0\farcs1$ for the SOFIA $7\mu m$ image and $0\farcs4$ at the longer wavelengths \citep[see][for further details]{2017ApJ...843...33D}.



G28.20-0.05 was also observed with {\it SOFIA}'s High-resolution Airborne Wideband Camera Plus (HAWC+) \citep[][]{2010SPIE.7735E..6HD,2018JAI.....740008H} at $53\:{\rm \mu m}$ (Band A) and $214\:{\rm \mu m}$ (Band E) on 7th Sept 2021 (epoch 2021.68) (Project ID 09\_0164, PI: C.-Y. Law). The full imaging data, including polarimetric imaging properties, of the source will be presented elsewhere (Law et al., in prep.). In this paper we use the fluxes derived from these images to further constrain the SED of the source.
The full width at half maximum (FWHM) at the Band A and Band E center wavelengths are $4\farcs85$ and $18\farcs2$. The observations were performed using the Nod-Match chop mode with a Lissajous scan pattern. The raw data were processed by the {\it SOFIA}/HAWC+ instrument team using the data reduction pipeline version 3.0.0. This pipeline includes different data processing steps, including corrections for dead pixels and the intrinsic polarization of the instrument and telescope \citep[][]{2018JAI.....740008H,2019ApJ...882..113S}. 


\subsection{Other Ancillary Data}\label{sec:ancillary}

The following archival imaging data for G28.20-0.05 were also retrieved and analyzed. \spitzer IRAC \citep[][]{2004ApJS..154....1W,2004ApJS..154...10F} data at $3.6, 4.5, 5.8$ and $8.0\:{\rm \mu m}$ from the Galactic Legacy Infrared Midplane Survey Extraordinaire (GLIMPSE) \spitzer legacy survey \citep[][]{2003PASP..115..953B,2009PASP..121..213C} were obtained and analyzed. The mean spatial resolutions are  $1\farcs6, 1\farcs7, 1\farcs8$ and $1\farcs9$, respectively \citep[][]{2004ApJS..154...10F}.

\herschel PACS \citep[][]{2010A&A...518L...2P} and SPIRE \citep[][]{2010A&A...518L...3G} images at $70, 160, 250, 350$ and $500\:{\rm \mu m}$ were obtained from the \herschel High-Level Images (HHLI)\footnote{https://irsa.ipac.caltech.edu/data/Herschel/HHLI/index.html} in the \herschel Science Archive \citep{https://doi.org/10.26131/irsa79}. These \herschel images are processed to the highest
level available through the Standard Product Generation pipeline (version 14.0). The image product level of the PACS and SPIRE data used is 2.5 or 3.0. The angular resolutions are $5\farcs2, 12^{\prime\prime}, 18^{\prime\prime}, 25^{\prime\prime}$ and $36^{\prime\prime}$. 

Archival VLA 1.3~cm data for G28.20-0.05 \citep{2011ApJS..194...44S} were retrieved and analyzed. The VLA observations were carried out on 14th March 2006 (epoch 2006.2) with the K-band A-array (Program ID AZ168). The angular resolution of the continuum image is $0\farcs09$.

\section{Characterizing the Protostar}\label{sec:g28-cont}

\subsection{Morphology\label{sec:morphology}}

\begin{figure*}
\includegraphics[width=\textwidth]{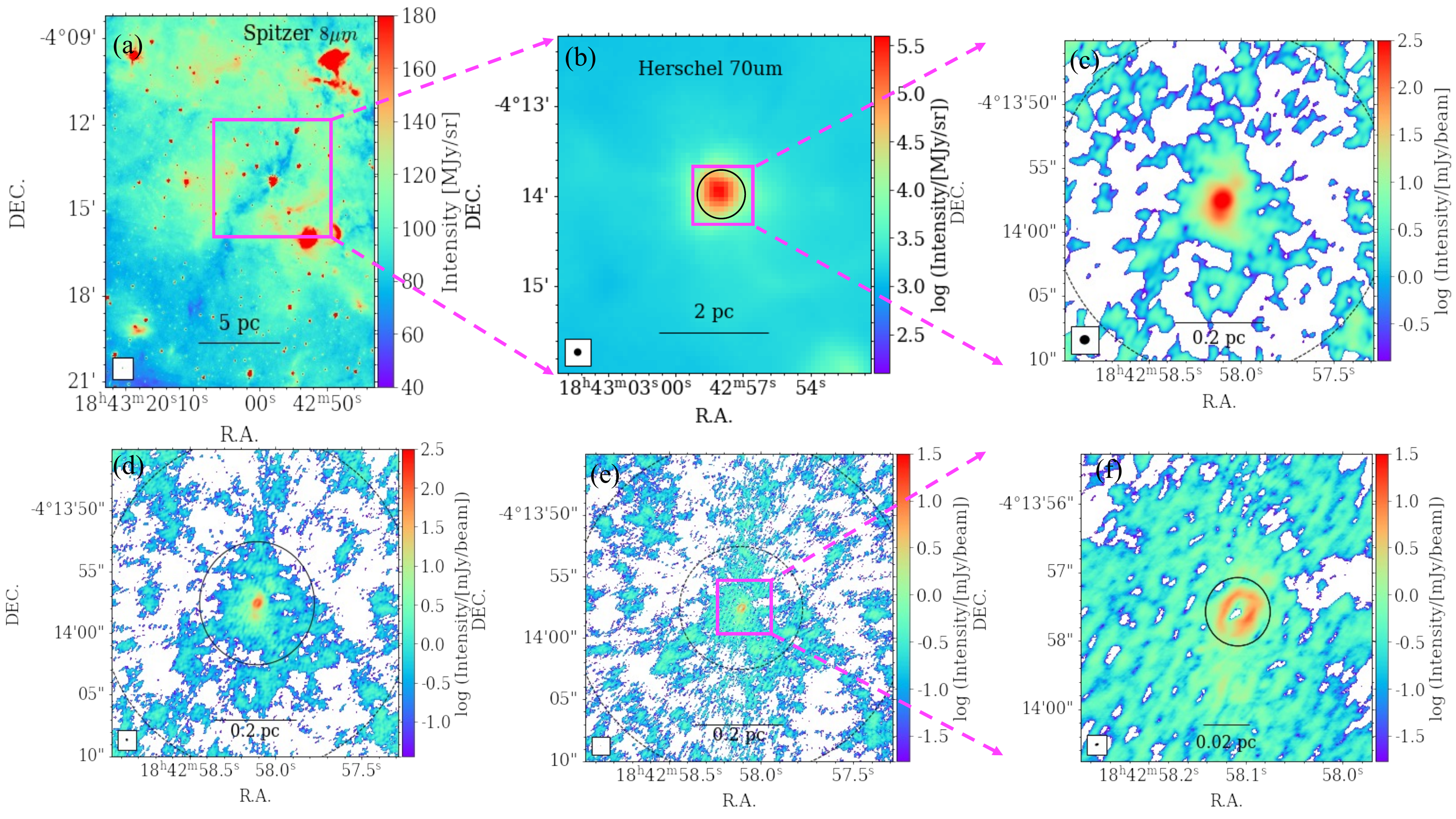}
\centering

\caption{Multi-scale views of the G28.20-0.05 massive protostar. 
{\it (a) Top left:} {\it Spitzer} $8\:{\rm \mu m}$ image ($1\farcs9$ resolution) of the $\sim 20$-pc scale region. A filamentary IRDC, from which the massive protostar appears to have formed, is visible as a dark shadow. {\it (b) Top middle:} {\it Herschel} $70\: {\rm \mu m}$ image ($5\farcs2$ resolution) of the $\sim 5$-pc scale region. The circle shows the aperture used for MIR to FIR SED photometry of the source (see text).
{\it (c) Top right:} \alma 1.3~mm compact (C) configuration only continuum map (beam size of $0\farcs813 \times 0\farcs735$), with intensity scale from $0.1 \sigma$ ($\sigma = 1.29\:$mJy~beam$^{-1}$) to $316~{\rm mJy~beam}^{-1}$. The dashed circle represents the ALMA 12-m primary beam.
{\it (d) Bottom left:} The same field of view as (c), but now showing the \alma 1.3 mm compact + intermediate (C+I) configurations continuum map (beam size of $0\farcs201 \times 0\farcs217$), with intensity scale from $0.1 \sigma$ ($\sigma = 0.358~\rm mJy\:beam^{-1}$) to $316~{\rm mJy~beam}^{-1}$. The solid circle shows a $5^{\prime\prime}$ radius aperture enclosing the main continuum structures, which is one scale used for flux measurements. 
{\it (e) Bottom middle:} As (c), but now showing the \alma 1.3 mm compact + intermediate + extended (C+I+E) configurations continuum map (beam size of $0\farcs060 \times 0\farcs036$), with intensity scale from $0.1 \sigma$ ($\sigma \sim 0.170~\rm mJy\:beam^{-1}$). 
{\it (f) Bottom right:} A zoom-in of panel (e) to the inner region of G28.20-0.05, which shows a ring-like structure. The solid circle shows a $0\farcs5$ radius aperture, which is used to measure the flux of the ring. 
}
\label{fig:cont-image}
\end{figure*}

Figure~\ref{fig:cont-image} presents an overview of the G28.20-0.05 protostellar system and its surroundings. Panel (a) shows the {\it Spitzer}-IRAC~$\rm 8\:\mu m$ image of the large-scale environment around the source, spanning $7$~pc by $10$~pc. The protostar is visible as a MIR-bright compact source that is embedded in a MIR-dark filament, i.e., an IRDC. We note that the G28.20-0.05 source is close in the sky position (about $6^\prime$) to the well-studied, massive IRDC~G028.37+00.07 \citep[also known as Cloud C in the sample of][]{2009ApJ...696..484B,2012ApJ...754....5B}, which has a similar estimated kinematic distance of 5~kpc. The uncertainties in kinematic distances are such that it is possible these sources could be in close proximity, with the projected separation being about 10~pc.


Figure~\ref{fig:cont-image}(b) presents the \herschel $70\:{\rm \mu m}$ continuum map of G28.20-0.05. The dynamic range in intensity of this image spans more than a factor of 1,000.
In this image the central source appears to be relatively isolated with no detection greater than $1\%$ of peak emission within a 2\arcmin\ (3.3~pc) radius around it.

Figure~\ref{fig:cont-image}(c) shows the primary beam corrected \alma 1.3~mm continuum image derived from the compact (C) configuration observation. As described above, the $1\sigma$ noise level in the central part of this image is $1.29~\rm mJy\:beam^{-1}$.
The intensity scale is set to have a minimum value of $0.1 \sigma$.
The image shows a central, compact source surrounded by a halo of fainter emission, but with an absence of other bright sources.

Figure~\ref{fig:cont-image}(d) shows the primary beam corrected 1.3~mm continuum image derived from the compact and intermediate combined (C+I) data, with $1\sigma$ noise level of $0.358~$mJy~beam$^{-1}$ in the central regions. Again, the intensity scale is set to have a minimum value of $0.1\sigma$. This image reveals finer details and substructure of the central source. However, again, there is no clear evidence of strong, compact secondary sources in the wider FOV. We return to this topic with a quantitative analysis of this image for the presence of secondary sources in \S\ref{sec:fragmentation}.

Figure~\ref{fig:cont-image}(e) shows the primary beam corrected 1.3~mm continuum image derived from all the configurations combined (C+I+E), while Figure~\ref{fig:cont-image}(f) presents a zoom-in view of the central source. The range of intensities shown extends down to 0.1$\sigma$, with the $1\sigma$ noise level being $0.170\:$mJy~beam$^{-1}$ in the central regions.
The image reveals a ring-like structure with a radius from its central minimum to its bright rim of $\sim 0.01\:$pc (2,000~au). Three peaks have been identified within the ring. The main peak of the continuum emission is on the SW side at R.A.~$=18:42:58.0997$, DEC.$=-4:13:57.636$. A secondary peak is found on the NE side, and a third relatively faint peak toward the northern part of the ring. Outside of the ring, more extended, fainter structures are visible, especially on each side that is aligned to the apparent long axis of the ring, i.e., NW to SE.

\subsection{Radio to mm Spectral Index to Probe Ionized and Dusty Gas\label{sec:radio_mm_sed}}

\begin{figure*}
\begin{minipage}{0.49\textwidth}
\includegraphics[width=\textwidth]{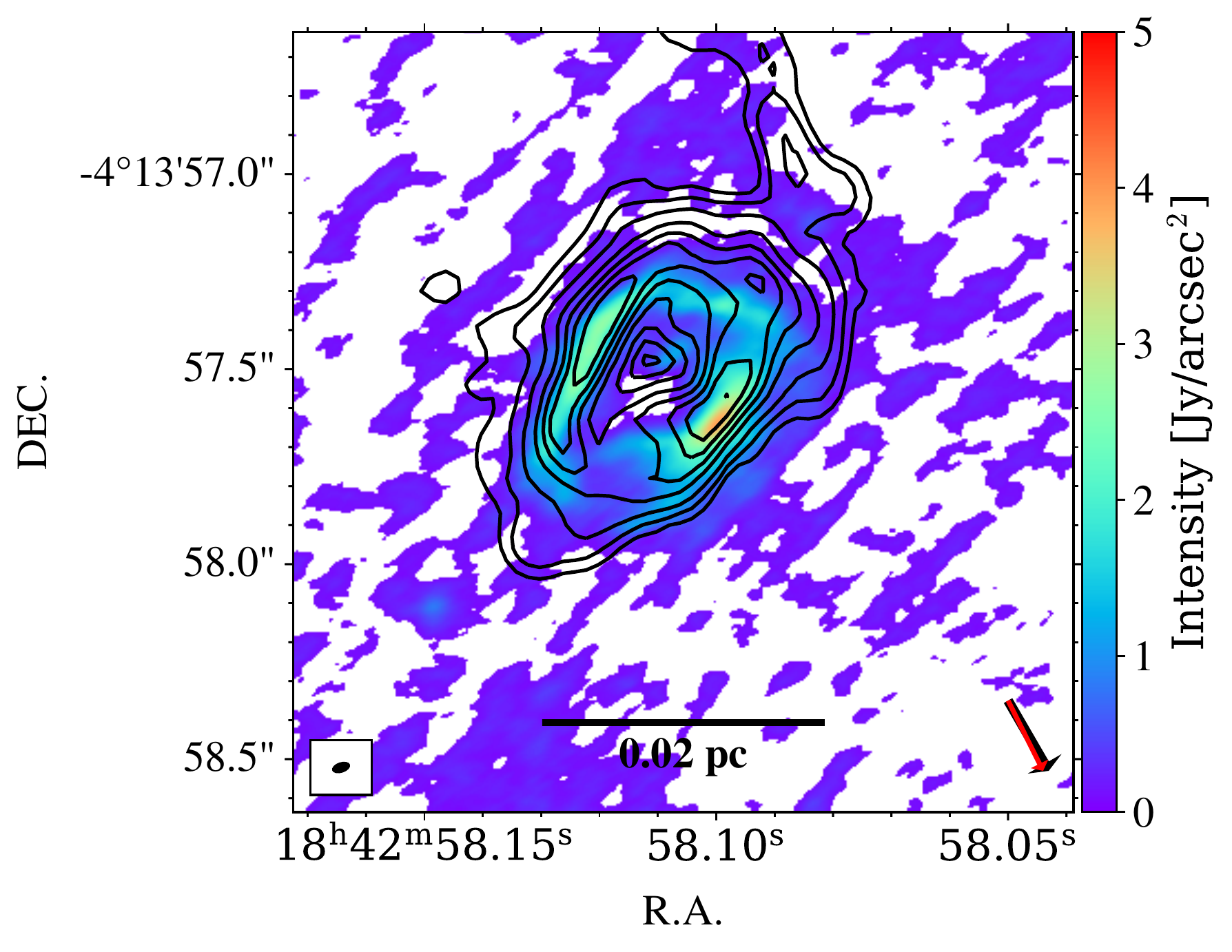}
\end{minipage}
\begin{minipage}{0.49\textwidth}
\includegraphics[width=\textwidth]{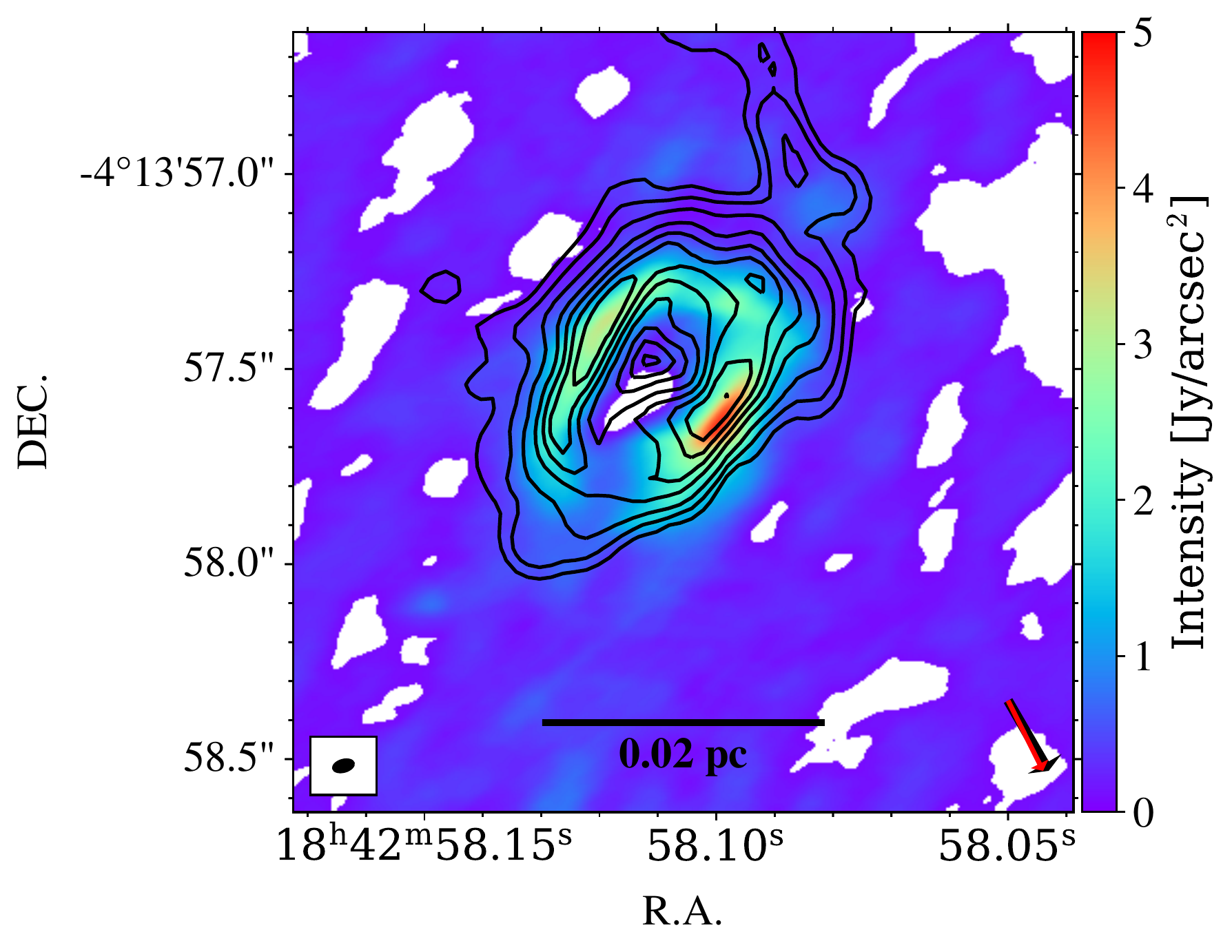}
\end{minipage}
\begin{minipage}{0.49\textwidth}
\includegraphics[width=\textwidth]{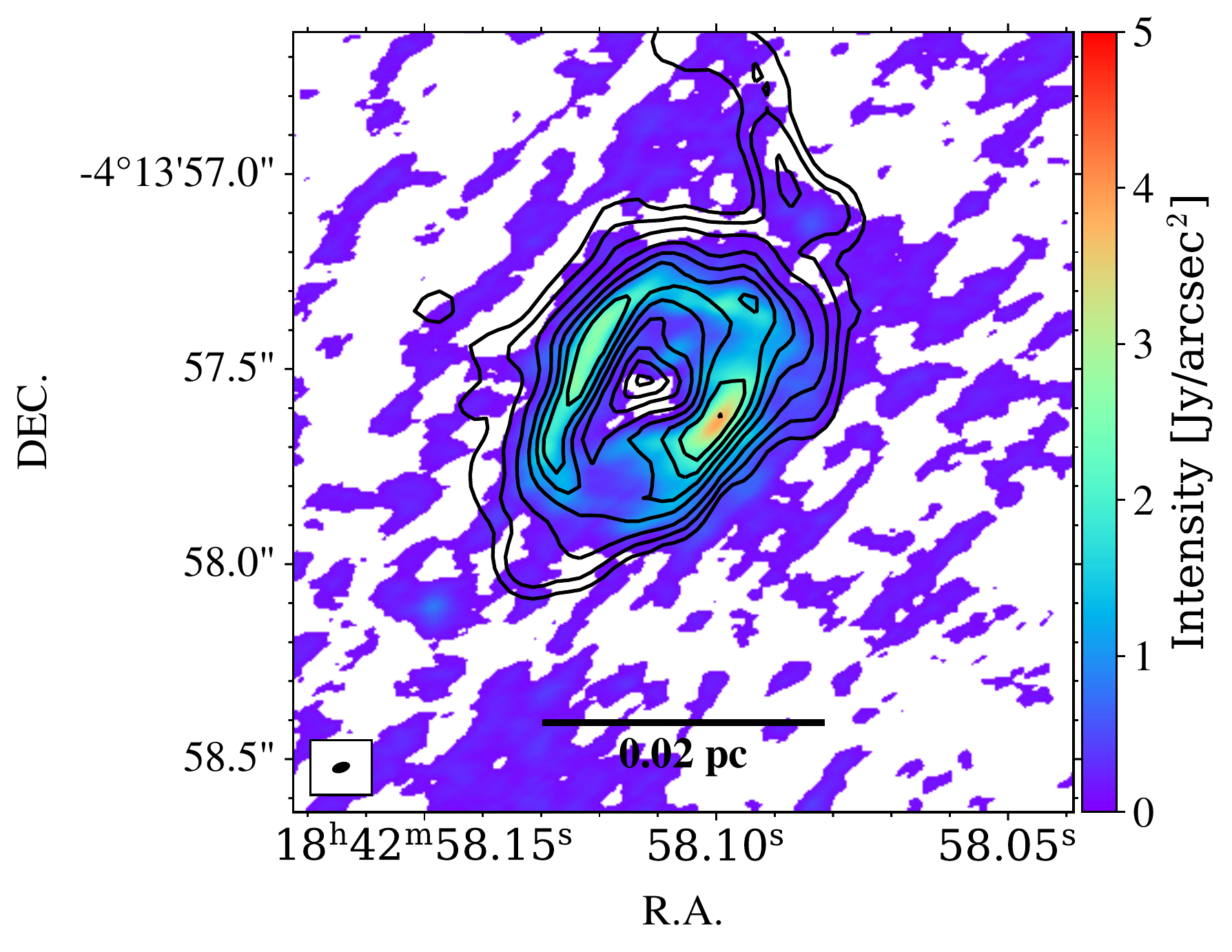}
\end{minipage}
\begin{minipage}{0.49\textwidth}
\includegraphics[width=\textwidth]{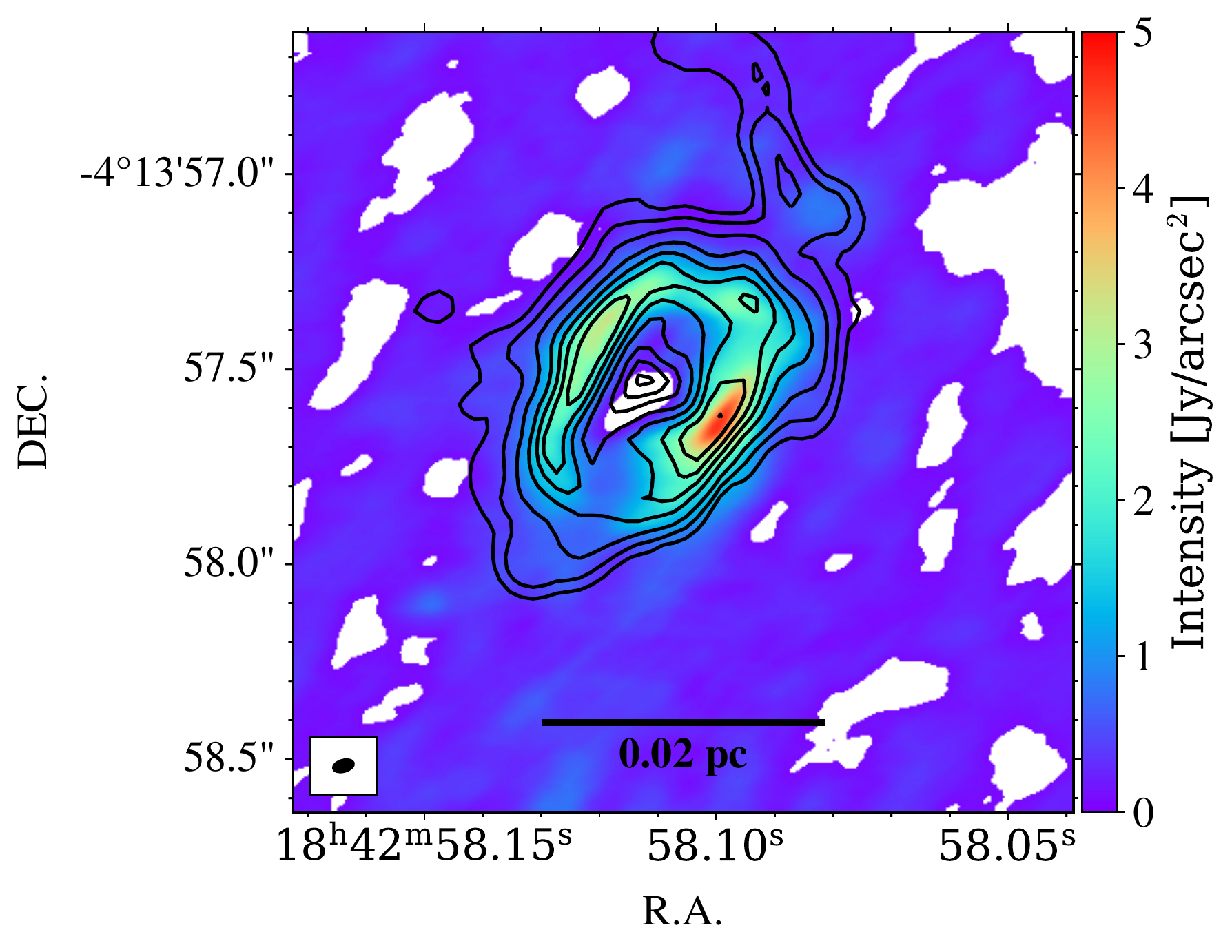}
\end{minipage}
\begin{minipage}{0.49\textwidth}
\includegraphics[width=\textwidth]{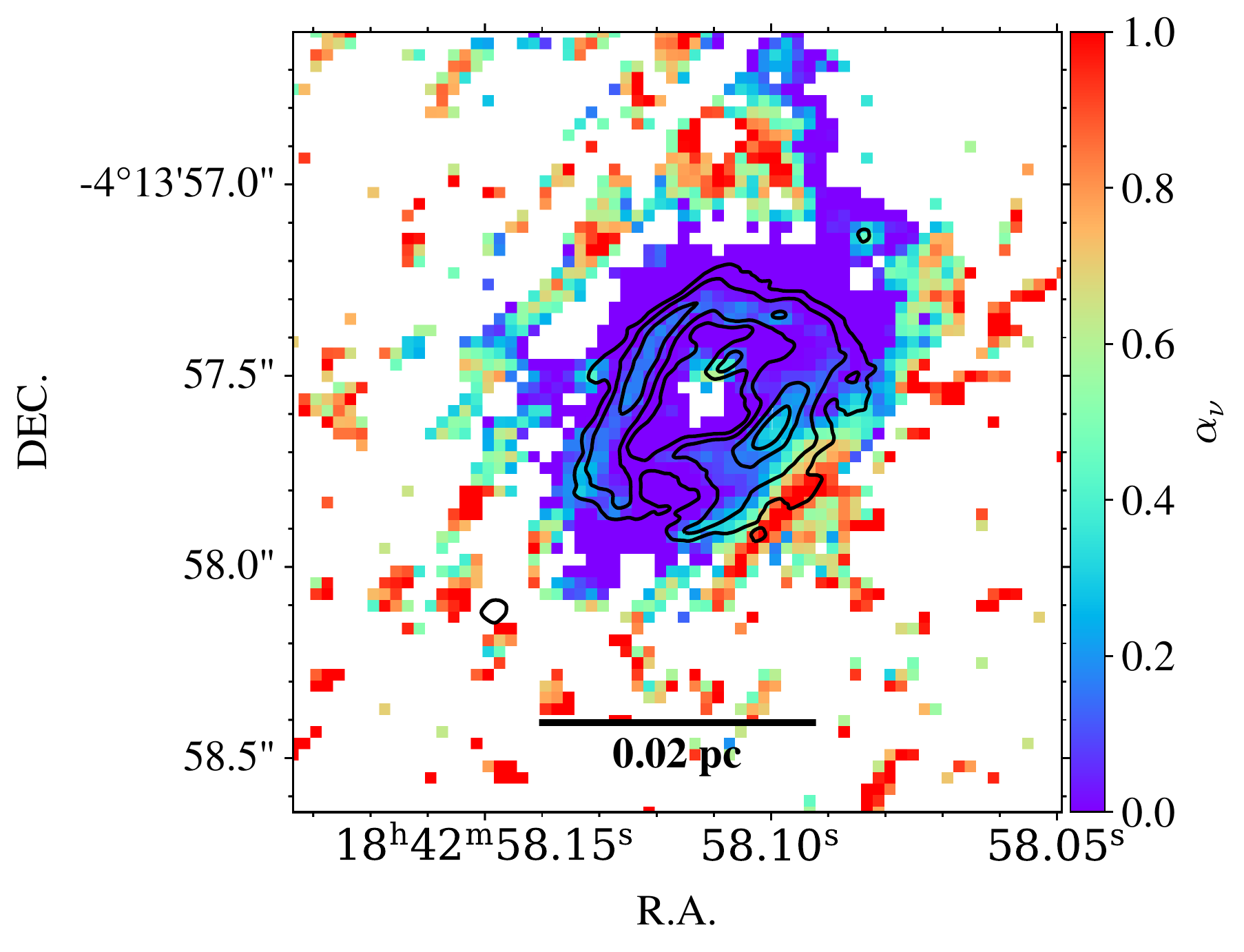}
\end{minipage}
\begin{minipage}{0.49\textwidth}
\includegraphics[width=\textwidth]{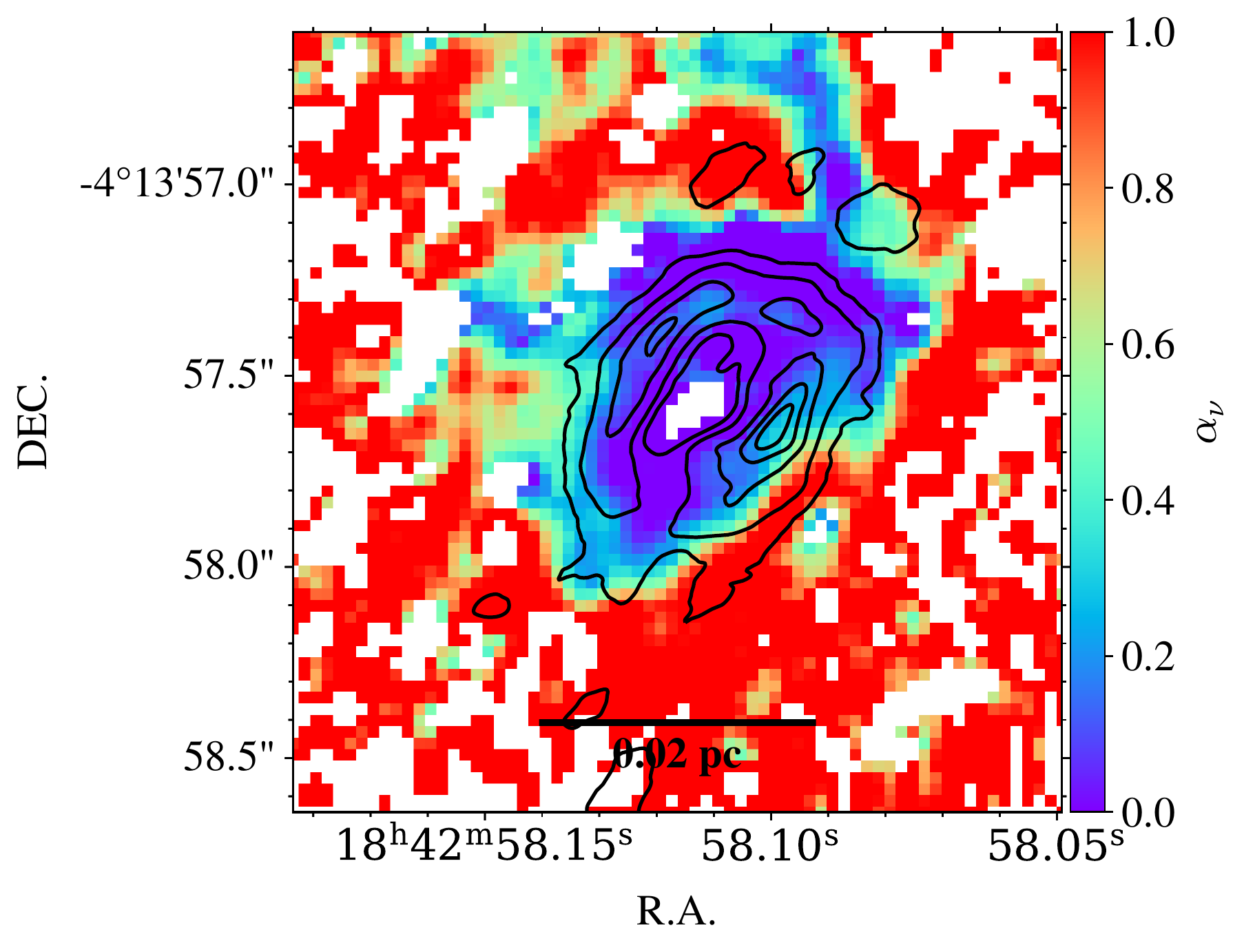}
\end{minipage}
\caption{\small {\it (a) Top left:} ALMA $1.3$~mm continuum image (E configuration only) of inner region of G28.20-0.05. The VLA $1.3$~cm continuum is shown with contours [$0.1, 0.2, 0.4, 0.8, 1.2, 1.6, 2.0, 2.4$~Jy/arcsec$^{2}$]. The beam size of 1.3~mm image is $0\farcs048 \times 0\farcs026$ (see inset), while that at 1.3~cm is $0\farcs09$. An offset of 34.8~mas in the direction of P.A.~$=216^\circ$ is seen between the peaks of the VLA and ALMA images (see red arrow in lower right, while black arrow, almost overlapping, shows the direction to Galactic centre).
%
{\it (b) Top right:} As (a), but now showing $1.3$~mm C+I+E combined image, with beam size $0\farcs060 \times 0\farcs036$ (see inset).
{\it (c) Middle left:} As (a), but now with the 1.3~cm image translated to align with the 1.3~mm image.
%
%
{\it (d) Middle right:} As (b), but now with the 1.3~cm image translated to align with the 1.3~mm image.
%
{\it (e) Bottom left:} Map of spectral index, $\alpha_\nu=\log \left(I_{\nu_{1}} / I_{\nu_{2}}\right) / \log \left(\nu_{1} / \nu_{2}\right)$, where $\nu_{1}=22.4~\mathrm{GHz}$ and $\nu_{2}=234~ \mathrm{GHz}$, i.e., evaluated using 1.3~cm and 1.3~mm (E-configuration) fluxes.
The contours are the $1.3$~mm continuum E configuration image, with levels at [0.5, 1, 1.5, 2, 2.5, 3, 3.5, 4, 4.5, 5, 5.5, 6] Jy/arcsec$^{2}$.
{\it (f) Bottom right:} As (e), but now using the 1.3~mm C+I+E configuration image. The overlaid contours levels are the same as in (e).}
\label{fig:int_flux_VLA_ALMA}
\end{figure*}

\citet{2008ApJ...681..350S,2011ApJS..194...44S} carried out high-resolution VLA observations toward G28.20-0.05
at $1.3$~cm ($22.4$~GHz, i.e., radio K-band) and detected a ring-like structure that is similar to the one we see in the 1.3~mm continuum. Figures~\ref{fig:int_flux_VLA_ALMA}(a) and (b) present the archival VLA $1.3$~cm continuum image overlaid with the E only and C+I+E combined $1.3$~mm continuum images, respectively. On first inspection the images show very similar morphology at these two wavelengths. Assuming the 1.3~cm continuum traces free-free emission from ionized gas, this suggests that a significant portion of the 1.3~mm continuum is also contributed by such emission. 

On closer examination of the VLA and ALMA images, we notice an apparent offset in peak positions and overall ring structure. Based on 2D Gaussian fits to the 3 peak positions, this offset has a magnitude of $(34.8\pm 9.5)\:{\rm mas}$ in a direction of P.A.~$=216^\circ$ (from VLA to ALMA). This offset is larger than the astrometric uncertainties of the VLA ($9$~mas)\footnote{\url{https://science.nrao.edu/facilities/vla/docs/manuals/oss/\\performance/positional-accuracy}} and ALMA ($3$~mas)\footnote{\url{https://help.almascience.org/kb/articles/what-is-the-astrometric-accuracy-of-alma}} observations. The VLA observations were carried out in March 14, 2006, i.e., 11.65 years before our ALMA observations (using the Nov 5, 2017 date of the longest E configuration observation). Thus, the observed offset corresponds to a proper motion of $(2.99\pm0.82)$~mas~$\rm yr^{-1}$, i.e., $(81\pm 22)\:\rm km\:s^{-1}$ at the 5.7~kpc distance of source.
The expected proper motion due to Galactic orbital motion (assuming, for simplicity, a constant rotation curve of amplitude 200~$\rm km\:s^{-1}$, a solar galactocentric distance of 8.0~kpc, and a kinematic distance to the source of 5.7~kpc in the direction of $l=28.2^\circ$) is 109~$\rm km\:s^{-1}$ in the direction of decreasing $l$, i.e., west in Galactic coordinates. The P.A. of this direction along the Galactic plane in R.A.-Dec. projection is 207$^\circ$. 
Thus, the observed proper motion, given the uncertainties, is consistent with being entirely in this direction along the Galactic plane. Additional velocity components of $\sim 10\:{\rm km\:s}^{-1}$ due to non-circular motions in the Galaxy, e.g., due to spiral arms or local turbulence, are also likely to be present, which can also help explain the difference between the observed motion and that predicted by the simple Galactic orbital model.
We note that if the source was at the far kinematic distance ($9.1~$kpc), then the expected motion would be larger, i.e., $243\:{\rm km\:s}^{-1}$, corresponding to 5.63~mas~yr$^{-1}$. Thus, overall, we conclude that the observed proper motion is consistent with that expected due to Galactic orbital motion and with the magnitude strongly favoring the source being at the near kinematic distance of 5.7~kpc.



We proceed by correcting for the apparent offset, i.e., by shifting the VLA image so that it best aligns with the ALMA image. These overlaid images, the equivalent of Fig.~\ref{fig:int_flux_VLA_ALMA}(a) and (b), are shown in Fig.~\ref{fig:int_flux_VLA_ALMA}(c) and (d). Apart from the general close agreement between the images, we also note the presence of an extended spur of emission in the 1.3~cm image extending from the north of the ring.

We next evaluate the spectral index, $\alpha_\nu$, map of the source based on the ratio of the intensities at 1.3~cm and 1.3~mm. The spectral index can help diagnose the physical processes responsible for the emission. In particular, regions where dust starts to make a dominant contribution to the 1.3~mm flux would have larger values of $\alpha_\nu$. We first re-grid the ALMA image to the VLA resolution (i.e., a pixel scale of $0.03\arcsec$) using the imregrid function in CASA. The spectral index is defined via
\begin{equation}
    \alpha_\nu =\log \left(I_{\nu_{1}} / I_{\nu_{2}}\right) / \log \left(\nu_{1} / \nu_{2}\right),
\end{equation}
where $\nu_{1}=22.4~\mathrm{GHz}$ and $\nu_{2}=234~\mathrm{GHz}$. When making the spectral index map, we only consider pixels that are 4 times the corresponding measured rms noise levels in both the ALMA and VLA images, i.e., 0.0689 Jy/arcsec$^{2}$ and 0.00679 Jy/arcsec$^{2}$, respectively. Figures~\ref{fig:int_flux_VLA_ALMA}(e) and (f) present the maps of $\alpha_\nu$ using the E and C+I+E ALMA images, respectively.

We see that $\alpha_\nu$ takes values of about 0.1 in the main ring structure, as based on VLA to ALMA C+I+E data. As expected, when only ALMA E configuration is used, smaller values of $\alpha_\nu$ are generally seen, which is likely due to missing flux at 1.3~mm in this case. We notice that toward the main 1.3~mm continuum peak there is a local enhancement of $\alpha_\nu$ to values of about 0.5. There are also larger values of $\alpha_\nu$, i.e., $\gtrsim 1$, seen immediately surrounding the ring.

\begin{figure}
    \includegraphics[width=1.1\columnwidth]{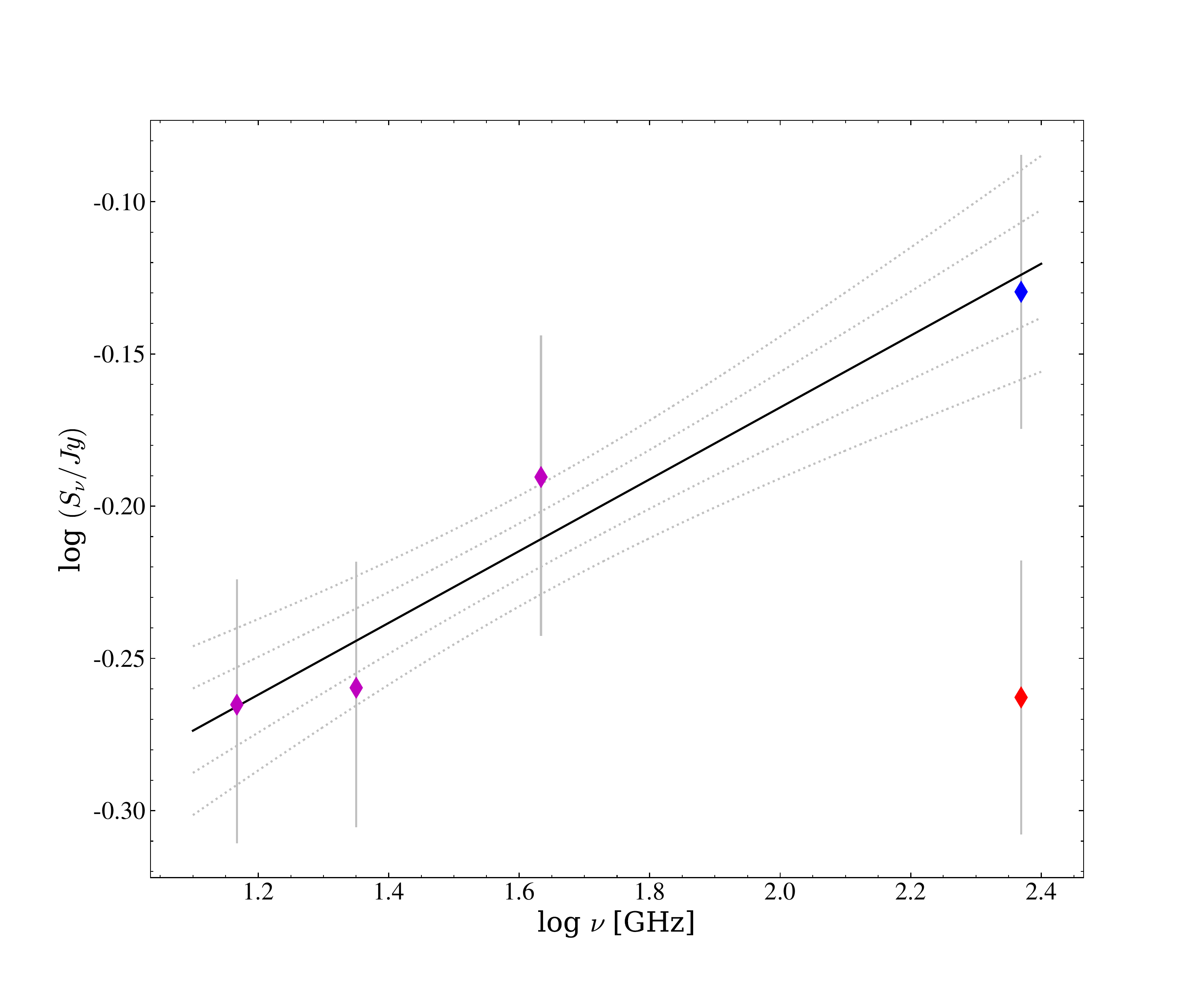}
    \includegraphics[width=1.1\columnwidth]{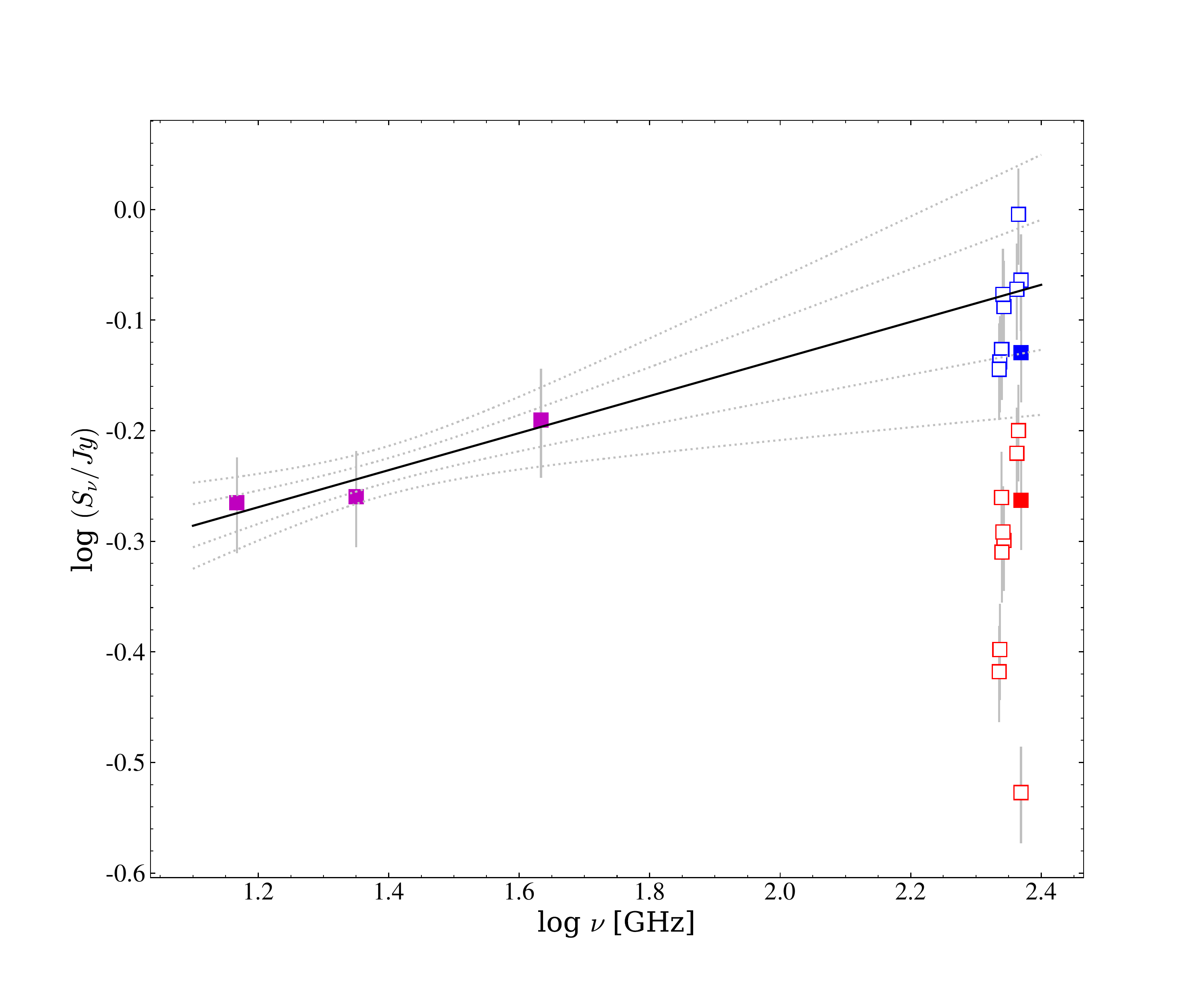}
    \caption{{\it (a) Top:} Radio to mm SED of the inner 0.5\arcsec\ (2,850~au) radius region of G28.20-0.05. Two data points are shown at 230~GHz (1.3~mm) from the ALMA observations. The higher (blue) point is based on the C+I+E combined image, while the lower (red) one is based on the E-only configuration image. Information about the other data points, which are from VLA observations, is given in Table~\ref{table:flux_wavelength}. A power-law fit to the SED (including ALMA C+I+E measurement) is shown, along with $1\sigma$ and $2\sigma$ confidence intervals, with derived spectral index of $\alpha_\nu = 0.118\pm0.020$.
    {\it (b) Bottom:} As (a), but now showing separate in-band ALMA measurements from 43 to 14.7~GHz (C+I+E - blue open squares; E - red open squares). Solid squares show the equivalent average ALMA fluxes with these configurations from (a). Now the power-law fit is only done to the VLA data points and then extrapolated to the ALMA frequencies. This power-law has $\alpha_\nu=0.168\pm0.058$.
    %
} 
    \label{fig:inband_sed_VLA_ALMA}
\end{figure}

To obtain an average value of the spectral index of the inner region we integrate the flux inside a radius of 0.5\arcsec. At 1.3~cm this flux is 0.550~Jy. At 1.3~mm the flux is 0.546~Jy in the E-configuration image and 0.742~Jy in the C+I+E image. Thus the average values of $\alpha_\nu$ of the inner region are $-0.00311\pm 0.06140$ and $0.128\pm 0.061$, respectively.
These data are shown in Figure~\ref{fig:inband_sed_VLA_ALMA}a, along with previous reported flux measurements from the VLA at 14.7~GHz (0.543~Jy with source size of $1.0\arcsec \times 0.7\arcsec$) and 43~GHz (0.645~Jy with source size of $0.9\arcsec$) \citep[][]{2011ApJS..194...44S} (see also Table \ref{table:flux_wavelength}). We see that the three VLA data points and the ALMA C+I+E data point can be well fit by a single power-law, i.e., $\alpha_\nu = 0.118\pm 0.020$. This suggests that a significant fraction of the 1.3~mm continuum flux on these scales is contributed by free-free emission from ionized gas, since approximately power-law behavior is often seen in the radio SEDs of ionized structures in the frequency range where they are transitioning from being partially optically thick to optically thin. However, it remains possible that the free-free emission spectrum deviates from this single power-law description, i.e., if it reaches the fully optically thin limiting value of $\alpha_\nu=-0.1$ by $\sim 100~$GHz. In this case a greater proportion of the 1.3~mm flux would be expected to be contributed by dust.

\begin{figure}
    \includegraphics[width=1.1\columnwidth]{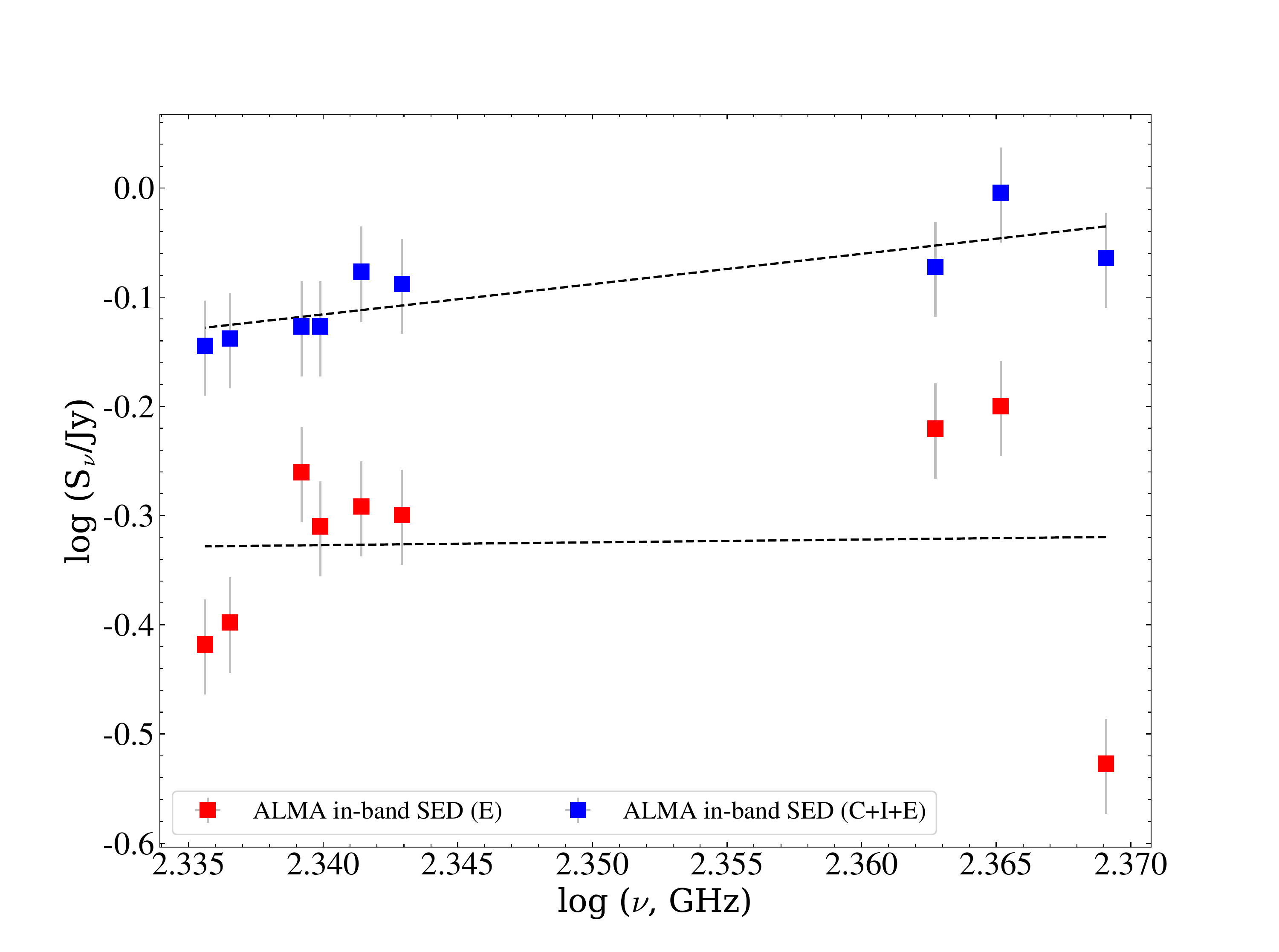}
    \caption{In-band 1.3~mm SED of the inner 0.5\arcsec\ (2,850~au) radius region of G28.20-0.05 based on E-only (red points) and C+I+E (blue points) configuration images. Power-law fits to these SEDs are shown with derived spectral indices of $\alpha_\nu = 0.255\pm2.930$ for E-only and $\alpha_\nu = 2.77\pm0.71$ for C+I+E}.
    
    \label{fig:inband_sed}
\end{figure}

As a further examination on the presence of dust in this region, we evaluate the ALMA in-band SED, i.e., within Band 6 from 216.575~GHz to 233.926~GHz. We make the same continuum measurements of the inner 0.5\arcsec\ region, but now carried out separately in the line-free regions of each of the spectral windows of the observation. These data are shown in Figure~\ref{fig:inband_sed_VLA_ALMA}b, as well as in more detail in Figure~\ref{fig:inband_sed}. Here we assume a $10\%$ calibration uncertainty in the integrated fluxes of each measurement, which dominates over other errors.
From the in-band data we measure the following spectral indices ($\alpha_\nu$), depending on which image is used: $0.255\pm 2.930$ (E); $2.77\pm0.71$ (C+I+E). We note the following results for other combinations: $1.65\pm 0.26$ (I); $1.43\pm 0.35$ (C+I).  These results suggest the potential presence of dust in the inner region, which leads to a steepening of the spectral index compared to the values seen at longer wavelengths. 

\begin{figure*}[t]
\begin{minipage}{0.49\textwidth}
\includegraphics[width=\textwidth]{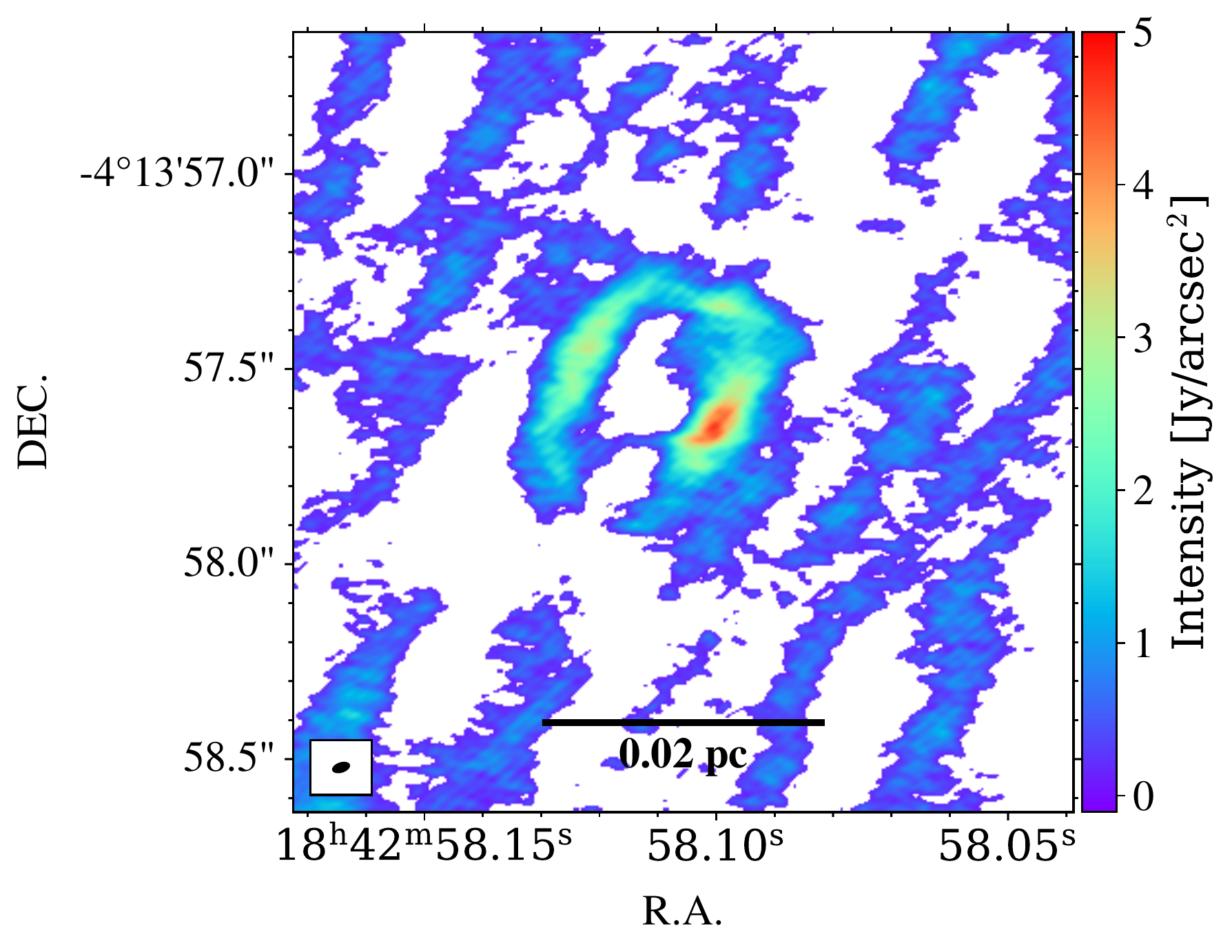}
\end{minipage}
\begin{minipage}{0.49\textwidth}
\includegraphics[width=\textwidth]{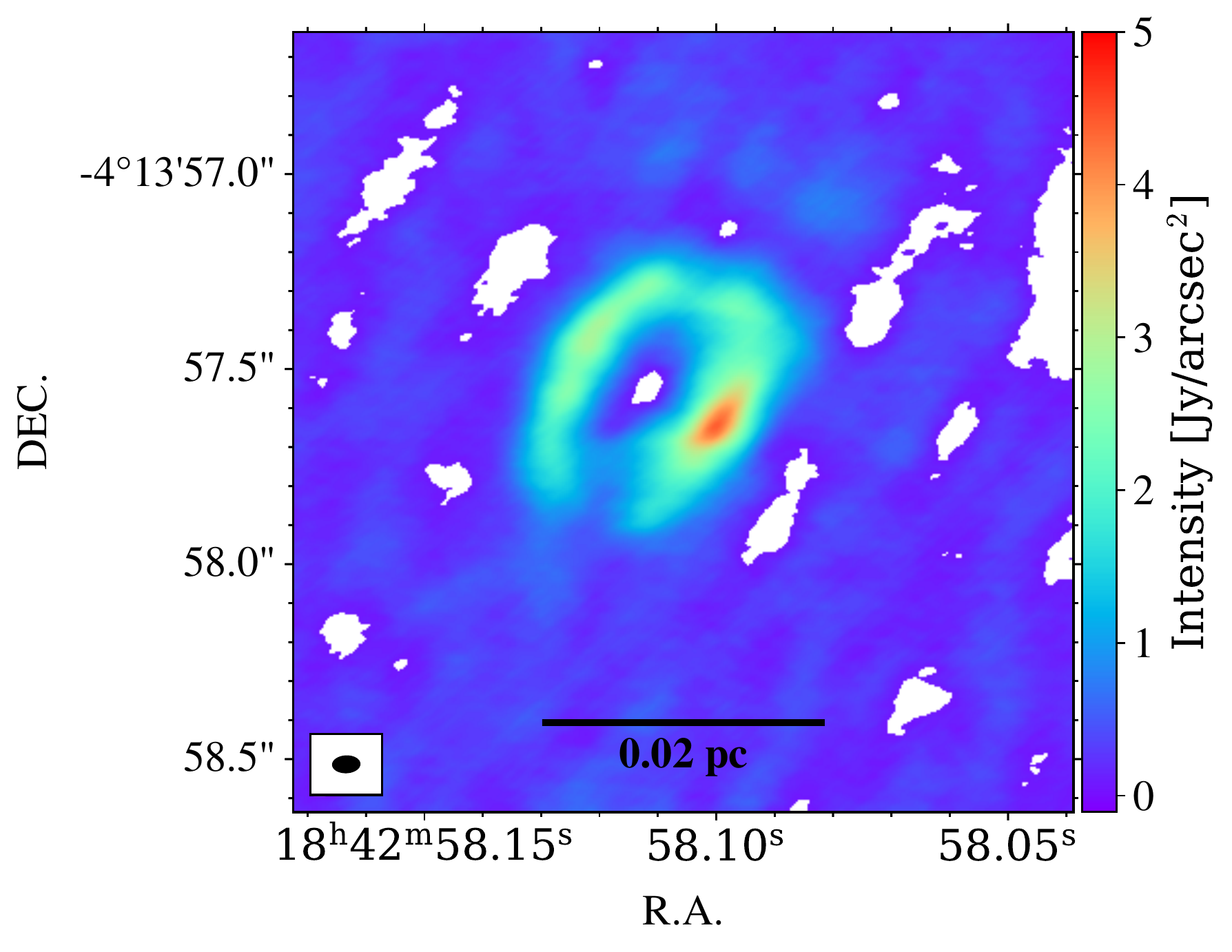}
\end{minipage}
\begin{minipage}{0.49\textwidth}
\includegraphics[width=\textwidth]{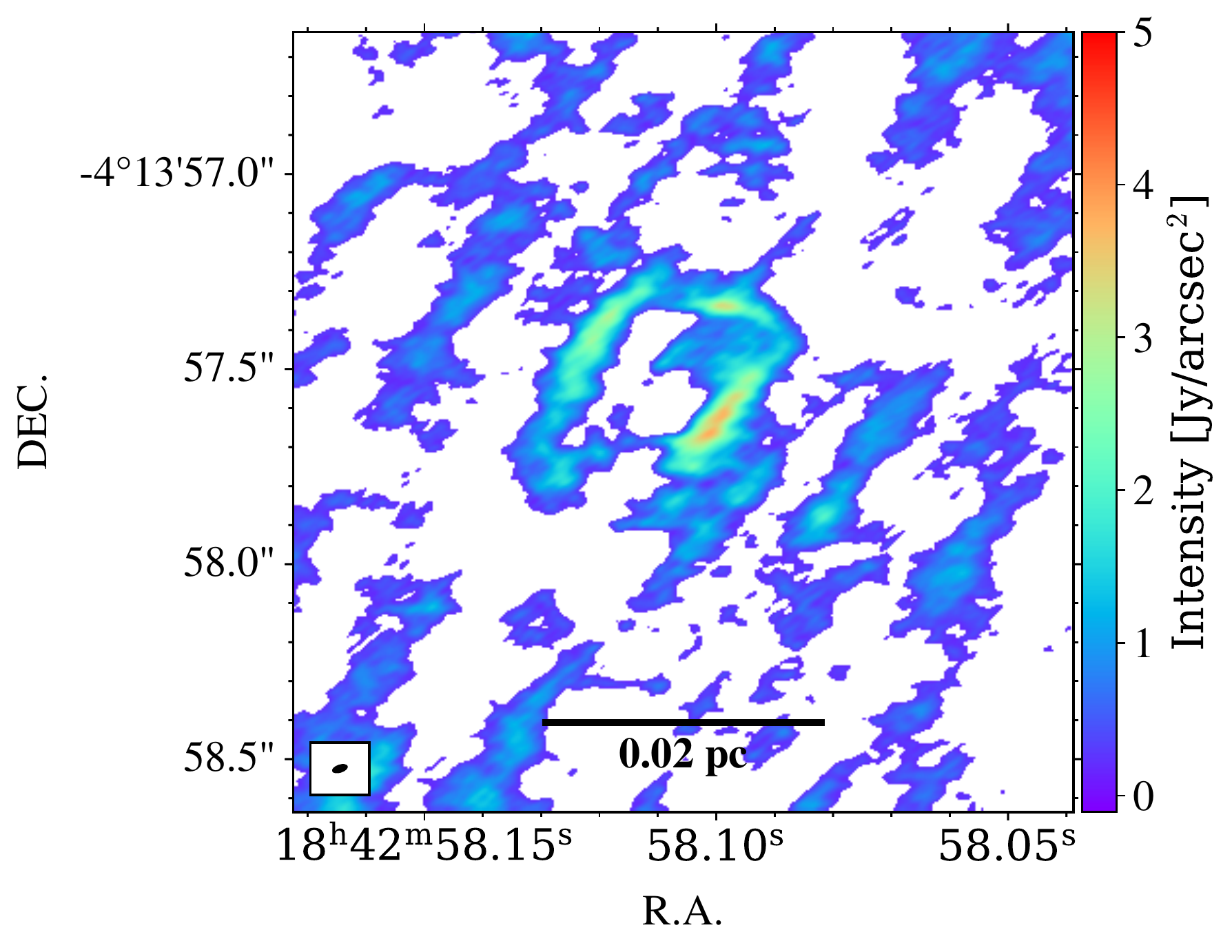}
\end{minipage}
\begin{minipage}{0.49\textwidth}
\includegraphics[width=\textwidth]{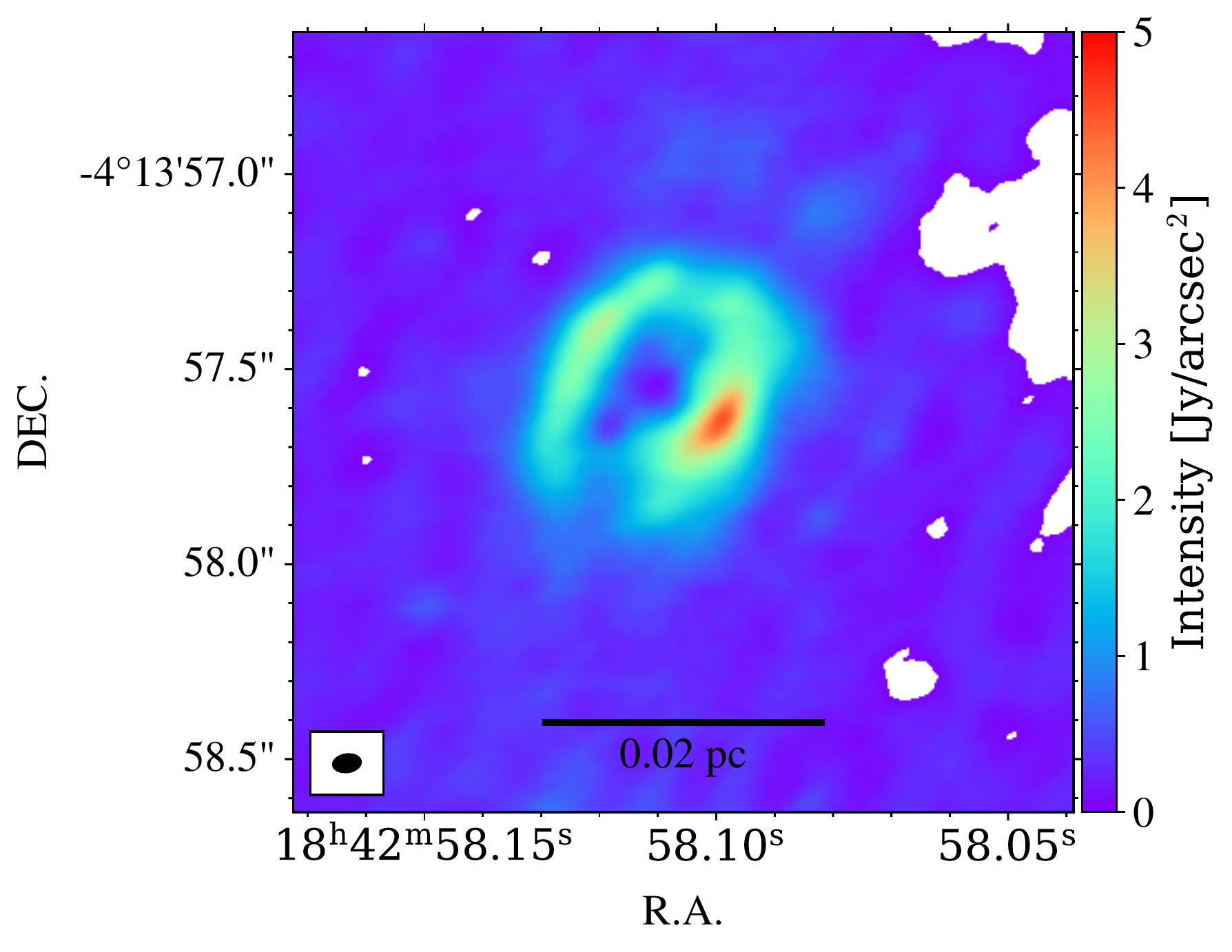}
\end{minipage}
\begin{minipage}{0.5\textwidth}
\includegraphics[width=\textwidth]{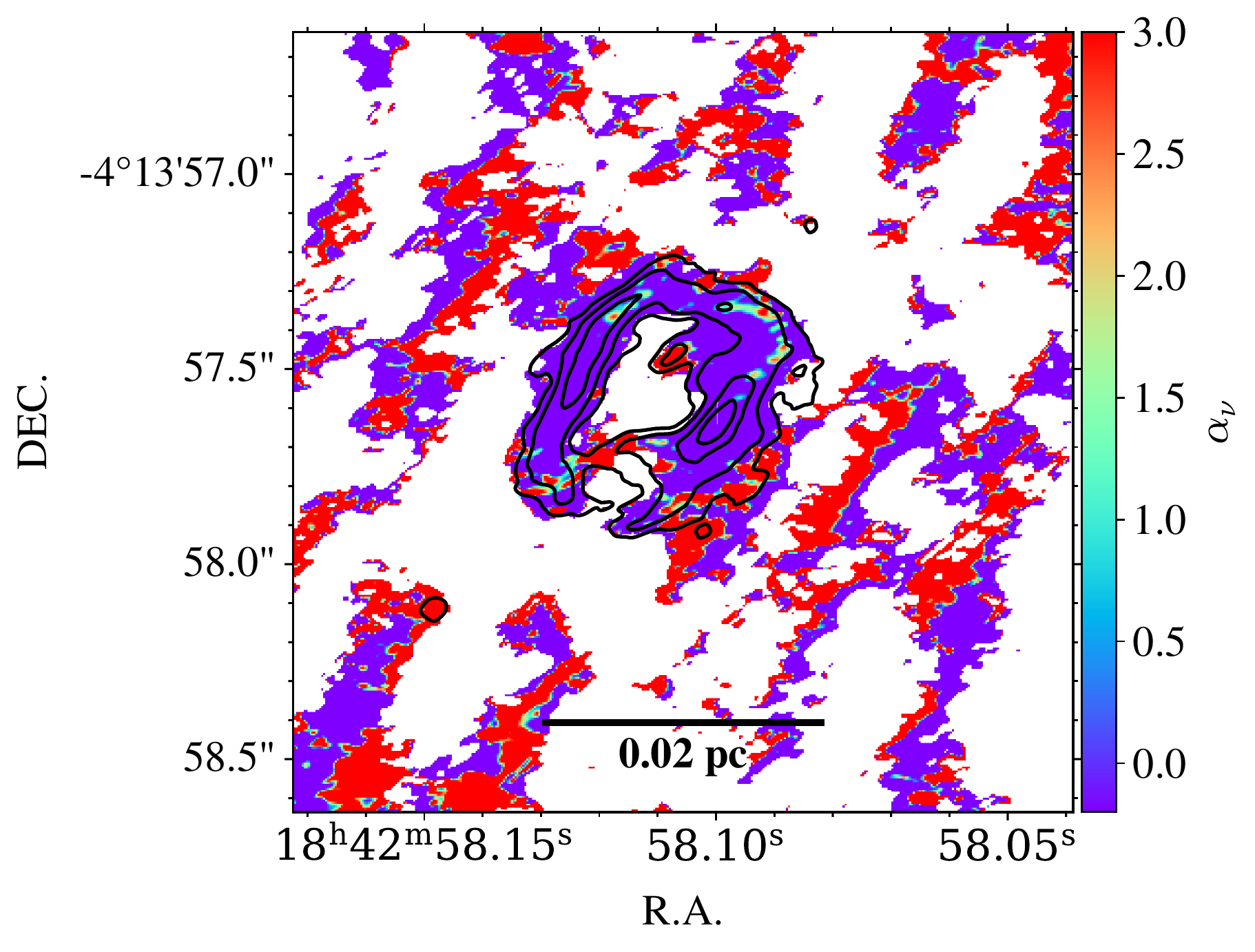}
\end{minipage}
\begin{minipage}{0.5\textwidth}
\includegraphics[width=\textwidth]{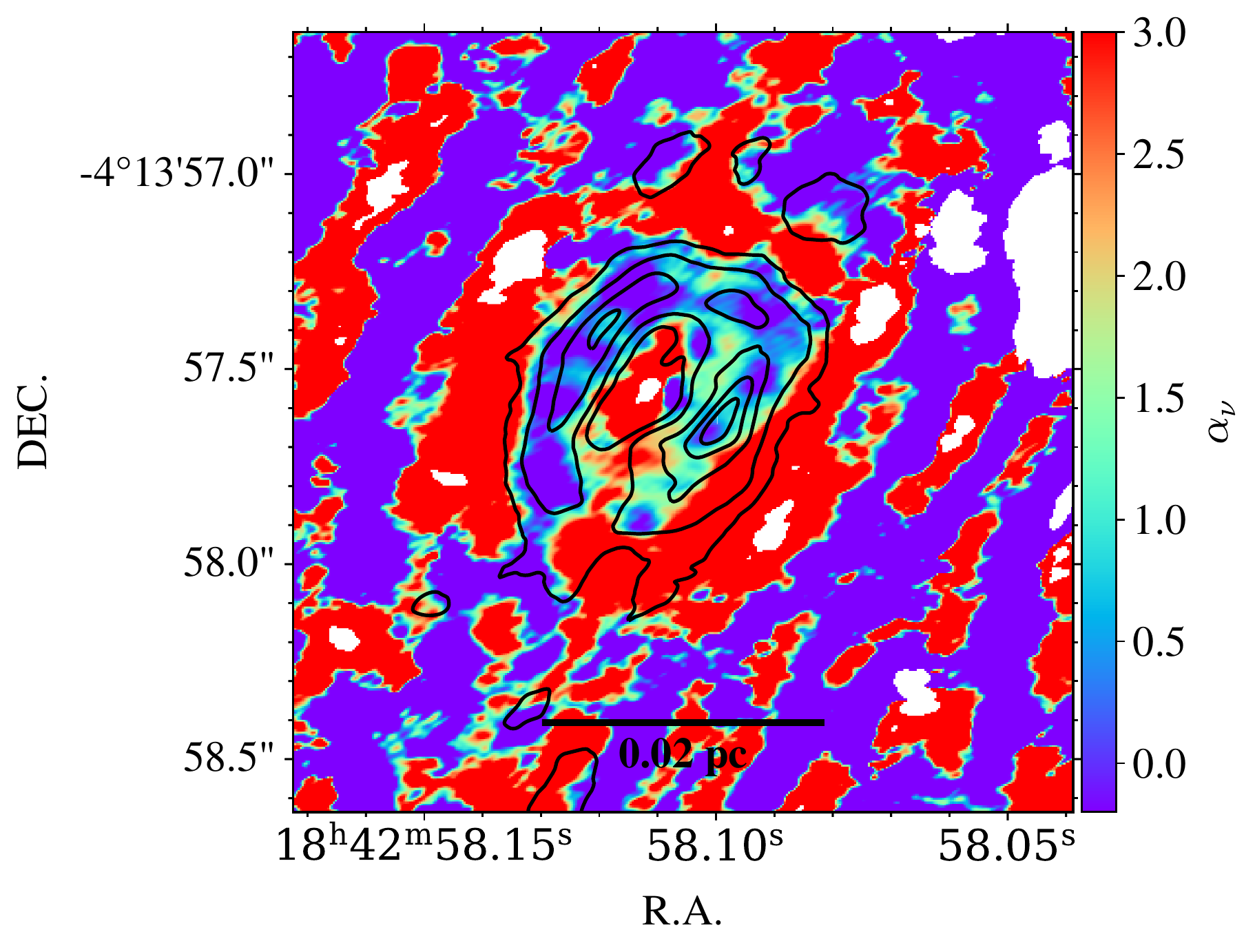}
\end{minipage}
\caption{{\it (a) Top left:} $216.575~$GHz (spw8) continuum image in E configuration only of the inner region of G28.20-0.05. The beam size is $0\farcs042 \times 0\farcs021$. {\it (b) Top right:} As (a), but now for C+I+E combined data. The beam size is $0\farcs073 \times 0\farcs046$. 
{\it (c) Middle left:} As (a), but for $233.926~$GHz (spw0). The beam size is $0\farcs048 \times 0\farcs027$. {\it (d) Middle right:} As (b), but for $233.926~$GHz (spw0). The beam size is $0\farcs077 \times 0\farcs051$. 
{\it (e) Bottom left:} E only spectral index map between spw8 and spw0, i.e., $\alpha_{\nu}=\log \left(I_{\nu_{1}} / I_{\nu_{2}}\right) / \log \left(\nu_{1} / \nu_{2}\right)$, where $\nu_{1}=216.575~\mathrm{GHz}$ and $\nu_{2}=233.926~ \mathrm{GHz}$. The overlaid continuum contours levels are the same as in Figure~\ref{fig:int_flux_VLA_ALMA}. {\it (f) Bottom right:} As (e), but for C+I+E combined data.}
\label{fig:int_flux_spw0_spw8}
\end{figure*}

If dust is contributing significantly, then we may expect local spatial variations to its contributions. In Figure~\ref{fig:int_flux_spw0_spw8} we present the continuum images (both for E-only and C+I+E) at 216.575~GHz and 233.926~GHz, i.e., from spectral windows 8 and 0, respectively. We also use these data to present in-band spectral index maps.

While the morphologies are generally quite similar, we notice a modest enhancement of the higher frequency emission in the vicinity of the main continuum peak. In this region of the ring, the in-band value of $\alpha_\nu$ has values $\sim 1$ to 2. Furthermore, the region just outside the ring also shows even larger values. These two features are consistent with those seen in the VLA to ALMA spectral index map. 

We thus draw the tentative conclusion that there is dust present in the inner region around the G28.20-0.05 protostar, especially around the main continuum peak and in surrounding regions outside of the ring. We will see below that there is evidence that the protostar is located at the position of the main continuum peak. However, these results motivate the need for high-resolution imaging at other frequencies, especially around $\sim 300$~GHz and higher to better confirm the presence of dust on these scales.

We proceed by making an approximate estimate of the 1.3~mm continuum flux from dust inside 0.5\arcsec. For this we simply take the difference in fluxes between 234~GHz and 217~GHz based on the in-band power-law fit, i.e., $0.94 - 0.75 = 0.19$~Jy. For optically thin dust emission, 1.3~mm flux density corresponds to a total (gas + dust) mass surface density of
\begin{equation}
\begin{aligned}
\Sigma_{\mathrm{mm}}=& 369 \frac{F_{\nu}}{\mathrm{Jy}} \frac{\left(1^{\prime \prime}\right)^{2}}{\Omega} \frac{\lambda_{1.3}^{3}}{\kappa_{\nu, 0.00638}} \\
& \times\left[\exp \left(0.111 T_{d, 100}^{-1} \lambda_{1.3}^{-1}\right)-1\right] \mathrm{g}\: \mathrm{cm}^{-2} \\
\rightarrow & 43.2 \frac{F_{\nu}}{\mathrm{Jy}} \frac{\left(1^{\prime \prime}\right)^{2}}{\Omega} \mathrm{g}\: \mathrm{cm}^{-2},
\end{aligned}\label{eq:dust}
\end{equation}
where $F_{\nu}$ is the total integrated flux over solid angle $\Omega$, $\kappa_{0.00638}$ is the dust absorption coefficient normalised to $0.00638~{\rm cm^{2}\:g}^{-1}$ \citep[e.g.,][]{2018ApJ...853..160C}. This fiducial value has been derived assuming an opacity per unit dust mass of $0.899~{\rm cm^{2}\:g}^{-1}$ \citep[i.e., from the moderately coagulated thin ice mantle model of][]{1994A&A...291..943O} and a gas-to-refractory-component-dust ratio of 141 \citep[][]{2011piim.book.....D}. We note that the mass surface density is sensitive to the temperature of the dust, with the above value normalized to a typical hot core temperature of 100\:K (see \S\ref{sec:hot_core_lines}), i.e., $T_{d,100}\equiv T_d/100\:{\rm K}=1$. The appropriate value of $T_d$ that should be used for this estimate is quite uncertain. Indeed, in reality there will be a range of temperatures along a given line of sight. From the types of hot core lines detected in the system (see \S\ref{sec:hot_core_lines}) and anticipating that gas and dust temperatures are well coupled in these high density conditions, we consider that a factor of two uncertainty in average line of sight temperature is reasonable. In this case, values of $T_d = 50$~K and 200~K would change the coefficient in equation (\ref{eq:dust}) by factors of 2.12 and 0.486, respectively. 

Applying equation~(\ref{eq:dust}) to the inner 0.5\arcsec\ circular aperture of G28.20-0.05, i.e., with $\Omega = 0.785\:{\rm arcsec}^2$, we estimate $\Sigma_{\rm mm} = 10.5\:{\rm g\:cm}^{-2}$ (averaged over this region). This corresponds to a total (gas + dust) mass of $30.3\:M_\odot$. 
If $T_d$ is in the range from 50 to 200~K, the mass would thus be in the range from about 60 to 15~$M_\odot$. 

\begin{figure*}
    \centering
        \includegraphics[width=\textwidth]{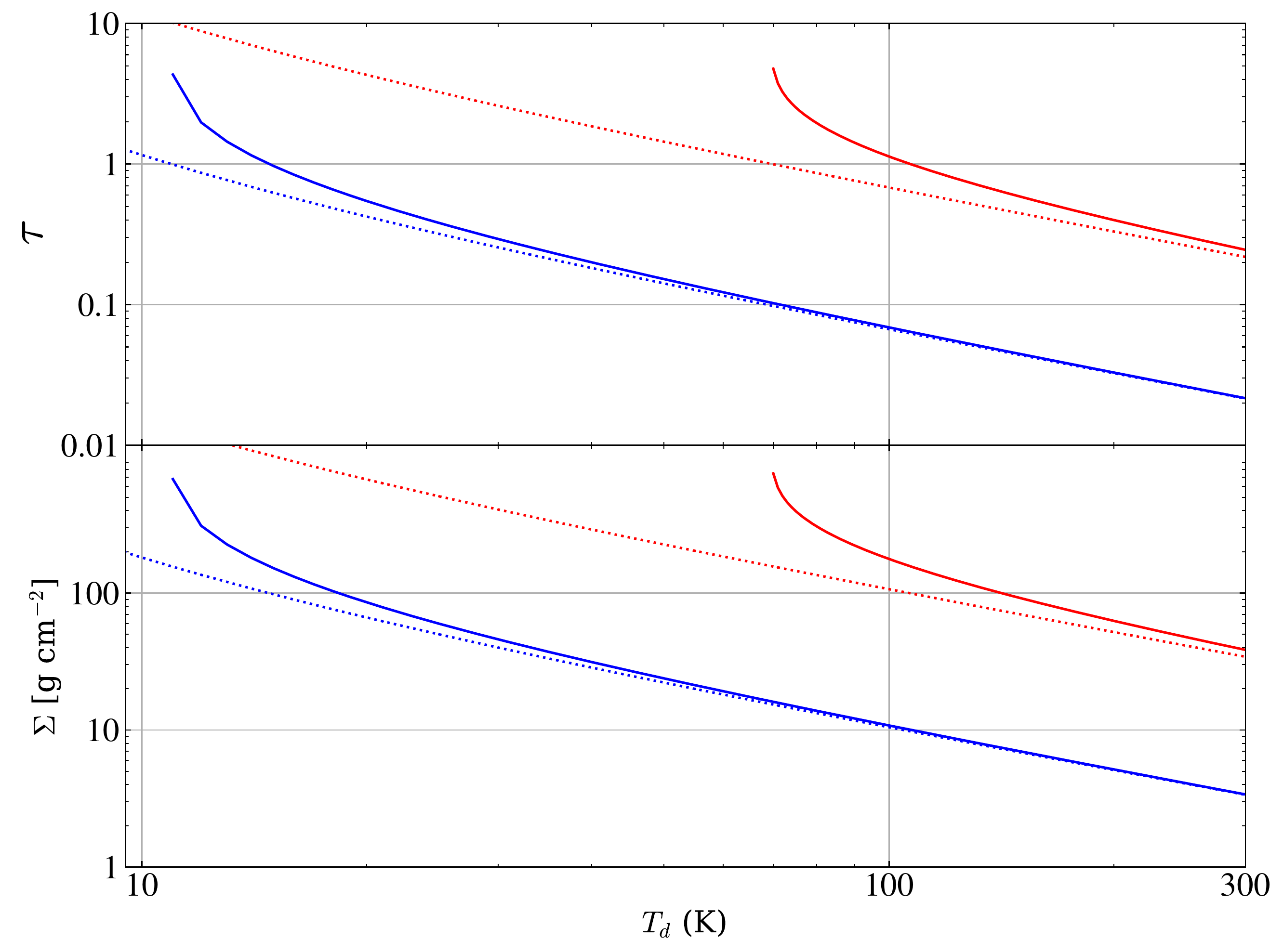}
     \caption{{\it (a) Top:} Optical depth ($\tau$) of inner-scale dust around G28.20-0.05, i.e., within a projected radius of 0.5\arcsec\ (2,850~au), versus assumed dust temperature. The blue dotted line shows $\tau$ evaluated from a uniform face-on slab of material in this region, with its mass surface density estimated assuming 0.19~Jy is due to optically thin dust emission at 1.3~mm. The solid blue line shows the equivalent $\tau$, but allowing for optical depth in the slab. The red dotted and solid lines show the equivalent cases when assuming this 1.3~mm continuum emission is concentrated in a 10 times smaller area, e.g., a uniform slab of radius 900~au.
     {\it (b) Bottom:} As (a), but now showing the implied mass surface densities, $\Sigma_{\rm mm}$, of the slabs.
}
\label{fig:tau_sigma}
\end{figure*}

It is possible that the optically thin assumption used for these mass estimates is not valid. To examine this possibility, we evaluate the dust optical depth $\tau = \kappa_{\nu} \Sigma_{\rm mm}$. For our fiducial estimate of $\Sigma_{\rm mm} = 10.5\:{\rm g\:cm}^{-2}$, we have $\tau = 0.067$, which implies the optically thin approximation is valid. As shown in Figure~\ref{fig:tau_sigma}, only if the dust temperature is as low as $\sim 20\:$K does $\tau$ start to become significant. However, as discussed below, such low temperatures are not expected to be realistic for this region that is so close to a massive protostar. On the other hand, these estimates assume the dust is spread out uniformly over the 0.5\arcsec\ scale region. The actual distribution is likely to show some spatial concentration and thus involves higher values of $\Sigma_{\rm mm}$. If the 0.19~Jy emission from dusty gas is concentrated in a region of 10 times smaller area, then Figure~\ref{fig:tau_sigma} shows that $\tau \sim 1$ for $T\lesssim 100\:$K and the method using the optically thin assumption would underestimate the mass by a significant factor. We will return to this mass estimate in \S\ref{sec:hot_core_lines} in the context of a dynamical mass estimate of the region.

\subsection{Hot Core Environment\label{sec:hot_core_lines}}

\begin{figure*}[t]
\begin{minipage}{0.49\textwidth}
\includegraphics[width=\textwidth]{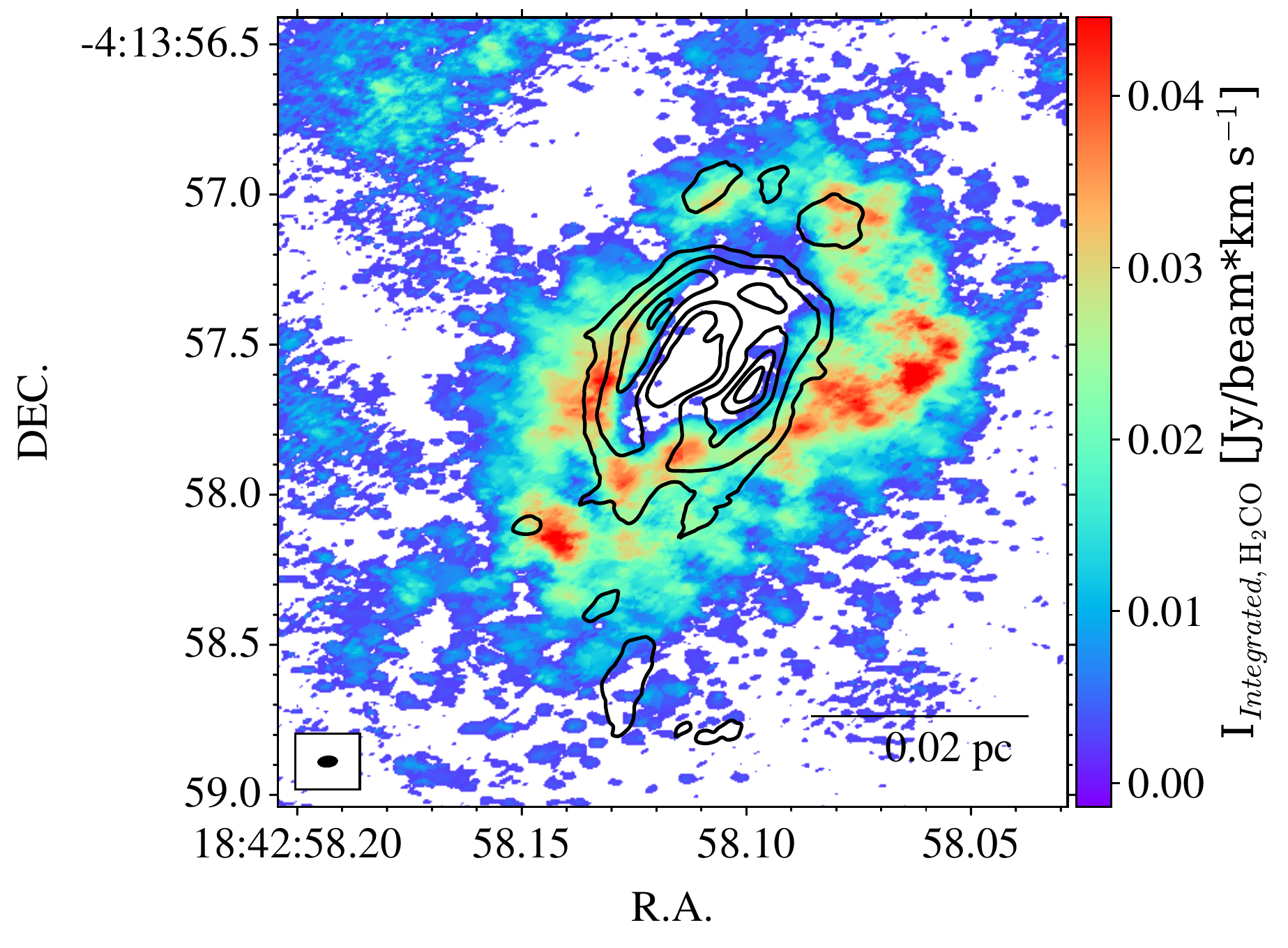}
\end{minipage}
\begin{minipage}{0.49\textwidth}
\includegraphics[width=\textwidth]{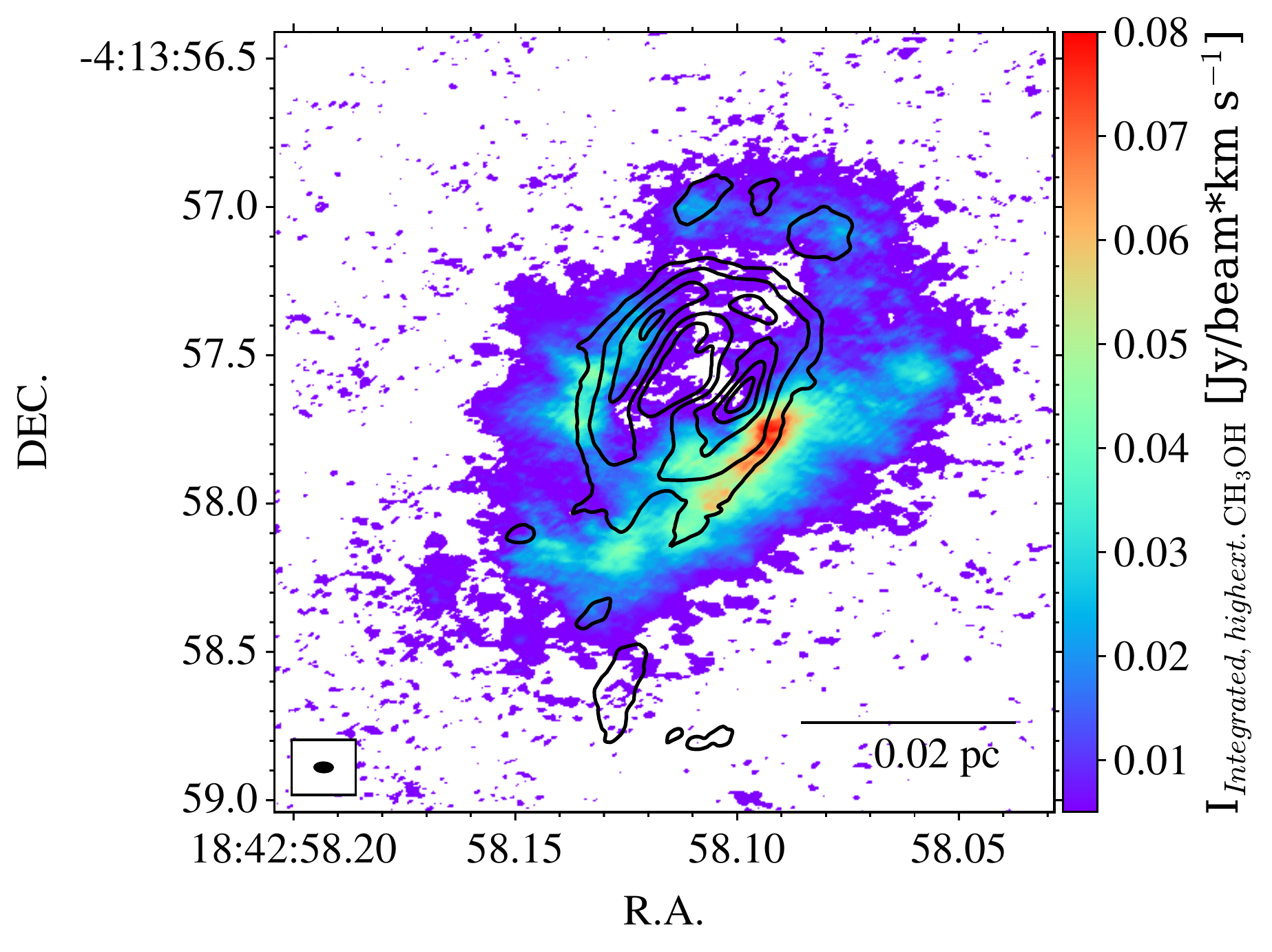}
\end{minipage}
\begin{minipage}{0.49\textwidth}
\centering
\includegraphics[width=\textwidth]{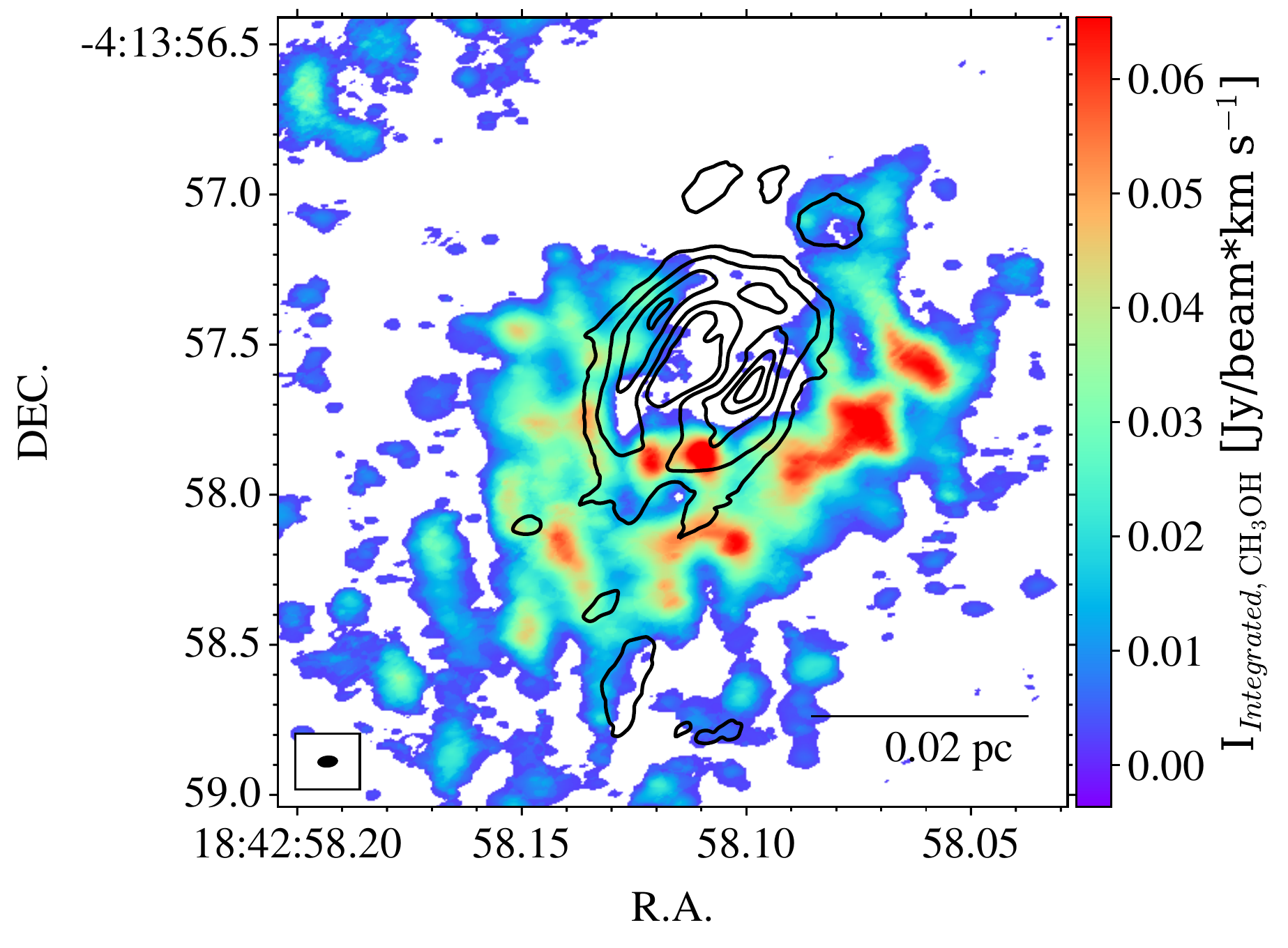}
\end{minipage}
\begin{minipage}{0.49\textwidth}
\includegraphics[width=\textwidth]{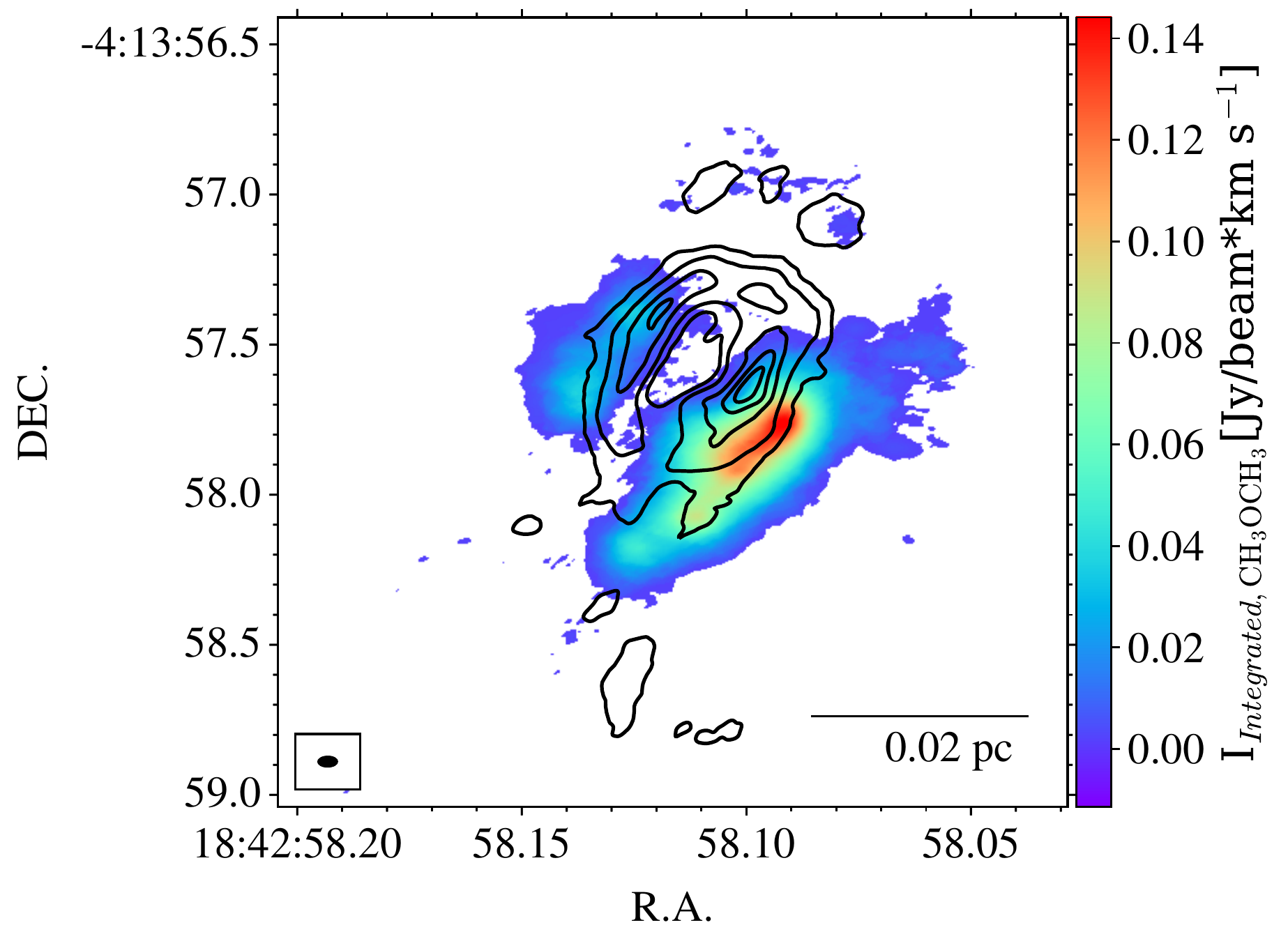}
\end{minipage}
\caption{{\it (a) Top left:} Moment 0 map of H$_{2}$CO$(\rm{3_{2,1}-2_{2,0}}$, $E_{\rm up} = 68.1\:$K) only considering cells above $1\sigma$ of the spectral rms measured over representative emission free channels. The black contours show the 1.3~mm continuum emission (0.5, 1, 1.5, 2, 2.5, 3, 3.5, 4, 4.5, 5, 5.5, 6 Jy/arcsec$^{2}$).
{\it (b) Top right:} As (a), but for H$_{2}$CO$(\rm{9_{1,8}-9_{1,9}}$, $E_{\rm up} =174\:$K).
{\it (c) Bottom left:} As (a), but for CH$_3$OH$(\rm{4_{-2,3}-3_{-1,2}}$, $E_{\rm up} =45.46\:$K).
{\it (d) Bottom right:} As (a), but for CH$_{3}$OCH$_{3}(\rm{22_{4,19} - 22_{3,20}}$, $E_{\rm up} =253.41\:$K).
}
\label{figure:h2co-ch3oh-moment0}
\end{figure*}

The \alma spectral set up includes various molecular lines (see Table~\ref{table:observ_sum_spectral}). The structure and kinematics of hot molecular core emission lines that trace dense and warm molecular gas can also be used to characterize the protostar.
Figure~\ref{figure:h2co-ch3oh-moment0} shows moment 0 maps of a lower excitation line of H$_{2}$CO($\rm{3_{2,1}-2_{2,0}}$, $E_{\rm up} = 68.1\:$K), a higher excitation line of H$_{2}$CO($\rm{9_{1,8}-9_{1,9}}$, $E_{\rm up} =174\:$K), CH$_3$OH$(\rm{4_{-2,3}-3_{-1,2}}$, $E_{\rm up} =45.46\:$K), and CH$_{3}$OCH$_{3}(\rm{22_{4,19} - 22_{3,20}}$, $E_{\rm up} =253.41\:$K).
We see that the two higher excitation species are concentrated in a region that is close to and overlapping with the main mm continuum peak, but with a slight offset of about 0.2\arcsec\ (i.e., $\sim 1,000$~au). Some emission from these species is also seen extending around and just exterior to the mm continuum ring. The two lower excitation species have a more extended distribution with their strongest emission just exterior to the mm continuum ring. These results indicate that there is dense, warm molecular gas present just outside the ring, but also even hotter gas near the main mm continuum peak and likely to be heated by a source at this location. The upper state energies of these transitions indicate that a typical value of $\sim 100\:$K for the dust temperature used above is a reasonable choice in the inner 0.5\arcsec\ scale region.



In addition, other molecules are detected in the spectra of G28.20-0.05, both relatively simple, such as SO$_{2}(\rm{22_{2,20}-22_{2,21}}$, $E_{\rm up} =248.44\:$K), H$_{2}$S$(\rm{2_{2,0}-2_{1,1}}$, $E_{\rm up} =83.98\:$K), to more complex hot core lines, such as 
C$_{2}$H$_{5}$CN($\rm{27_{1,27} - 26_{1,26}}$, $E_{\rm up} =157.73\:$K). Thus, G28.20-0.05 appears to be a relatively chemically rich massive protostar, e.g., compared to other sources studied with the same spectra set up, such as IRAS 07299-1651 \citep[][]{2019NatAs...3..517Z}, G339.88-1.26 \citep[][]{2019ApJ...873...73Z}, or G35.20-0.74N \citep[][]{2022arXiv220711320Z}. A detailed chemical characterisation of G28.20-0.05 will be presented in a companion paper to this one (Gorai et al., in prep.).

\begin{figure}[t]
\centering
\includegraphics[width=\columnwidth]{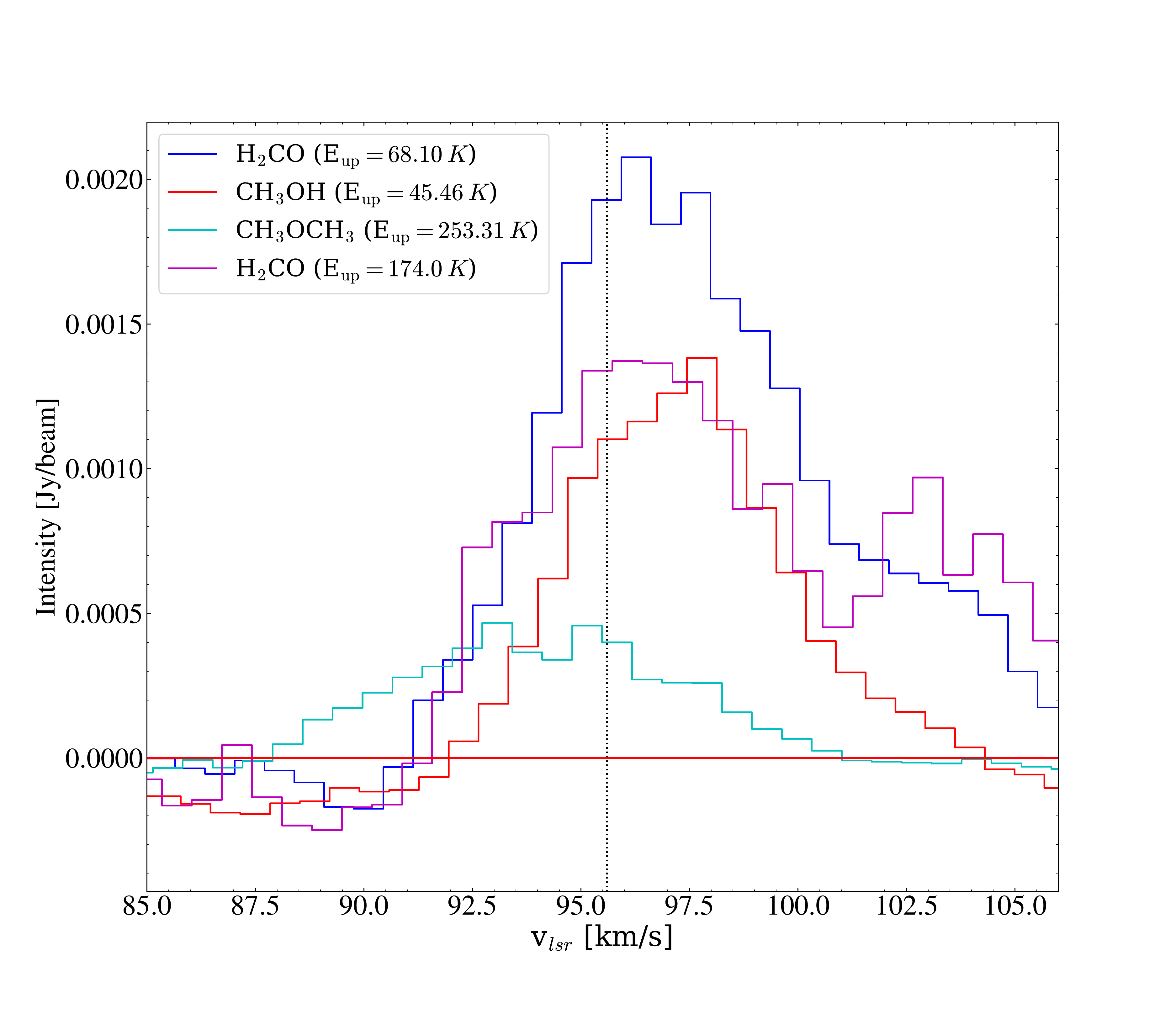}
\caption{Averaged H$_{2}$CO$(\rm{3_{2,1}-2_{2,0}}$, $E_{\rm up} = 68.1\:$K) (blue), CH$_3$OH$(\rm{4_{-2,3}-3_{-1,2}}$, $E_{\rm up} =45.46\:$K) (red), H$_{2}$CO$(\rm{9_{1,8}-9_{1,9}}$, $E_{\rm up} =174\:$K) (magenta), and CH$_{3}$OCH$_{3}(\rm{22_{4,19} - 22_{3,20}}$, $E_{\rm up} =253.41\:$K) (cyan) spectrum of G28.20-0.05 over an aperture $3^{\prime\prime}$ radius. We notice the lines both show good Gaussian-like single peak shape. The black dotted line shows the systemic velocity from the literature at $95.6~{\rm km~s}^{-1}$, which is consistent to the peaks of both lines. Hence, we adopt the literature value as the systemic velocity of the source and used through this work. } 
\label{figure:systemic_velocity}
\end{figure}

In Figure~\ref{figure:systemic_velocity} we show the average spectra of the four lines shown in Figure~\ref{figure:h2co-ch3oh-moment0}. These spectra exhibit a central main Gaussian peak, but with evidence of high-velocity line wings, especially to more redshifted velocities.
The lines peak at velocities close to the reported literature source systemic velocity of $95.6~{\rm km~s}^{-1}$ (see \S\ref{sec:intro}). Hence, we adopt this value as the systemic velocity of the source throughout this work. 

\begin{figure*}
    \centering
        \begin{minipage}{\textwidth}
        \includegraphics[width=0.32\textwidth]{G28_CH3OCH3_mom0.pdf}
        \includegraphics[width=0.32\textwidth]{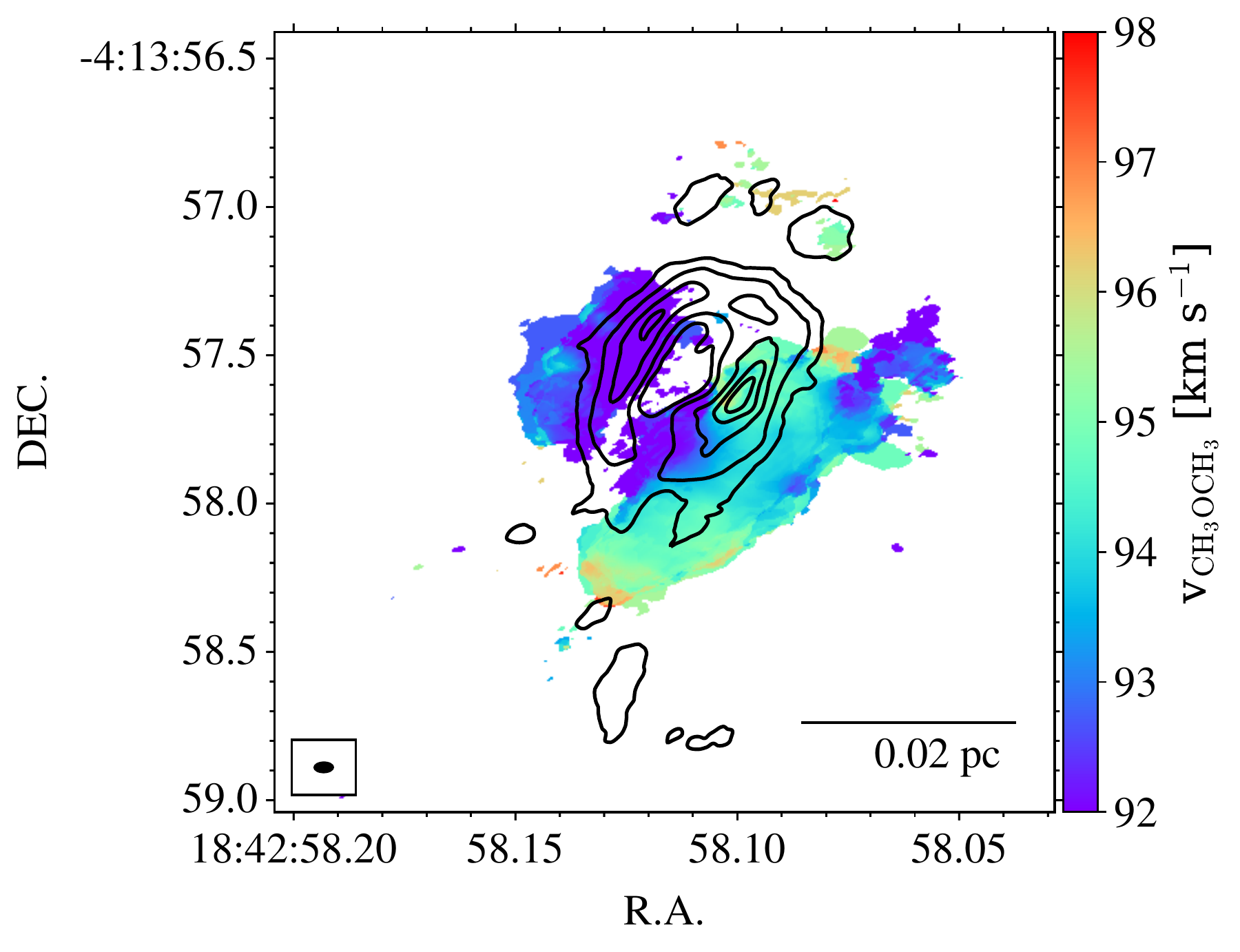}
        \includegraphics[width=0.32\textwidth]{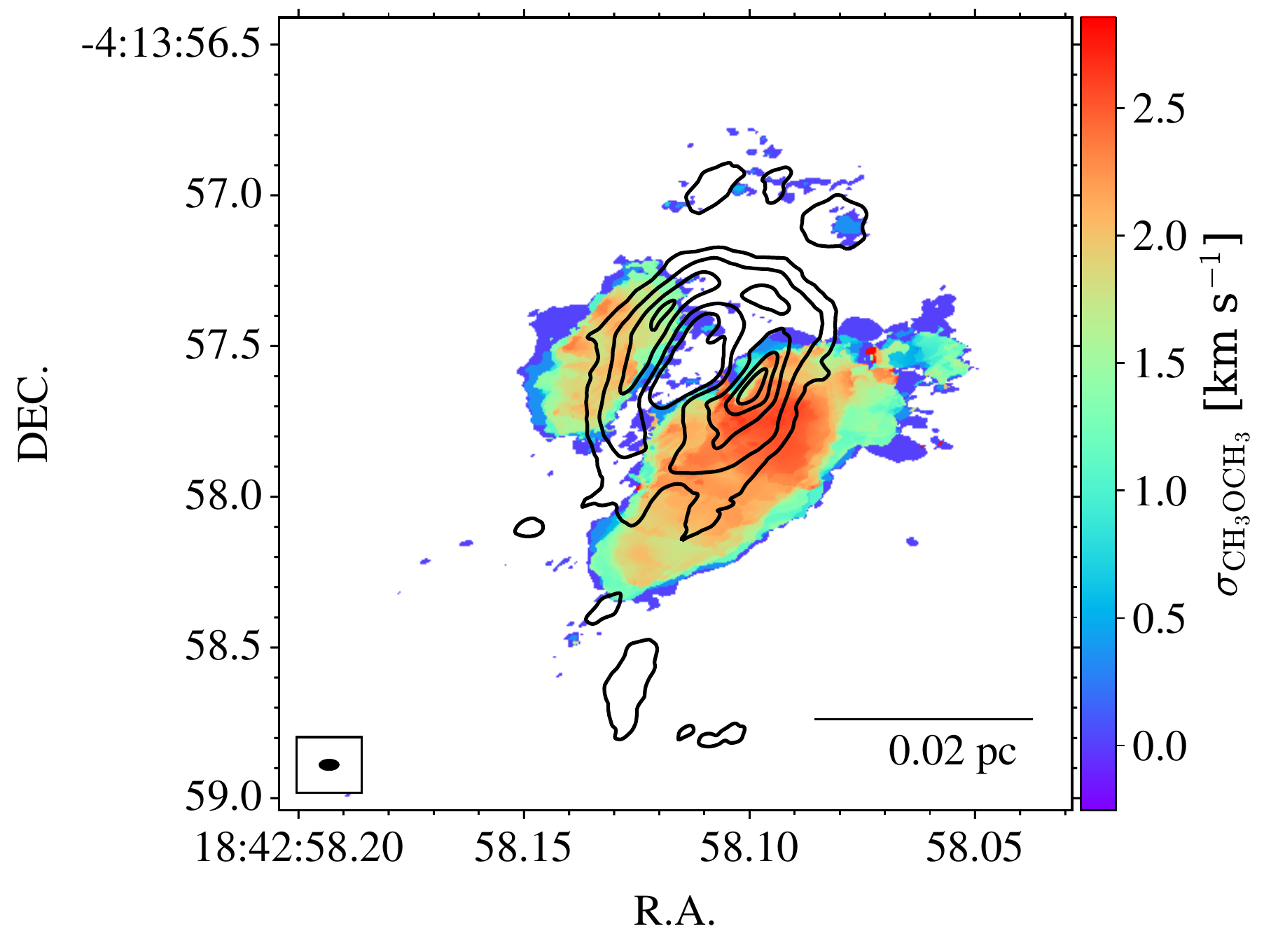}
        \end{minipage}
\caption{Moment 0, 1 and 2 maps (left to right) of CH$_{3}$OCH$_{3}(\rm{22_{4,19} - 22_{3,20}}$, $E_{\rm up} =253.41\:$K) emission, only including pixels that are above $1\sigma$ spectral rms (see Table 2 spw7).
The black contours show the 1.3~mm continuum emission (0.5, 1, 1.5, 2, 2.5, 3, 3.5, 4, 4.5, 5, 5.5, 6 Jy/arcsec$^{2}$).
}
\label{figure:CH3OCH3}
\end{figure*}

In Figure~\ref{figure:CH3OCH3} we present the moment 0, 1 and 2 maps of the CH$_{3}$OCH$_{3}$ line. The velocities near the main mm continuum peak are seen to be close to the systemic velocity of $+95.6\:{\rm km\:s}^{-1}$, but become blueshifted by several $\rm km\:s^{-1}$ as one moves around the ring. The moment 2 map, which shows the estimate of the 1D line of sight velocity dispersion, $\sigma$, exhibits values as high as 2.5~${\rm km\:s}^{-1}$ near the main mm continuum peak.


We now use the velocity dispersion of the highest excitation species, i.e., the CH$_{3}$OCH$_{3}$ line, to estimate a dynamical mass of the protostar assuming it traces virialized motions of a region extending out to radius, $R = 1,700$~au (0.3\arcsec). This radius is justified as being the approximate extent of the emission from the main mm continuum peak. The measured 1D velocity dispersion in this region is $\sigma = 2.95\:{\rm km\:s}^{-1}$. Thus the dynamical mass assuming simple virial equilibrium ignoring magnetic fields and surface pressure terms \citep[see, e.g.,][]{1992ApJ...395..140B} is 
\begin{equation}\label{eq:mdyn}
M_{\rm dyn} = 5 \sigma^2 R / G \simeq 84\:M_\odot,
\end{equation}
with this evaluation further assuming that the gas is distributed as an uniform sphere. We consider that the uncertainty in this mass estimate is at least $\sim 20\%$ due to a combination of kinematic distance uncertainty to the source, choice of radius of region traced by CH$_{3}$OCH$_{3}$ emission, and simplifying assumptions in application of the virial theorem to the region, such as density structure, surface pressure terms and effects of magnetic fields.
Nevertheless, we see that the dynamical mass estimate is comparable to the previous estimate of dusty gas mass (see \S\ref{sec:radio_mm_sed}), but is about a factor of two larger. The dynamical mass is expected to be larger since it probes the potential of the total mass enclosed in the region, i.e., of the gas and the protostar.



\subsection{H30$\alpha$ emission}

\begin{figure*}[t]
\begin{minipage}{0.49\textwidth}
\includegraphics[width=\textwidth]{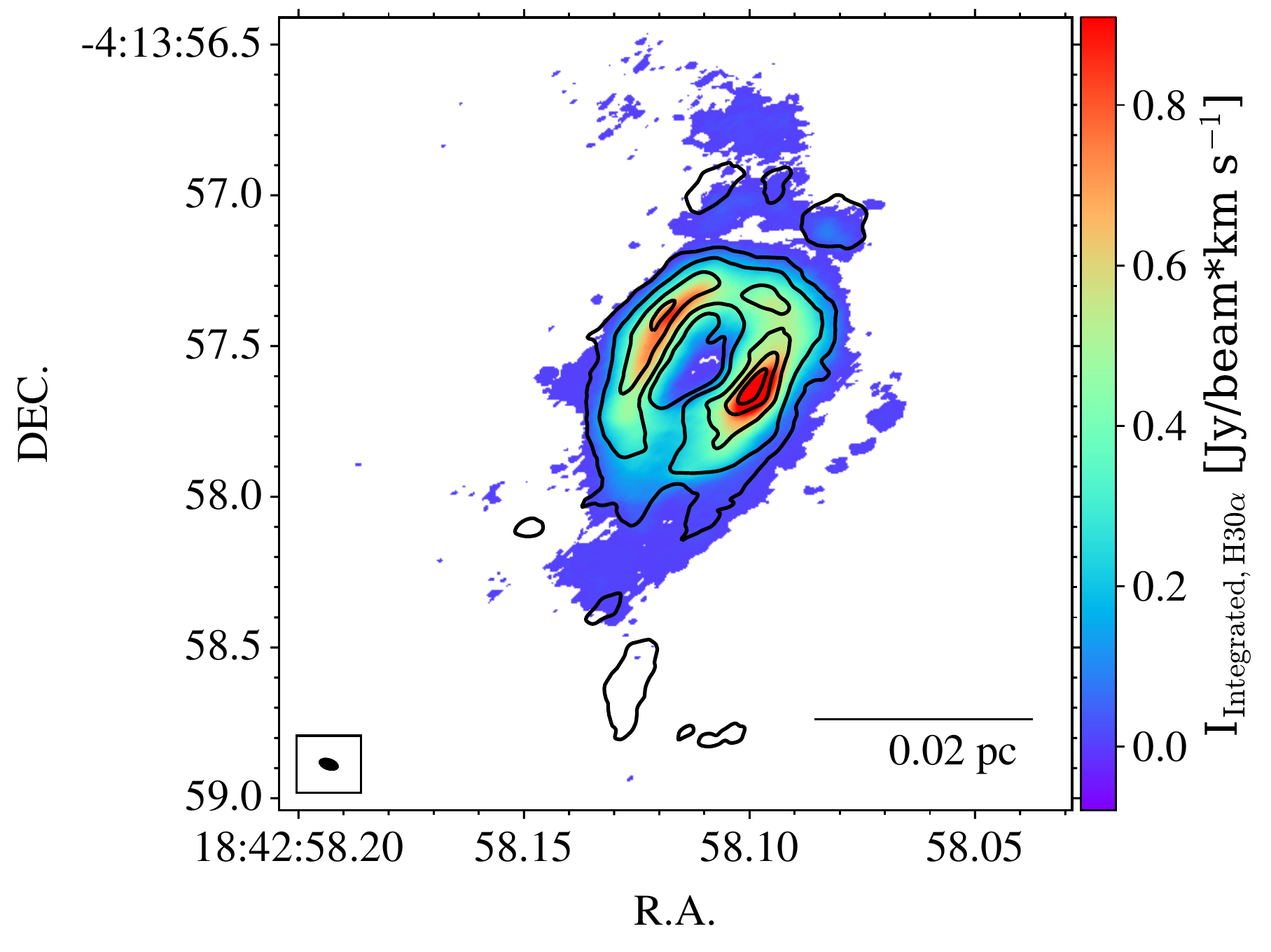}
\end{minipage}
\begin{minipage}{0.49\textwidth}
\includegraphics[width=\textwidth]{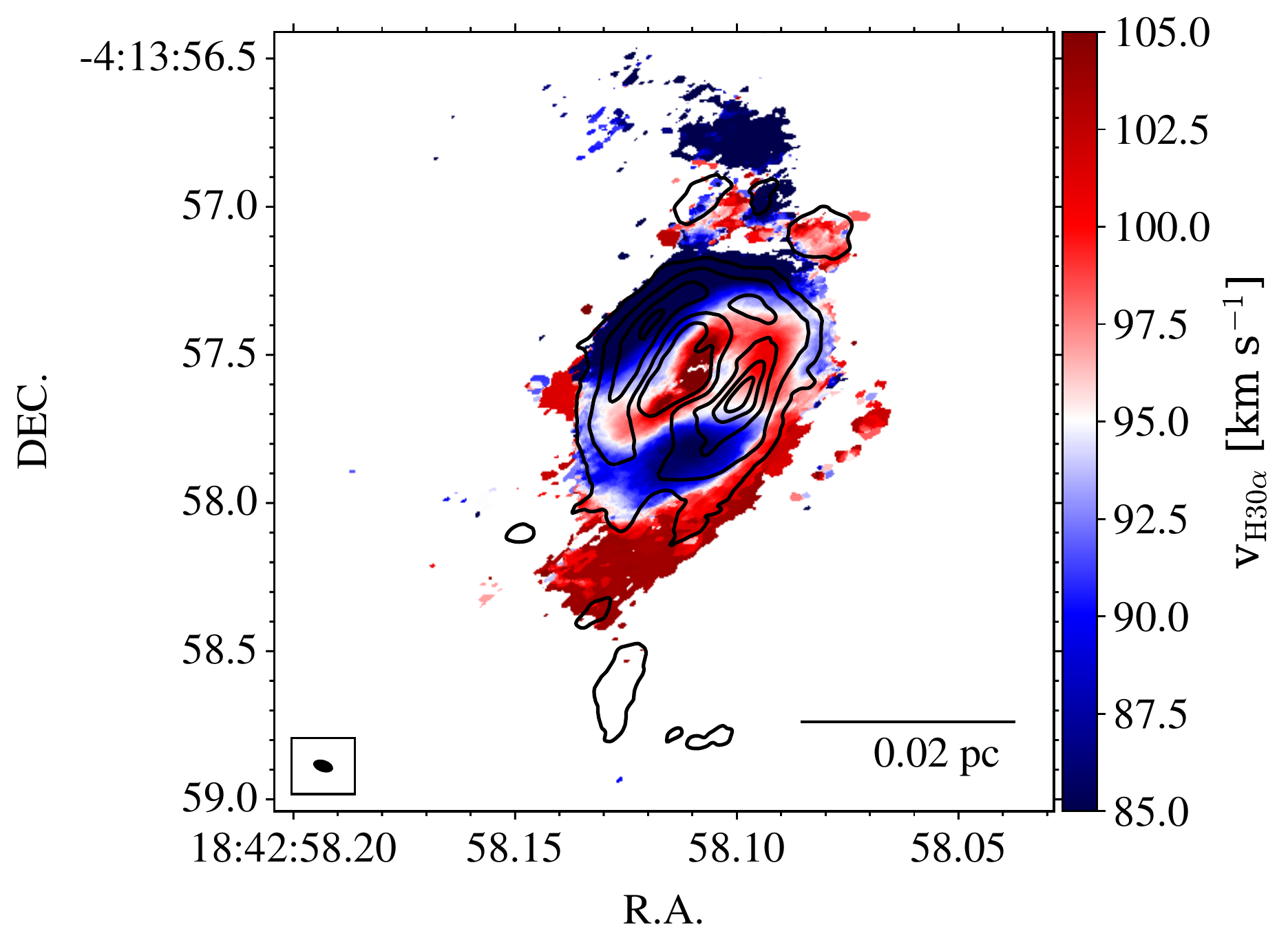}
\end{minipage}
\begin{minipage}{0.49\textwidth}
\includegraphics[width=\textwidth]{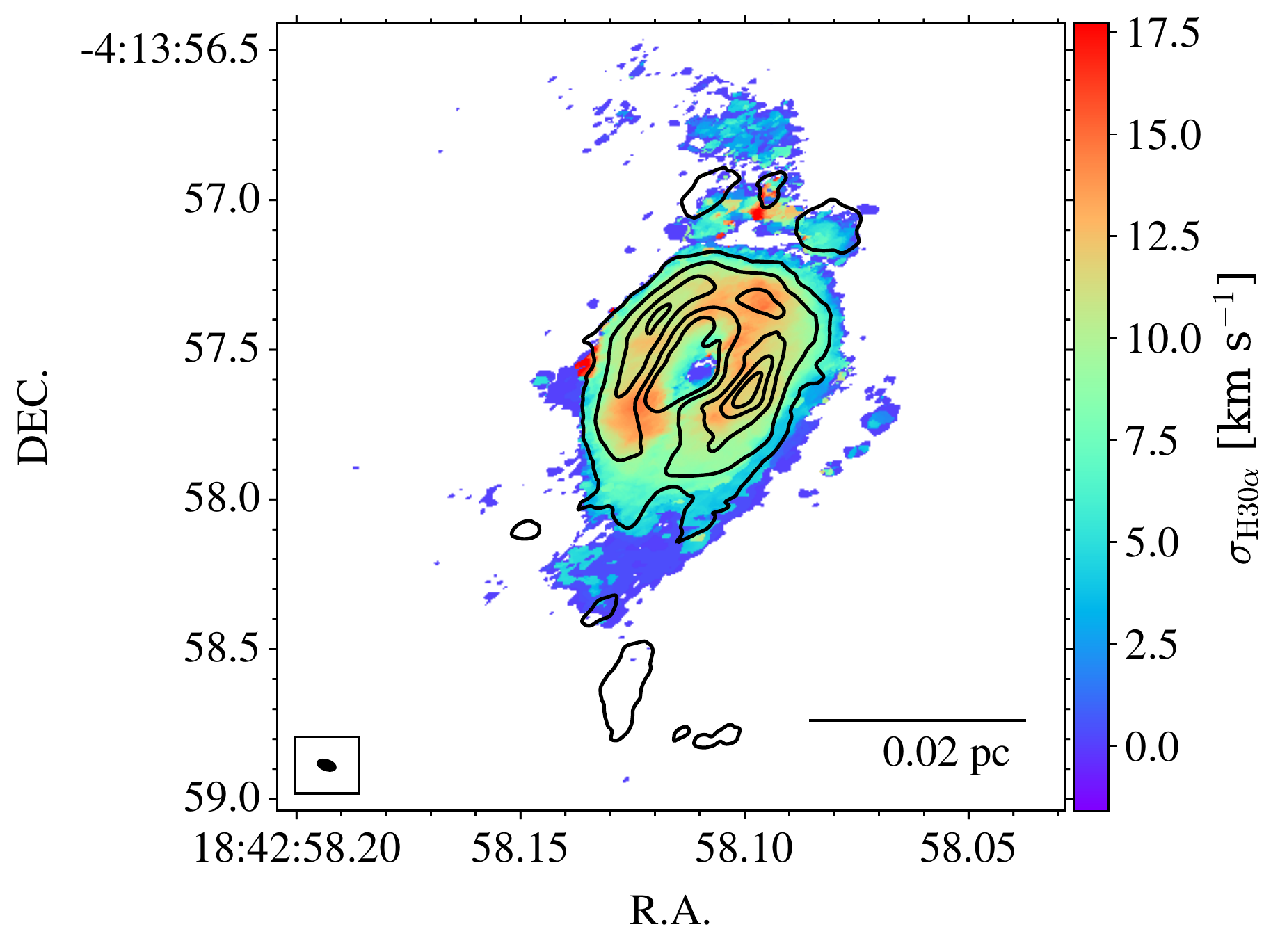}
\end{minipage}
\begin{minipage}{0.49\textwidth}
\includegraphics[width=\textwidth]{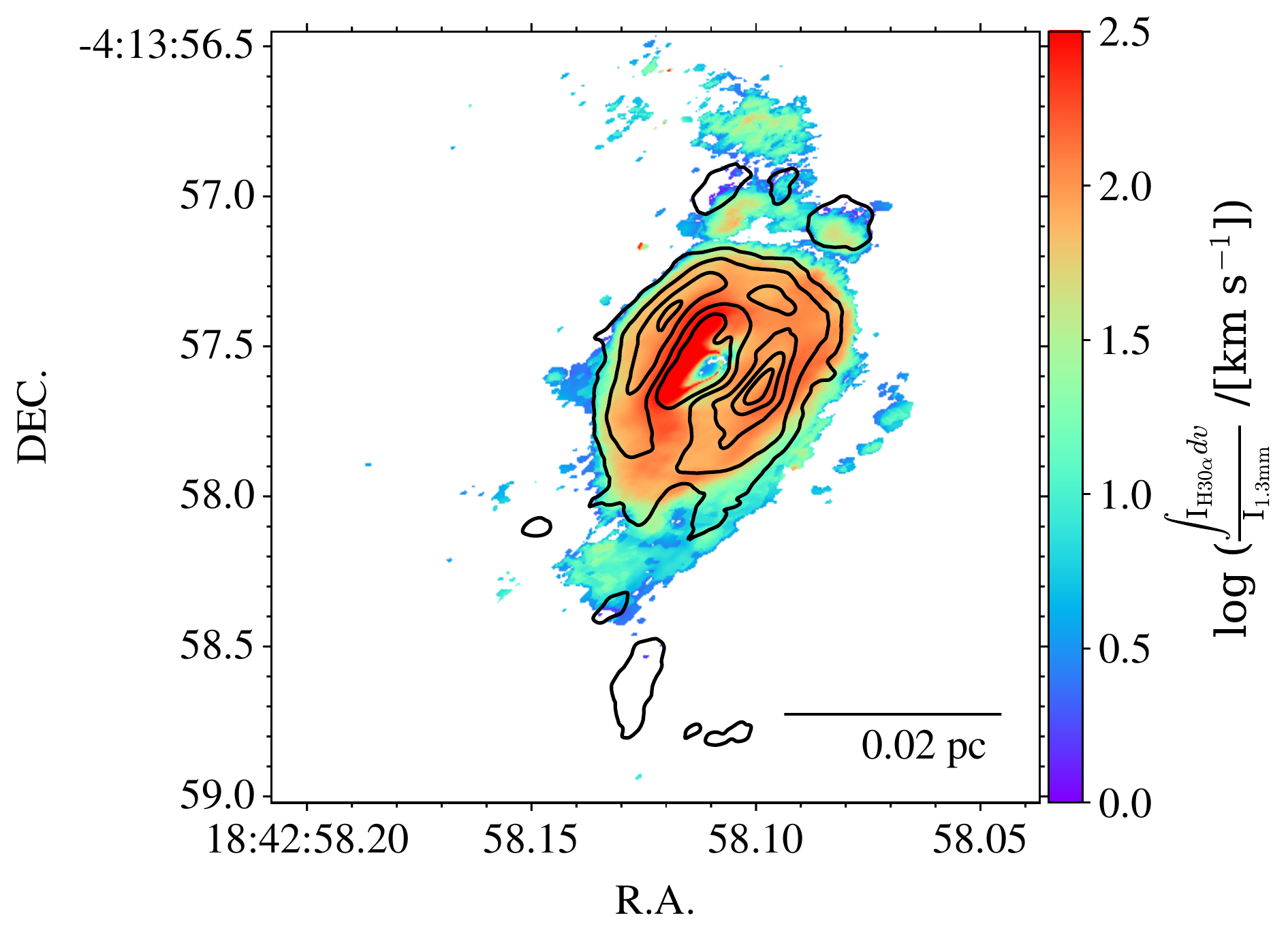}
\end{minipage}
\caption{{\it (a) Top left:} 
H$30\alpha$ moment zero map, only considering pixels with integrated intensity above $5\sigma$ of the spectral rms measured over emission free velocity channels in the averaged velocity spectrum 
(note, $1 \sigma \simeq 0.8\:{\rm mJy\:beam^{-1}}$). The black contours show the 1.3~mm continuum emission (0.5, 1, 1.5, 2, 2.5, 3, 3.5, 4, 4.5, 5, 5.5, 6 Jy/arcsec$^{2}$).  
{\it (b) Top right:} As (a), but now showing the H$30\alpha$ moment one map, i.e., averaged line-of-sight velocity.
{\it (c) Bottom left:} As (a), but now showing the H$30\alpha$ moment two map, i.e., the 1D velocity dispersion along the line-of-sight of the H$30\alpha$ emission.
{\it (d) Bottom right:} As (a), but now showing the ratio between the integrated intensity of H$30\alpha$ and the continuum intensity. 
}
\label{figure:h30alpha-moment0}
\end{figure*}

\begin{figure*}[t]
\centering
\includegraphics[width=\textwidth]{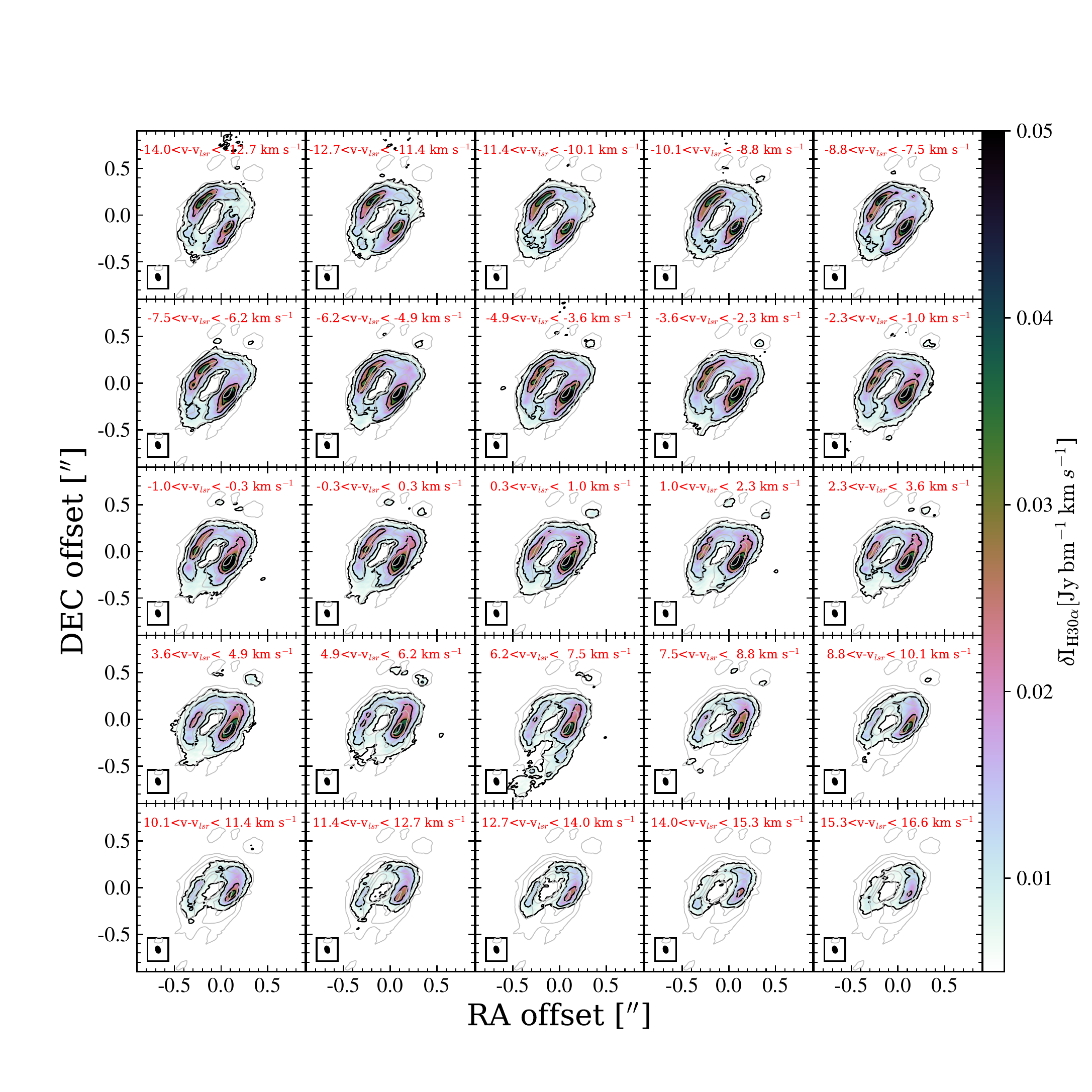}
\caption{Channel maps of  H$30\alpha$ emission based on  C+I+E configurations. Each panel shows a moment 0 map integrated over the labelled velocity range. The synthesized beam is shown at the lower left corner of each panel. The overlaid C+I+E continuum contours (gray contours) have intensities 0.5, 1, 1.5, 2, 2.5, 3, 3.5, 4, 4.5, 5, 5.5, 6 Jy/arcsec$^{2}$.} 
\label{fig:h30a-channelmap}
\end{figure*}

The \alma spectral set-up also includes the H$30\alpha$ recombination line that traces ionized gas. Figure~\ref{figure:h30alpha-moment0} presents the moment 0, 1 and 2 maps of H$30\alpha$, only including pixels with values five times larger 
than the root mean square noise measured from line-free channels in the neighborhood of the spectral line. Figure~\ref{fig:h30a-channelmap} presents channel maps of the H$30\alpha$ emission.
The moment 0 map shows a structure, including ring and extended NW-SE emission, that has close correspondence to the 1.3~mm continuum emission. 
This is additional evidence that a large fraction of the 1.3~mm continuum emission is tracing ionized gas, as already concluded from the 1.3~cm to 1.3~mm spectral index analysis.

The moment 1 map reveals a very strong velocity gradient towards the main mm continuum peak, which is also clearly seen in the channels maps (Figure~\ref{fig:h30a-channelmap}). At the location of the peak, the velocity is close to the $+95.6\:{\rm km\:s}^{-1}$ systemic velocity of the protostar inferred from molecular lines (see above). Then, in the direction of elongation of the mm continuum source there is an ordered, relatively smooth gradient to blueshifted velocities in the SE and redshifted velocities in the NW, with velocities differences of up to $\pm 10\:{\rm km\:s}^{-1}$ being observed. We will see later in \S\ref{sec:outflow} that this direction of the H30$\alpha$ velocity gradient is perpendicular to a large-scale CO outflow from the region. This fact suggests that rotation, either in a disk or in a disk wind, plays a role in setting this kinematic structure, which we discuss in more detail below. Other features seen in the moment 1 map include that the NE side of the ring and northern spur have blueshifted velocities, again by about $10\:{\rm km\:s}^{-1}$ from the systemic. The southern spur, which connects to the main mm continuum peak, shows redshifted velocities by up to about $10\:{\rm km\:s}^{-1}$ from the systemic.
The moment 2 map shows that 1D velocity dispersions can exceed $10\:{\rm km\:s}^{-1}$ in the ring, but have much lower values in the northern and southern spurs.



Figure~\ref{figure:h30alpha-moment0} bottom right panel presents a map of the ratio of H$30\alpha$ integrated intensity to 1.3~mm continuum.
The ratio between H$30\alpha$ integrated intensity and free-free continuum intensity for optically thin Local Thermodynamic Equilibrium (LTE) conditions is \citep[see, e.g.,][]{2019ApJ...886L...4Z}
\begin{equation}
   \frac{\int I_{\mathrm{H} 30 \alpha} d v}{I_{1.3 \mathrm{mm}}}  =  \frac{4.678 \times 10^{6}\:{\rm km\:s}^{-1}(T_{\rm e}/{\rm K})^{-1}}{ [1.5 {\rm ln} (T_{\rm e}/{\rm K})-8.443] (1+N_{\rm He^+}/N_{\rm H^+}) }.
\end{equation}
For a fiducial ionized gas temperature of $T_{\rm e} =10^{4}\:$K and $N_{\rm He^+}/N_{\rm H^+}=0.1$, 
we obtain a reference value for the ratio of $79\:{\rm km\:s}^{-1}$. If the temperature is as low as 5,000~K, then the value increases to about $200\:{\rm km\:s}^{-1}$.

Most of the H30$\alpha$ emitting region shows line-to-continuum ratios $\lesssim 200\:{\rm km\:s}^{-1}$, which could thus be consistent with optically thin LTE conditions. Some relatively low values could be due to the presence of dust contributions to the mm continuum, e.g, in the regions just outside the ring. 



\begin{figure}[t]
\centering
\includegraphics[width=\columnwidth]{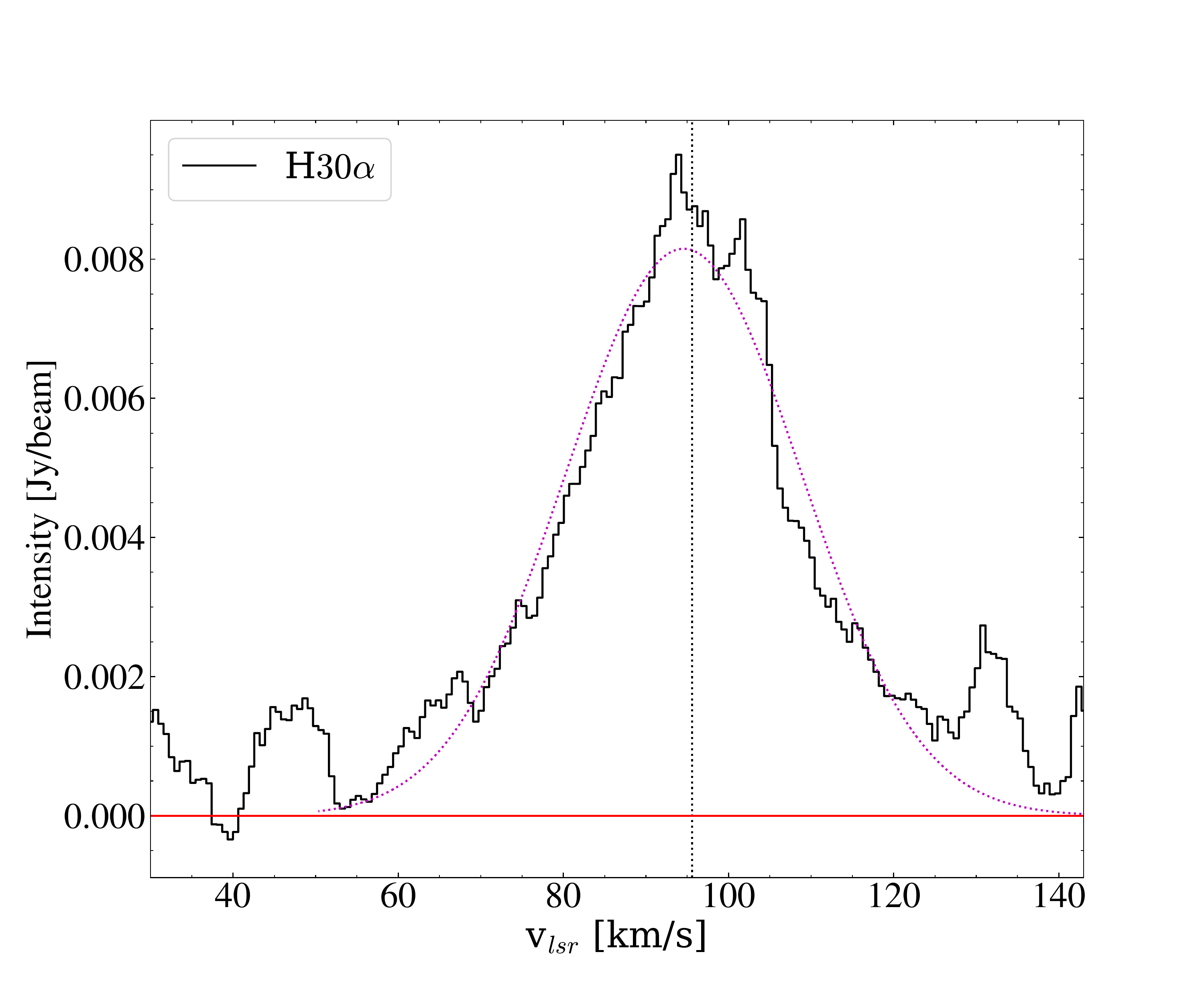}
\caption{H$30\alpha$ averaged spectrum toward a circular aperture of $0.3\arcsec$ radius from the main mm continuum peak.
The red dotted line is the Gaussian fit to the spectrum. The corresponding dispersion is $14.2~km~s^{-1}$. The vertical black dotted line represents the systemic velocity of the source.
} 
\label{fig:h30a-spectrum}
\end{figure}

We next consider if the implications of the kinematics of the H30$\alpha$ emission for the dynamics of the system. The spectrum of this emission extracted from a region with radius of 0.3\arcsec\ around the main mm continuum peak is shown in Figure~\ref{fig:h30a-spectrum}. If we attempt a virial analysis based on the velocity dispersion within this region, as was done above for CH$_{3}$OCH$_{3}$ emission, we find that the 1D velocity dispersion is 14.2~$\rm km\:s^{-1}$ and so $M_{\rm dyn}$ given by equation (\ref{eq:mdyn}) is $\simeq 2,000\:M_\odot$, i.e., $>20\times$ larger than that inferred from CH$_{3}$OCH$_{3}$. We conclude that the H30$\alpha$ emission is most likely to be tracing an ionized wind that is escaping from the massive protostar. For example, this may be the ionized base of a rotating magneto-centrifugally launched disk wind. Such winds typically achieve speeds of order the escape speed from their launching radii, $r_{\rm dw}$, i.e.,
\begin{eqnarray}
    v_{\rm w,esc} & = & (2 G m_{*d} / r_{\rm dw})^{1/2}\\
    & \rightarrow & 23.0 \left(\frac{m_{*d}}{30\:M_\odot}\right)^{1/2} \left(\frac{r_{\rm dw}}{100\:{\rm au}}\right)^{-1/2}\:{\rm km\:s^{-1}},\nonumber
\end{eqnarray}
where $m_{*d}$ is the mass of the star and disk enclosed within $r_{\rm dw}$. 
Inspection of Figure~\ref{fig:h30a-spectrum} reveals that there is high-velocity H30$\alpha$-emitting gas out to at least $30\:{\rm km\:s}^{-1}$ to both redshifted and blueshifted velocities. The models of rotating ionized disk winds of \citet[][e.g., see their Figures~15 and 16]{2016ApJ...818...52T} appear to be highly relevant to explain the general features of broad line-width with the large-scale velocity gradient that we see in the H30$\alpha$ emission from G28.20-0.05.

\subsection{Outflows traced by CO, SiO and NIR emission}\label{sec:outflow}

\begin{figure*}[t]
\begin{minipage}{0.49\textwidth}
\includegraphics[width=\textwidth]{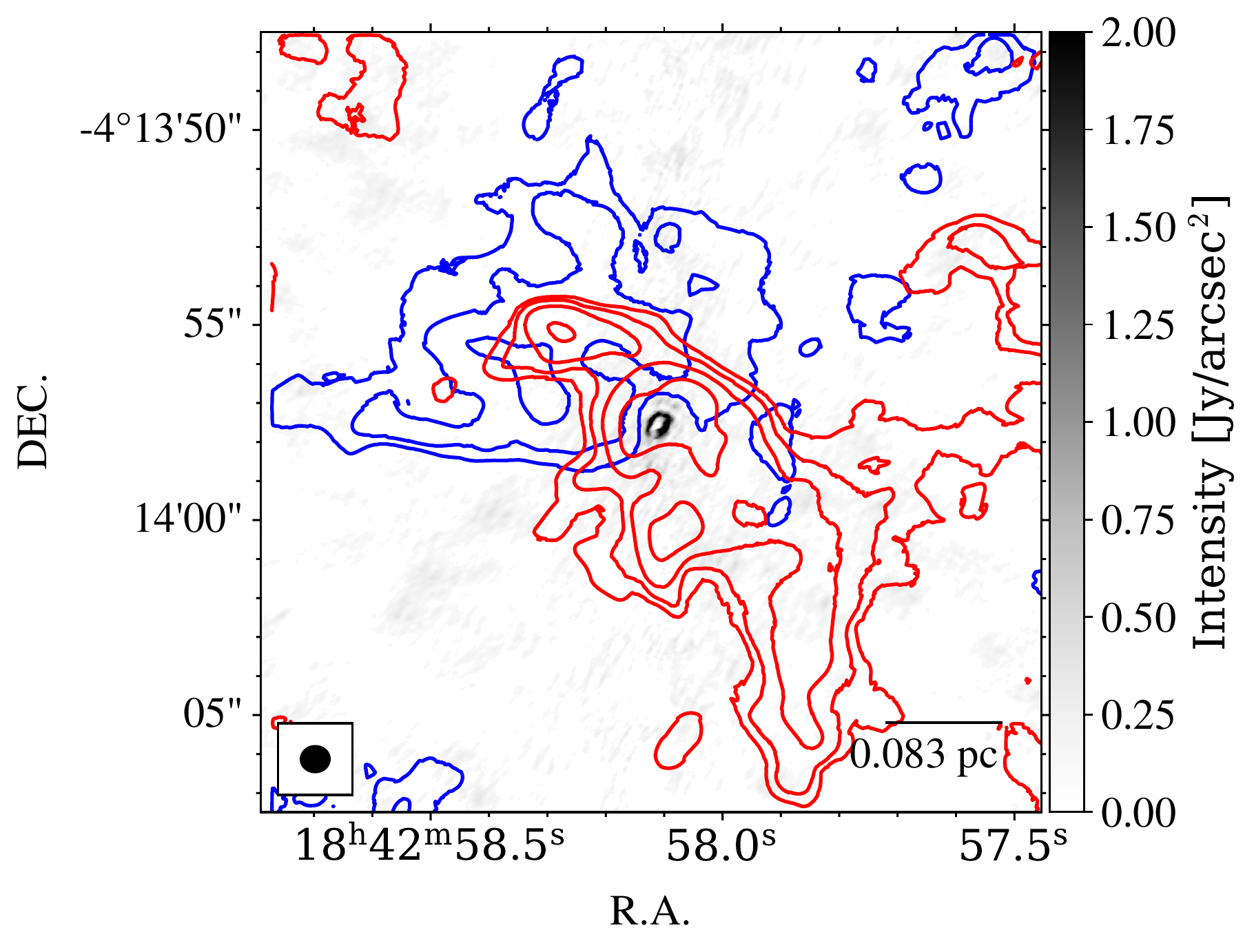}
\end{minipage}
\begin{minipage}{0.49\textwidth}
\includegraphics[width=\textwidth]{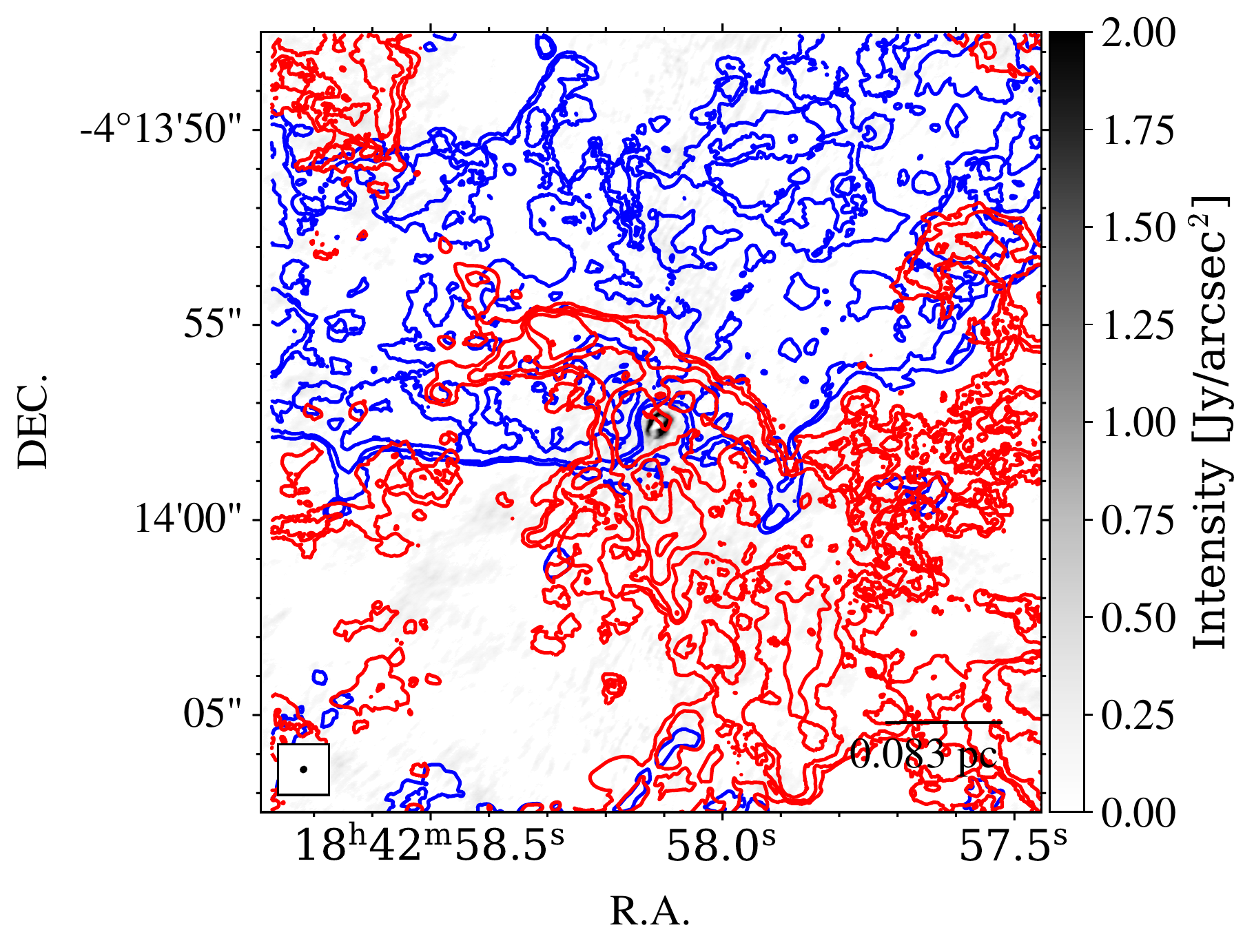}
\end{minipage}
\begin{minipage}{0.49\textwidth}
\includegraphics[width=\textwidth]{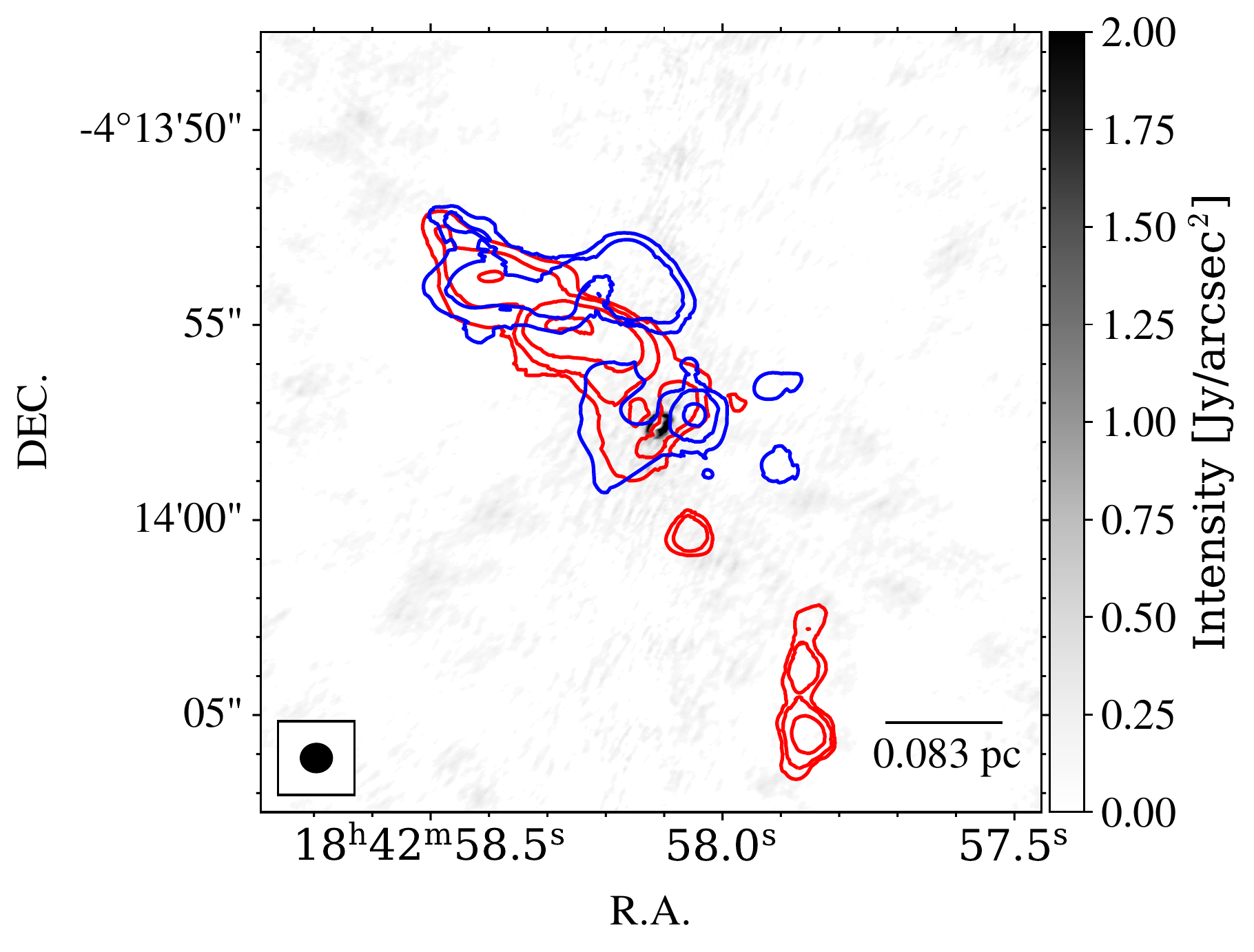}
\end{minipage}
\begin{minipage}{0.49\textwidth}
\includegraphics[width=\textwidth]{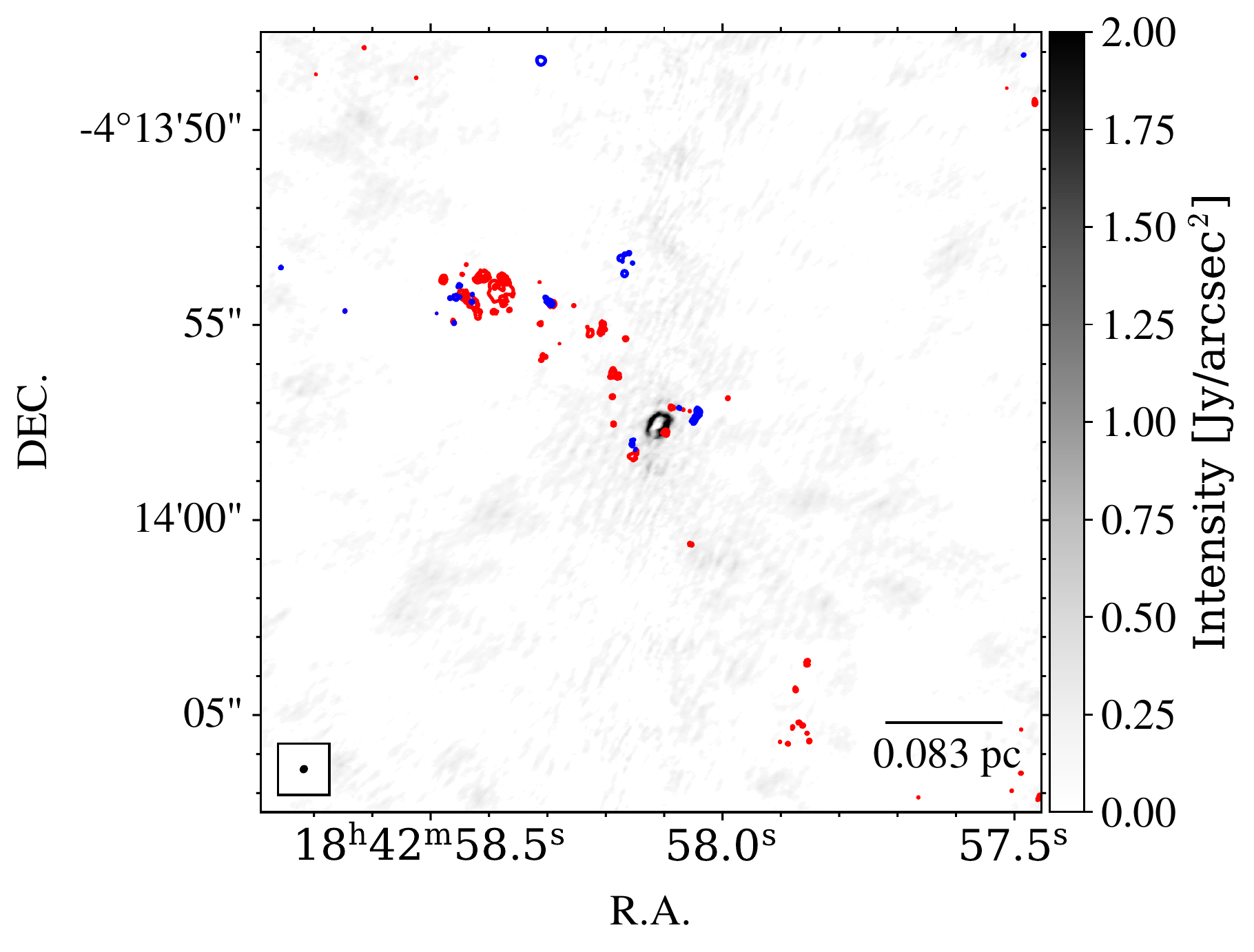}
\end{minipage}
\caption{{\it (a) Top Left:} Integrated intensity maps of CO(2-1) emission tracing outflowing gas as observed in C configuration. The blue contours show blueshifted emission from +80 to $+95\:{\rm km\:s}^{-1}$. The contours levels are [1.28, 2.5, 5, 10, 20]~{$\rm Jy\:beam^{-1}\:{\rm km\:s}^{-1}$}. The red contours show redshifted emission from +96 to $+115\:{\rm km\:s}^{-1}$. The contours levels are [1.28, 2.5, 5, 10, 20]~{$\rm Jy\:beam^{-1}\:{\rm km\:s}^{-1}$}. Only pixels that are above $1\sigma$ spectral rms ($=0.31~\rm Jy\:beam^{-1}$) are included. The grey scale shading shows the 1.3~mm continuum image (C+I+E).
{\it (b) Top Right:} As (a), but for C+I combined configurations of $^{12}$CO(2-1) emission. The contours levels for the blueshifted and the redshifted emissions are [0.16, 0.32, 0.64, 1.28, 2.5, 5]~$\rm Jy\:beam^{-1}\:{\rm km\:s}^{-1}$.
{\it (c) Bottom Left:} As (a), but now showing integrated intensity maps of SiO(5-4) emission (C configuration). The blue contours show blueshifted emission from +80 to $+95\:{\rm km\:s}^{-1}$. 
The red contours show redshifted emission from +96 to $+115\:{\rm km\:s}^{-1}$. The contours levels are [0.16, 0.32, 0.64, 1.28, 2.56, 5.12]~ $\rm Jy\:beam^{-1}\:{\rm km\:s}^{-1}$. Only pixels that are above $1\sigma$ spectral rms ($=0.279~\rm mJy\:beam^{-1}$) are included.
{\it (d) Bottom Right:} As (c), but now for C+I combined configurations of SiO(5-4) emission. The contours levels are [0.01, 0.02, 0.04, 0.06, 0.08, 0.1, 0.12]~$\rm Jy\:beam^{-1}$. 
}
\label{figure:sio-moment0}

\end{figure*}

\begin{figure*}[t]
\centering
\includegraphics[width=\textwidth]{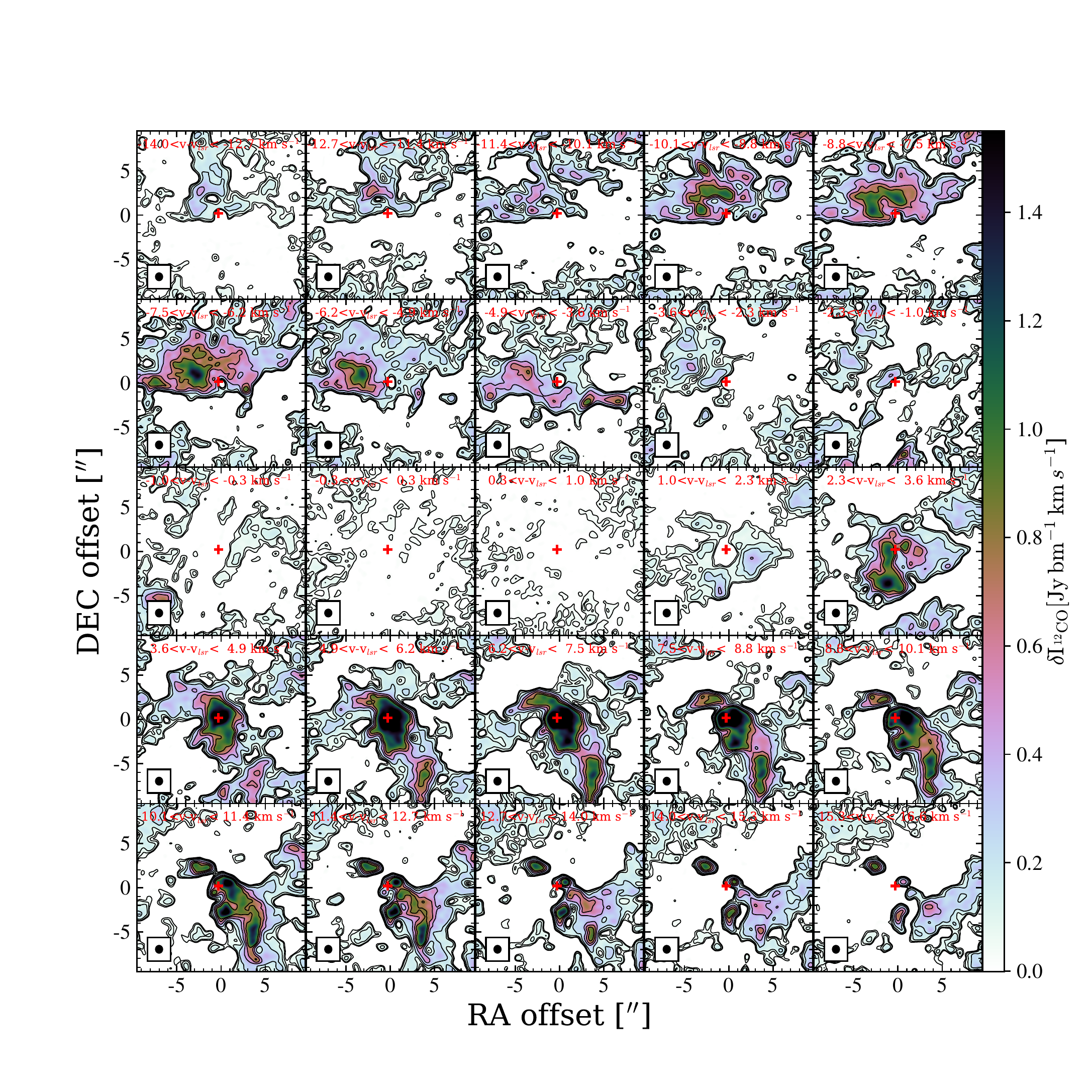}
\caption{The C-only configuration channel maps of $^{12}$CO(2-1) emission. Each panel shows the moment 0 map calculated within the labelled velocity range. The synthesized beam is shown in the lower-left corner of each panel. The [0,0] point corresponds to the main continuum peak location (RA: 18:42:58.09979, DEC: -04:13:57.64121) and is marked by the red '$+$' symbol.}
\label{fig:12CO-channelmap}
\end{figure*}

The \alma spectral windows also include $^{12}$CO(2-1), which we use to trace the presence of outflows. Figure~\ref{figure:sio-moment0} presents the moment 0 maps of $^{12}$CO(2-1) emission: the blueshifted emission is integrated from +80 to $+95\:{\rm km\:s}^{-1}$; the redshifted emission is integrated from $+96$ to $+115\:{\rm km\:s}^{-1}$. Figure~\ref{fig:12CO-channelmap} presents channel maps of this $^{12}$CO(2-1) emission. 
The overall morphology is that expected from a wide-angle bipolar outflow, with the P.A. of the near-facing blueshifted outflow axis being in the NE direction (i.e., ${\rm PA} \sim 45\degree$), 
consistent with being perpendicular to both the major axis of the main mm continuum peak and the direction of maximum velocity gradient of H30$\alpha$ emission. Note that there is some redshifted CO(2-1) emission in the NE direction (and some blueshifted emission in the SW direction), as would be expected in a wide-angle outflow. From the morphology shown in Figure~\ref{figure:sio-moment0}a and b, i.e., the projected lateral extent of the emission relative to the protostellar position, we estimate an opening angle (from outflow axis to outer extent of outflow projected on the sky) of the blueshifted outflow to be about $40^\circ$. This bipolar morphology is consistent withthat reported in the study of \citet{2011A&A...530A..53K}.

SiO line emission, also included in the \alma spectral set-up, is another tracer of outflows from massive protostars \citep[e.g.,][]{2013A&A...550A..81C,2013A&A...554A..35L}. It is expected to be strong in regions of faster flows and/or strong shocks that may lead to destruction or sputtering of dust grains, which then enhances the gas phase abundance of SiO. Figure~\ref{figure:sio-moment0} shows blueshifted ($+80$ to $+95\:{\rm km\:s}^{-1}$) and redshifted (+96 to $+115\:{\rm km\:s}^{-1}$) integrated intensity maps of SiO(5-4) emission. This emission is much sparser than the high-velocity CO(2-1) emission. The blueshifted SiO is again found mostly in the NE direction and at a narrower range of position angles from the protostar, i.e., apparently tracing the more central part of the cavity. Some redshifted emission is also seen in this region. The SiO emission is much weaker towards the SW side, where there it is dominated by a modest knot of redshifted emission.

\begin{figure}[t]
\centering
\begin{minipage}{0.49\textwidth}
\includegraphics[width=\columnwidth]{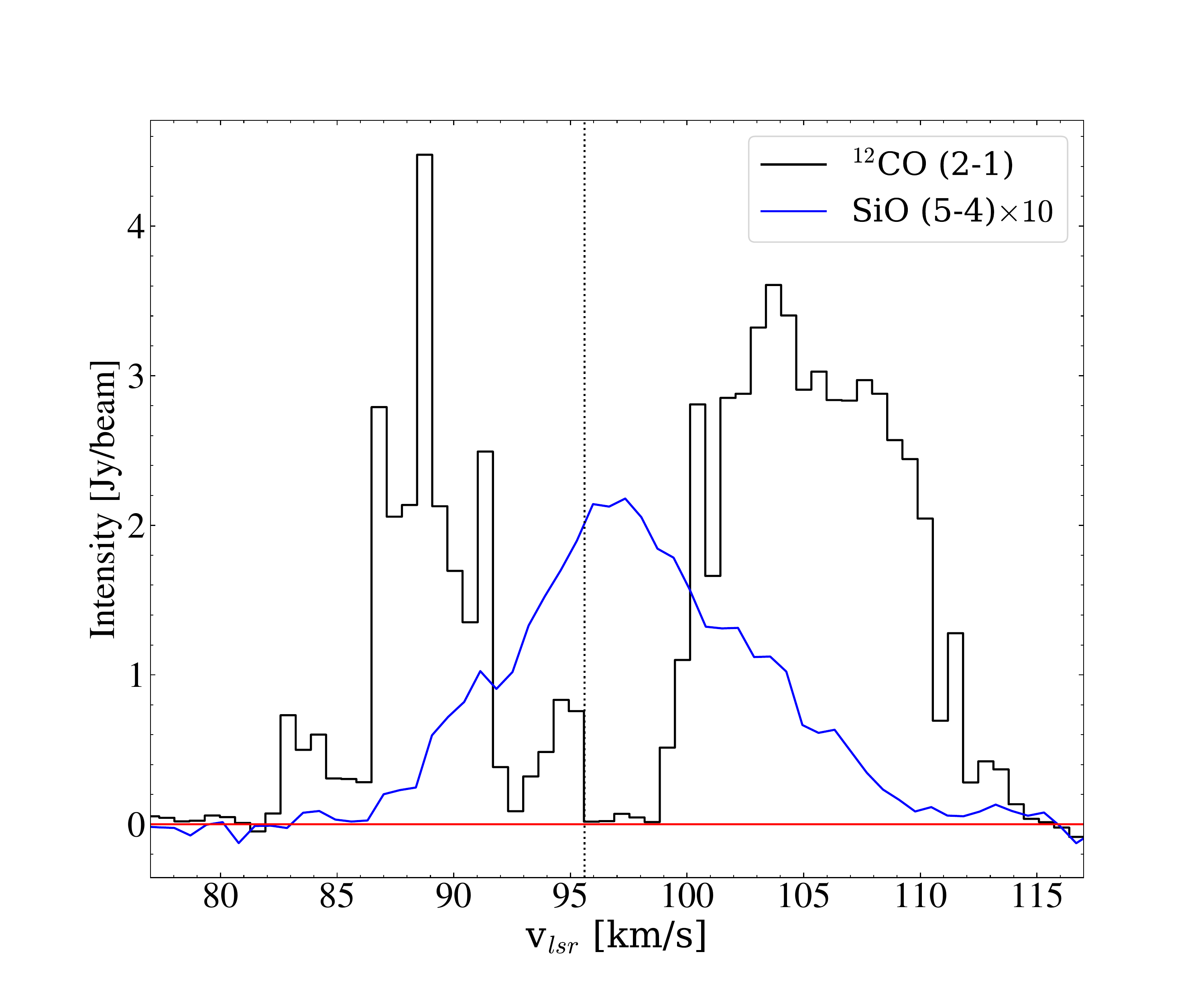}
\end{minipage}
\caption{CO(2-1) and SiO(5-4) averaged spectra extracted from a circular aperture of radius $10^{\prime\prime}$. 
} 
\label{fig:CO&SiO-spectrum}
\end{figure}

Figure~\ref{fig:CO&SiO-spectrum} presents spectra of CO(2-1) and SiO(5-4) extracted from a region of radius 10\arcsec\ centered on the protostar (based on C configuration data). In the CO spectrum, there is an absence of signal near the systemic velocity, likely due to absorption from ambient gas. CO emission is seen to extend to velocities that are up to about $\pm 25\:{\rm km\:s}^{-1}$ from the systemic velocity. On the other hand, the SiO(5-4) spectrum peaks near the systemic velocity, but also extends out to cover the same velocity range seen in CO(2-1).



\begin{figure*}[t]
    \centering
    \vskip -2.0cm 
       \begin{minipage}{0.6\textwidth}
        \includegraphics[width=\textwidth]{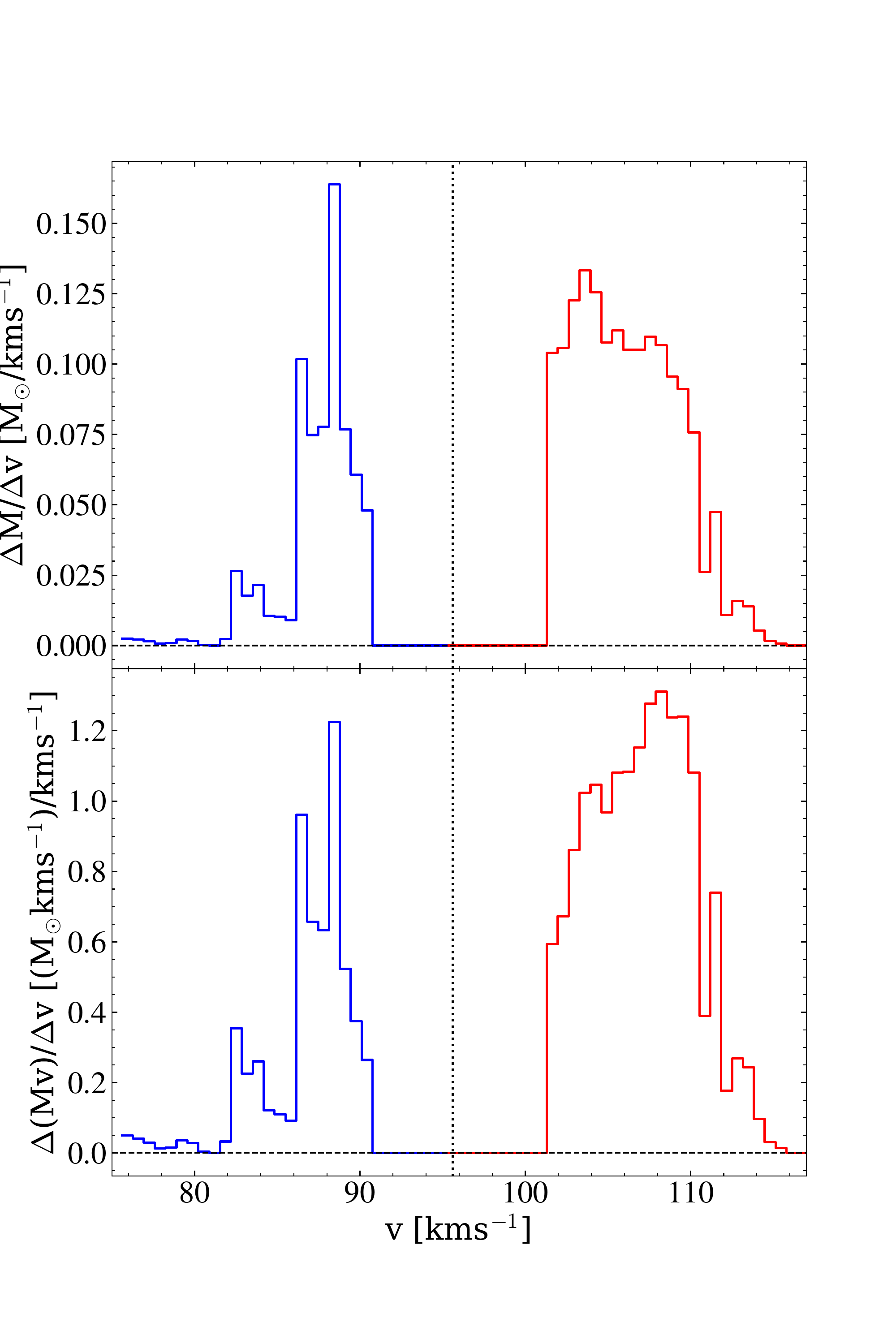}
        \end{minipage}
    \vskip -1.6cm 
       \begin{minipage}{0.6\textwidth}
        \includegraphics[width=\textwidth]{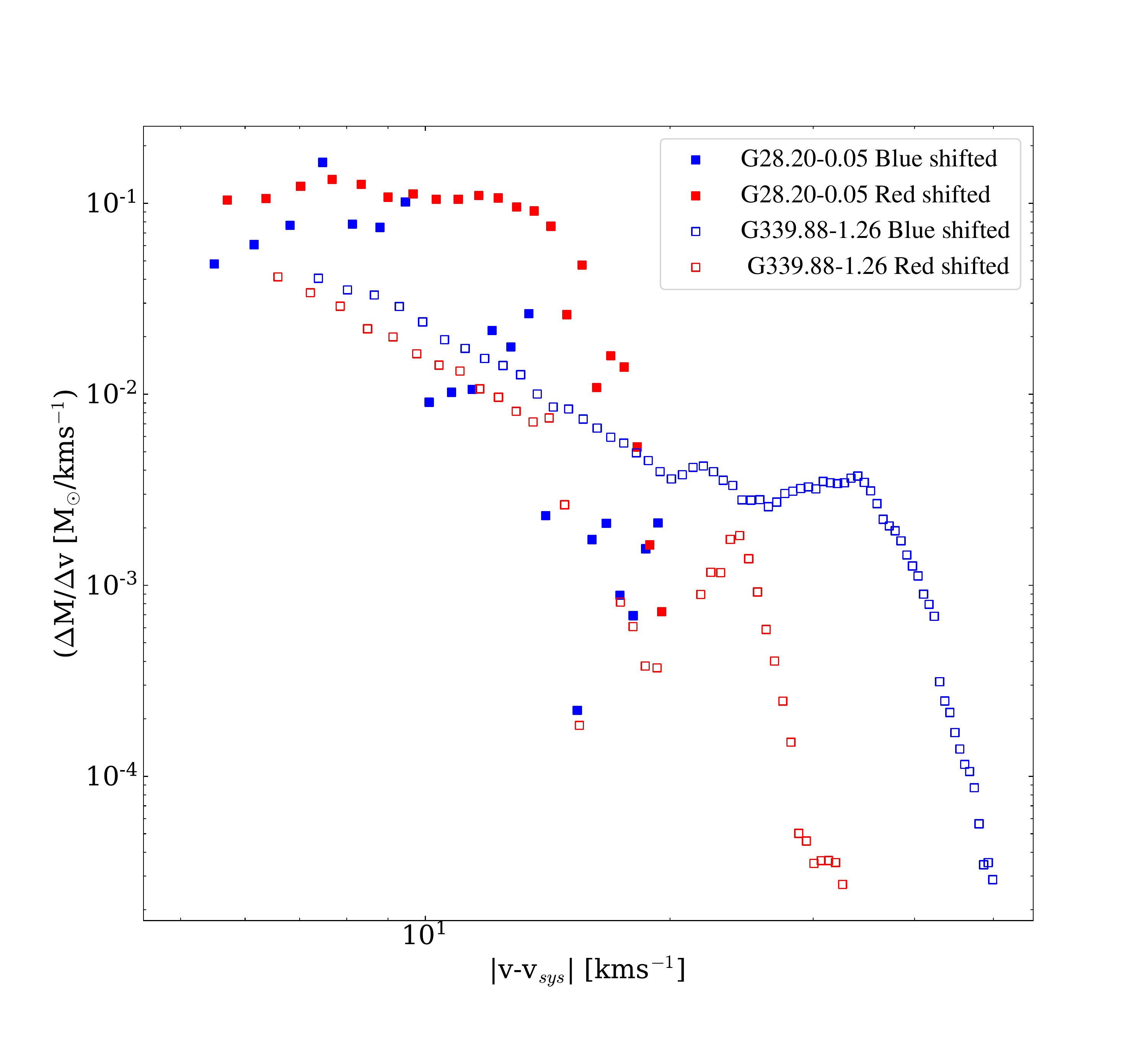}
        \end{minipage}
\caption{{\it (a) Top:} Distribution of outflow mass inferred from CO(2-1) versus velocity. {\it (b) Middle:} Distribution of outflow momentum inferred from CO(2-1) versus velocity. {\it (c) Bottom:} Mass spectra of blue and redshifted outflow components. The equivalent data from the G339.88-1.26 massive protostar \citep{2019ApJ...873...73Z} are also shown.
}
\label{figure:12C0-outflow}
\end{figure*}

Based on this extracted spectrum, Figure~\ref{figure:12C0-outflow} presents an analysis of the CO-traced outflow mass and momentum, following the methods and assumptions of \citet{2019ApJ...873...73Z} that were applied to similar data for the G339.88-1.26 massive protostar. The key assumptions of this method are the choice of CO abundance, i.e., $X_{\rm CO}= [{ }^{12} {\rm CO} / {\rm H}_{2}]=10^{-4}$, the choice of excitation temperature, i.e., $T_{\rm ex}=10-50\:$K \citep[see also ][]{2014ApJ...783...29D} and the assumption that the emission is optically thin. We follow \citet{2019ApJ...873...73Z} to adopt a fiducial excitation temperature of $17.5~$K, which minimizes the mass estimate from the (2-1) transition. 
A choice of $T_{\rm ex} = 50$~K would increase the mass by a factor of 1.5. 

From this analysis, we obtain the following fiducial estimates that should be regarded as minimum values. We find masses of outflowing gas of 0.464~$M_\odot$ and 1.06~$M_\odot$ in the blue and redshifted components, respectively. These components have total momenta of 3.93~$M_\odot\:{\rm km\:s}^{-1}$ and 10.8~$M_\odot\:{\rm km\:s}^{-1}$, respectively. The mean (mass-weighted) velocities of the components are thus 8.48~${\rm km\:s}^{-1}$ and 10.2~${\rm km\:s}^{-1}$, respectively. We associate the outflows with a size of 10\arcsec, i.e., $L_{\rm out}=$~57,000~au, and so the outflow timescales are $3.19 \times 10^{4}$~yr and $2.65\times 10^{4}$~yr, respectively. Thus the mass flow rates are $1.46 \times 10^{-5}$~$M_\odot\:{\rm yr}^{-1}$ and $4.00 \times 10^{-5}$~$M_\odot\:{\rm yr}^{-1}$ and the momentum injection rates are $1.23 \times 10^{-4}$~$M_\odot\:{\rm km\:s^{-1}\: yr}^{-1}$, and $4.07 \times 10^{-4}$~$M_\odot\:{\rm km\:s^{-1}\: yr}^{-1}$, respectively.
%

Figure~\ref{figure:12C0-outflow} also shows the distribution of mass with velocity. Such distributions are important diagnostics that can be compared with theoretical models of massive protostellar outflows \citep[e.g.,][]{2019ApJ...882..123S}. In principle, such comparisons allow constraints to be placed on the evolutionary stage, the inclination of outflow axis to the line of sight, and other properties of the protostellar core.

The same figure also presents the mass spectrum of the outflow with a log-log scaling. In this panel, we also compare with the outflow mass spectra of G339.88-1.26 \citep{2019ApJ...873...73Z}, which is more collimated and thus likely to be at an earlier evolutionary stage. We see that G28.20-0.05 has a more powerful low-velocity outflow than G339.88-1.26, but the latter has a larger amount of mass at high velocities. Some of these differences could be a result of varying degrees of inclination to the line of sight. However, we suspect that much of the fast outflowing gas that is closer to the outflow axis may have been photodissociated in G28.20-0.05, since this source is already starting to create a HC HII region. In this case, it is predicted that relatively broad and strong tracers of atomic outflow components are present, such as [OI] and [CII] lines.

\begin{figure*}[t]
\begin{minipage}{0.49\textwidth}
\includegraphics[width=\textwidth]{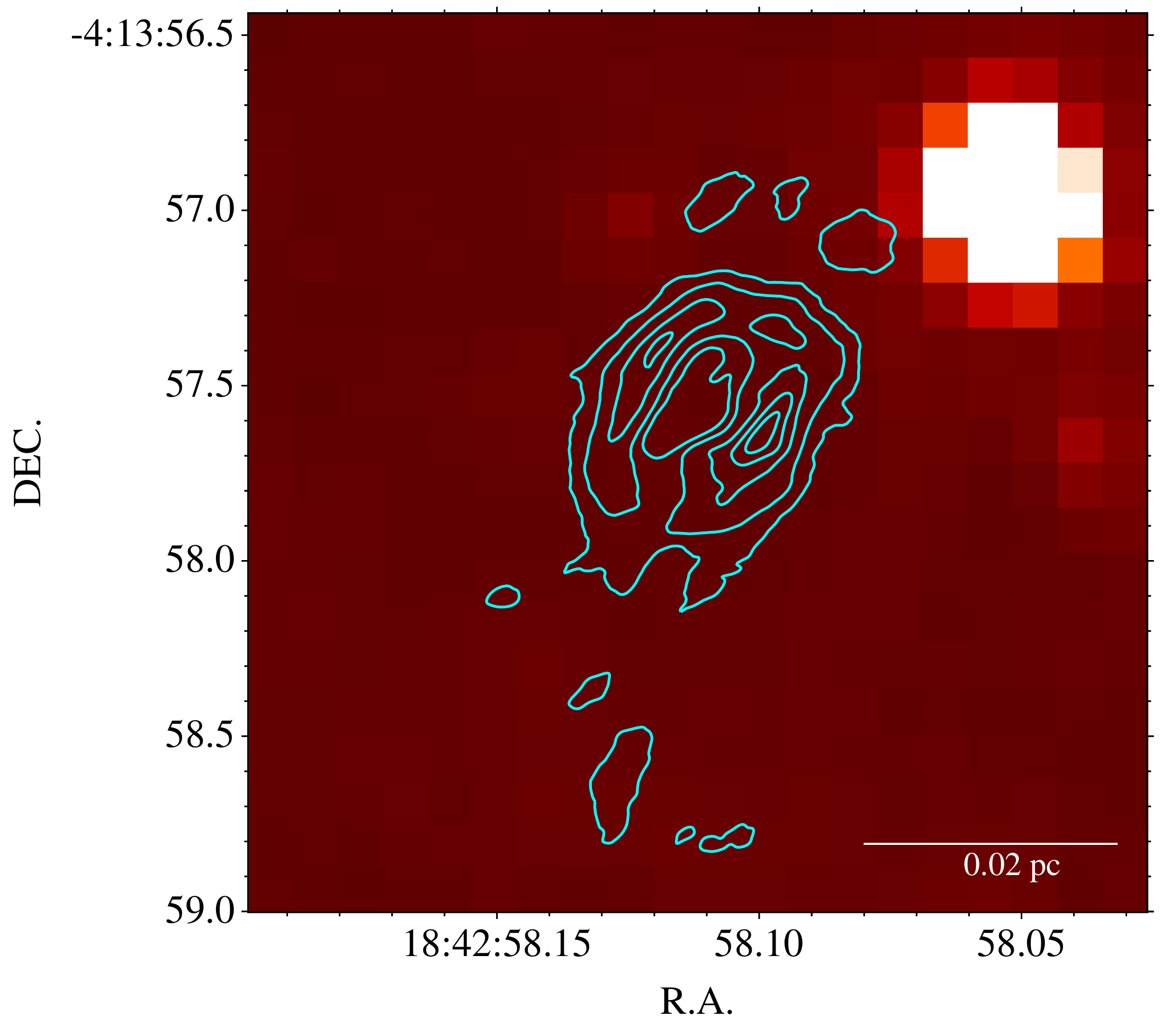}
\end{minipage}
\begin{minipage}{0.49\textwidth}
\centering
\includegraphics[width=\textwidth]{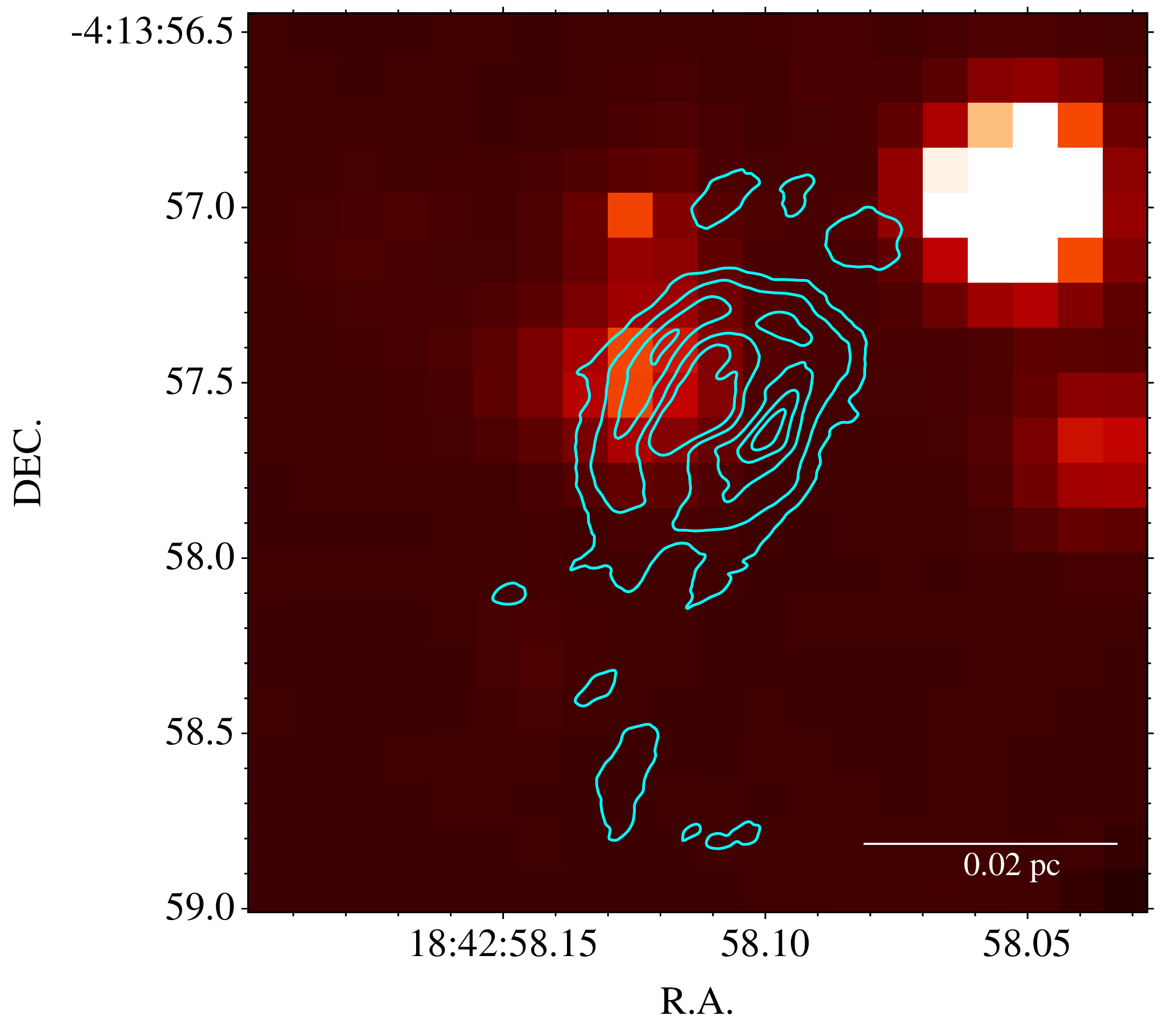}
\end{minipage}
\caption{
{\it (a) Left:} {\it HST} J band ($1.1\,\mu$m) image of G28.20-0.05. Cyan contours show the ALMA 1.3~mm continuum image (same contour levels as in Figure~\ref{figure:h30alpha-moment0}).
{\it (b) Right:} As (a), but now showing the {\it HST} H band ($1.6\,\mu$m) image.
%
%
}
\label{figure:hubble_ALMA}
\end{figure*}

To investigate if there is any further evidence of outflow activity, we observed the NIR emission in the region. Figure~\ref{figure:hubble_ALMA} shows the HST images in the J band (left panel) and in the H band (right panel) together with the ALMA band 6 continuum as cyan contours. The inner region near the mm continuum ring does not show significant emission at the shorter wavelengths, i.e., $\sim1.1\,\mu$m. 
However, emission is clearly seen in the H band, i.e., at $\sim1.6\,\mu$m, peaking at a position that is to the NE of the main mm continuum peak by about 0.5\arcsec\ (i.e., $\sim 3000\:$au). This is the direction of the near-facing, mainly blueshifted outflow. Thus, one interpretation of the H band emission is that it is scattered light from the massive protostar and/or inner accretion disk, which is able to reach us via a relatively low extinction path through the near-facing outflow cavity. 


We use the HST images to measure/constrain the NIR fluxes from the source. This was done by performing circular aperture photometry using the python package \textit{photutils}\footnote{\url{https://photutils.readthedocs.io/en/stable/}}\citep[][]{larry_bradley_2020_4044744}
in both bands with an aperture size equivalent to $\sim4$ times the FWHM, which is about $\sim0\farcs15$, i.e., 6 pixels. We subtract the local background emission by measuring the median value of an annulus with inner and outer radii of 10 and 15 pixels, respectively. We centered the apertures at the peak of the emission in the H band and used the same location for the J (since no emission was found in this band). We applied a correction factor for the IR encircled flux\footnote{\url{https://www.stsci.edu/hst/instrumentation/wfc3/data-analysis/photometric-calibration/ir-encircled-energy}} of 0.893 and 0.863 for the J and H bands, respectively, needed for the extracted aperture of $\sim0\farcs6$. We measure magnitudes in the HST Vega system of 23.48 and 18.98 for the J and H bands, respectively, although in the case of the J band image only an upper limit was retrieved as we have a non detection. Using the method described in \citet{2017A&A...602A..22A}, we transform the HST VEGAMAG system to 2MASS system yielding magnitudes of $22.28\pm0.14$ and $18.04\pm0.14$, for the J and H bands, respectively. The uncertainties in the magnitudes are dominated by the uncertainties in the transformation to 2MASS \citep[see, e.g.,][]{2017A&A...602A..22A}.

A lower limit on the amount extinction to the source based on the J and H magnitudes was calculated using the extinction law of \citet{1985ApJ...288..618R}. This law relates the extinction in J ($A_J$) and in H ($A_H$) with the visual extinction ($A_V$) through $A_J = 0.282A_V$, $A_H = 0.175A_V$. If we assume an intrinsic colour of $-0.164$ for an O9V type star taken from Table 5 of \citet{2013ApJS..208....9P}, this results in a lower limit in the visual extinction of $A_V>41$\,mag, corresponding to mass surface density $\Sigma = 1.83\times 10^{-4}\:{\rm g\:cm}^{-2}$ assuming a conversion factor from $A_V$~(mag) to $N_{\rm H}^{\rm NIR}$~(cm$^{-2}$) of $1.9\times 10^{21}$~cm$^{-2}\:(A_V /{\rm mag})$ \citep[][]{1978ApJ...224..132B} and $\Sigma = 1.4 m_{\rm H} N_{\rm H}^{\rm NIR}\:{\rm g\:cm}^{-2}$.




The {\it HST} observations also included the F128N and F164N narrow band filters designed to detect Pa$\beta$ ($1.28\:{\rm \mu m}$) emission from ionized gas and [FeII] ($1.64\:{\rm \mu m}$), which is a tracer of outflow shocks \citep[e.g.,][]{2019NatCo..10.3630F}. However, no significant emission was detected in these continuum-subtracted images toward the protostar nor in the larger scale outflow. One possible explanation for this is the relatively large amount of extinction of this region.

\section{Protostellar Properties from SED Modeling}\label{sec:SED}

\begin{figure*}
    \centering
    \includegraphics[width=\textwidth]{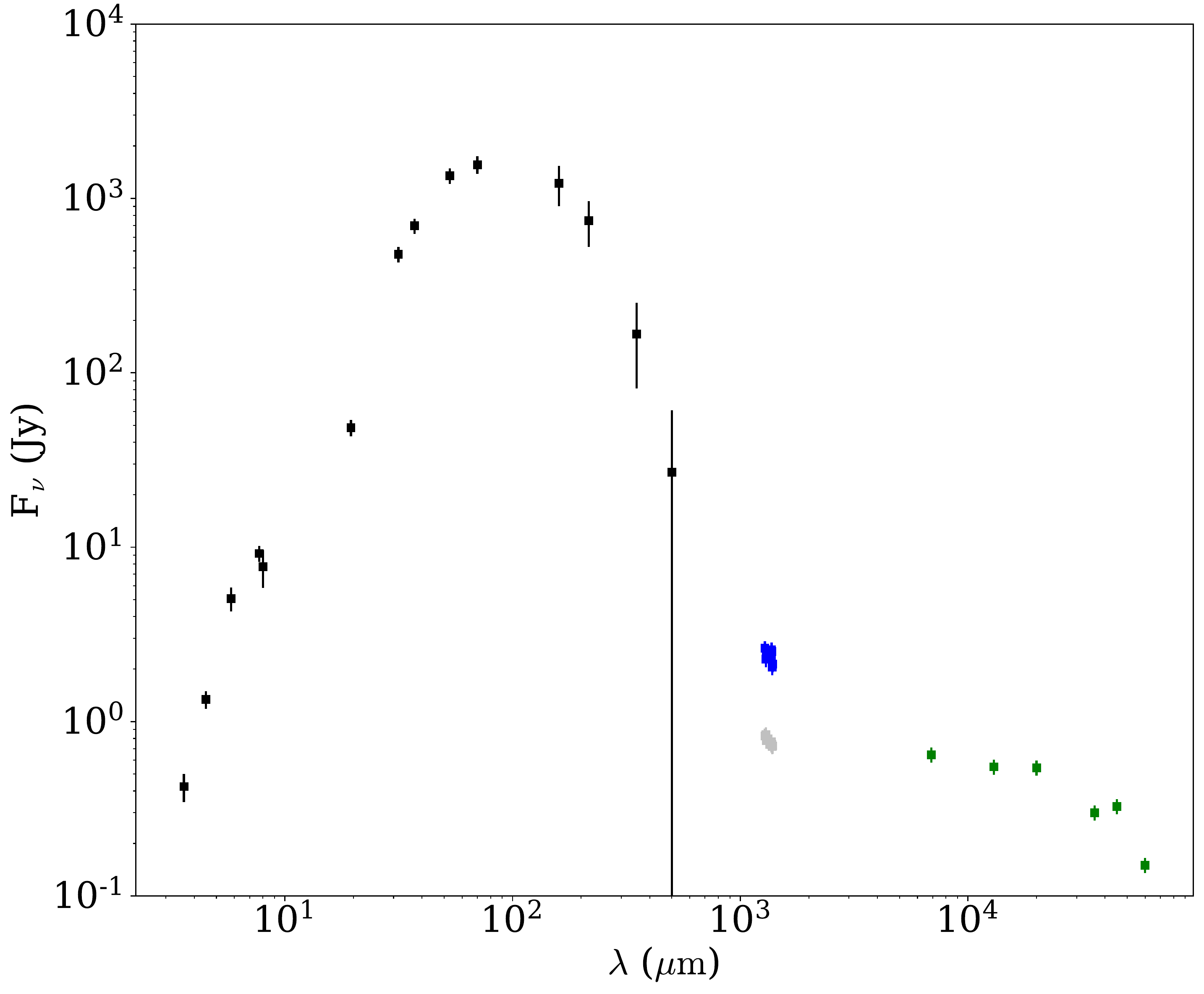}
    \caption{Spectral energy distribution (SED) of G28.20-0.05 from $3.6~{\rm \mu m}$ to $6$ cm. 
    See Table~\ref{table:flux_wavelength} for the detailed information about the data. 
    Black squares and error bars show the fluxes from near-infrared to sub-mm wavelengths. Green squares and error bars show the fluxes in radio wavelengths from 0.7~cm to 6~cm. The blue and grey squares and error bars show the fluxes of C+I ALMA configurations measured with assumed $10\%$ error over a circular aperture of $5.0^{\prime\prime}$ and $0.5^{\prime\prime}$, respectively.} 
    \label{fig:sed_full}
\end{figure*}

\begin{figure*}
    \centering
    \includegraphics[width=1.05\columnwidth]{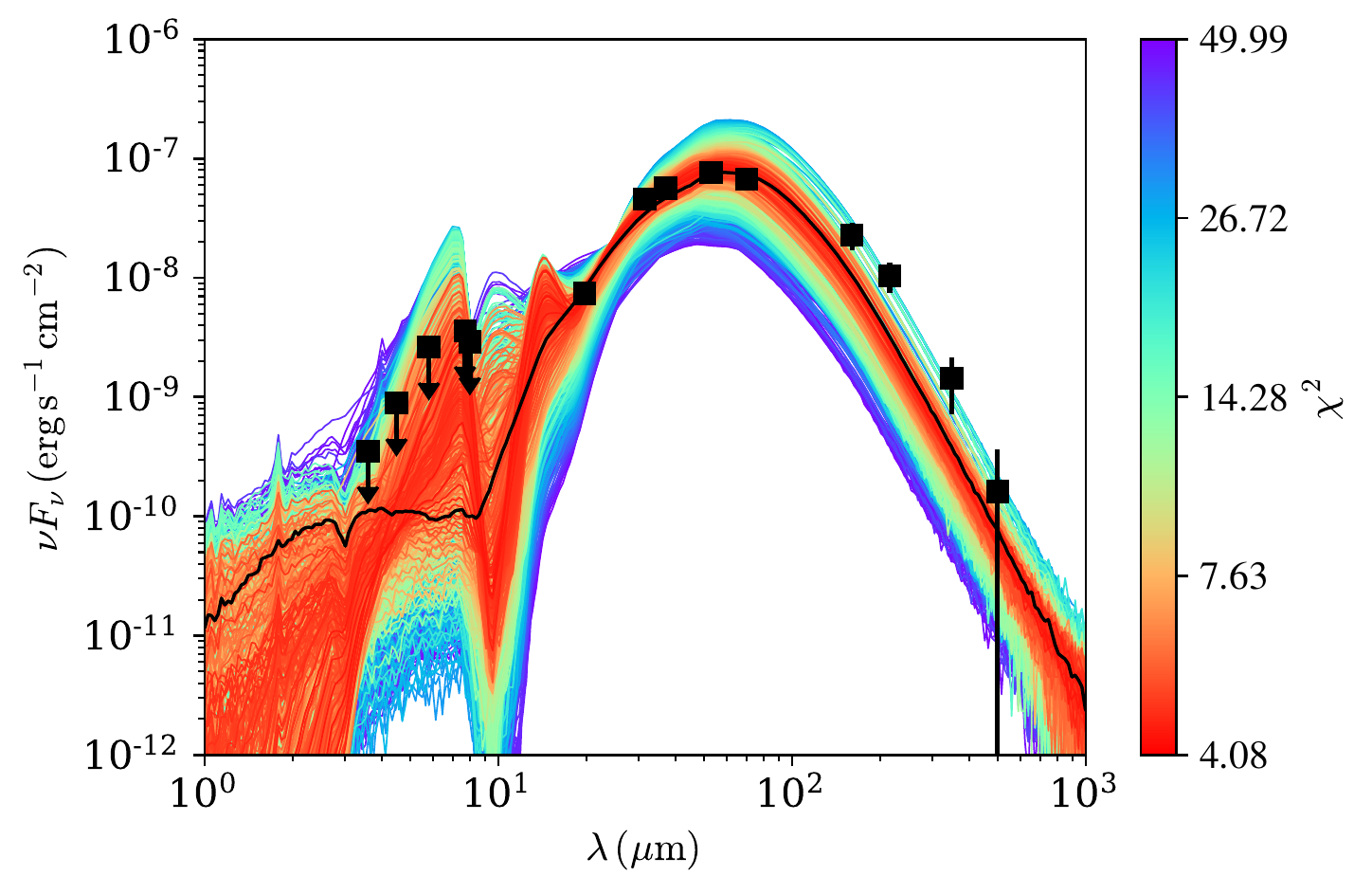}
    \includegraphics[width=1.05\columnwidth]{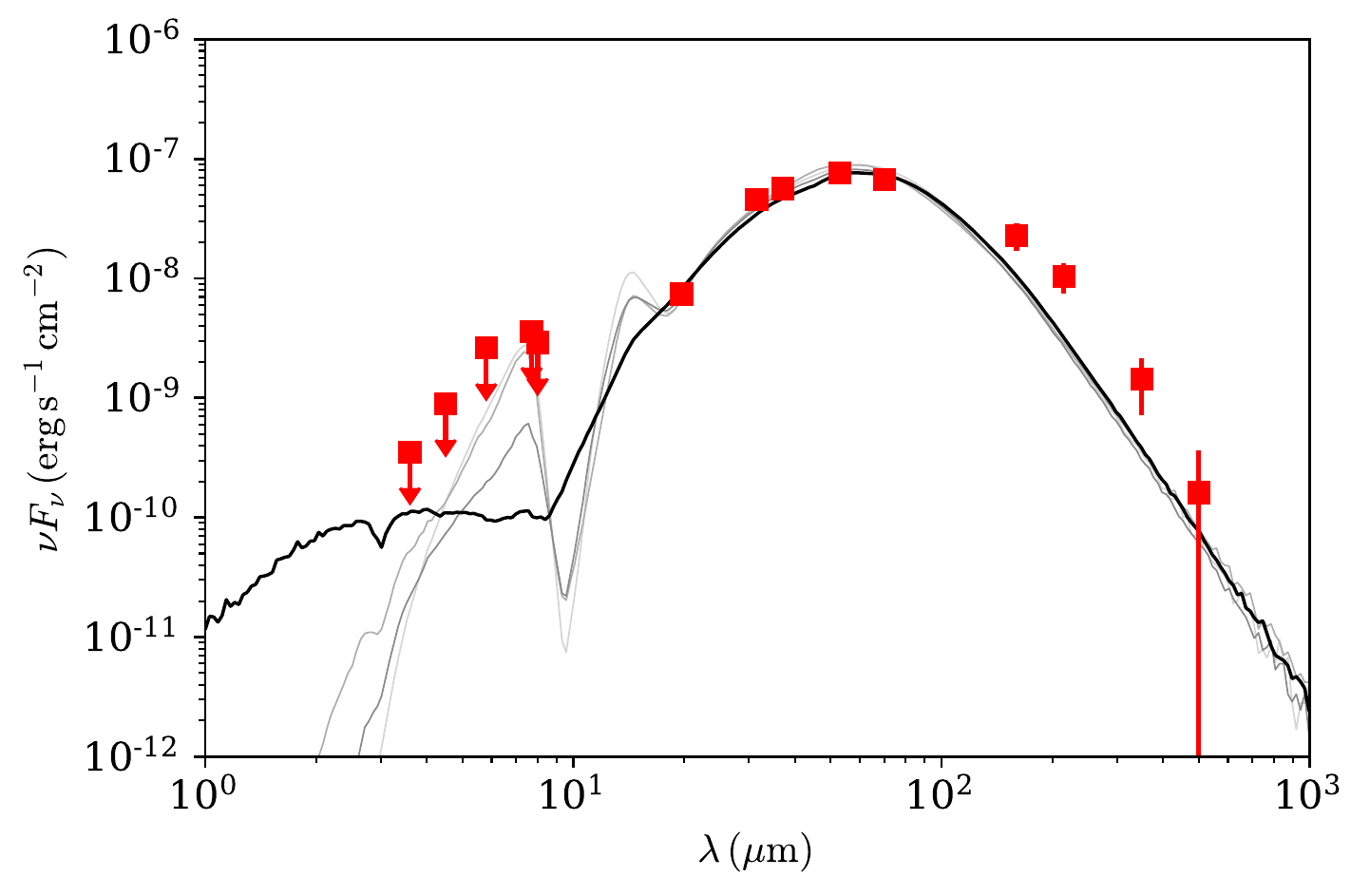}
    \caption{{\it (a) Left panel:} Best fitting SEDs of G28.20-0.05 based on Zhang \& Tan (2018) model grid based on the indicated data (see text). 
    {\it (b) Right panel:} As (a), but only showing the best model (black line) and next four best models (gray lines).
    } 
    \label{fig:SED_mode_range}
\end{figure*}
\begin{figure*}
\centering
    \includegraphics[width=\textwidth]{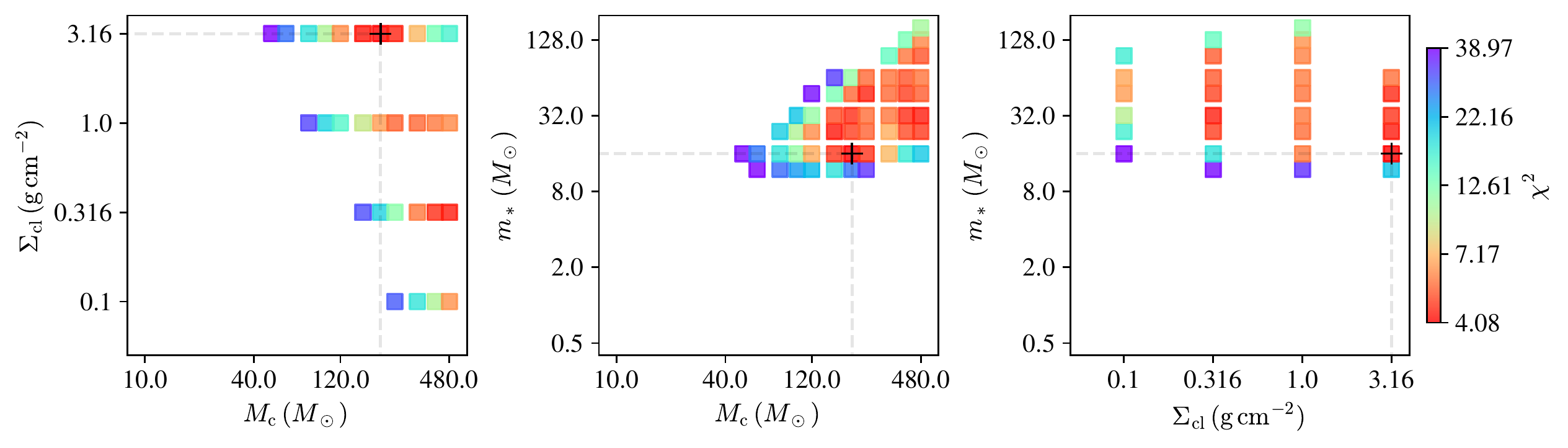}
    \caption{Constrained parameter space ($M_c$, $\Sigma_{\rm cl}$ and $m_*$) of the SED analysis. The color indicates the $\chi^{2}$ parameter. The black plus sign indicates the best model.} 
    \label{fig:SED_summary_plot}
\end{figure*}

Light from the protostellar photosphere is expected to be mostly absorbed by dust in the surrounding disk and infall envelope and then be reprocessed into the infrared. The MIR to FIR SED can thus be used to constrain protostellar properties. The most direct observable is the bolometric flux ($F_{\rm bol,iso}$), i.e., integrating over the SED. Then, given the distance to the source and accounting for foreground extinction, one can estimate the isotropic bolometric luminosity ($L_{\rm bol,iso}$), i.e., assuming the protostar emits isotropically.

In the context of core accretion models for massive star formation, \citet{2018ApJ...853...18Z} have presented a grid of model protostellar SEDs that depend on the initial mass of the core ($M_c$), the mass surface density of the clump environment ($\Sigma_{\rm cl}$) and the  evolutionary stage as parameterized via the current protostellar mass ($m_*$).
Two additional parameters that influence the SED are the viewing angle to the outflow axis ($\theta_{\rm view}$) and the level of foreground extinction ($A_V$). All other core properties, such as initial core radius ($R_c$), current accretion rate ($\dot{m}_*$), intrinsic bolometric luminosity ($L_{\rm bol}$) and outflow opening angle ($\theta_{\rm w,esc}$), are determined from $M_c$, $\Sigma_{\rm cl}$ and $m_*$. Here we determine the protostellar model parameters that best match the SED of G28.20-0.05.


To constrain the protostellar MIR to FIR SED we utilize data from {\it Spitzer}-IRAC, {\it SOFIA}-FORCAST/HAWC+ and {\it Herschel}-PACS/SPIRE (see Table~\ref{table:flux_wavelength}). We note that the IRAC fluxes are used only as upper limit constraints, since the models do not include PAH emission or emission from transiently heated small grains. We note also that mm to cm fluxes from {\it ALMA}, {\it VLA} and {\it ATCA} are not used here to constrain the protostellar models. We follow methods of SED determination and fitting developed for the SOFIA Massive (SOMA) star formation survey \citep{2017ApJ...843...33D,2019ApJ...874...16L,2020ApJ...904...75L,2022arXiv220511422F}. The latest version of these methods involve choosing the radius of a circular aperture ($R_{\rm ap}$) for the source objectively by examination of the {\it Herschel} 70~$\rm \mu m$ image (when available). The radius is set at the point at which a further increase of 30\% in radius leads to the background-subtracted flux increasing by $<10\%$. For background estimation, the method evaluates the average intensity of emission in an annulus from $R_{\rm ap}$ to $2 R_{\rm ap}$ and assumes this applies over the area of the source aperture in order to derive the background-subtracted flux. The uncertainties in the fluxes are assumed to be a combination of 10\% systematic uncertainty, e.g., due to flux calibration, and a contribution from the background, which here is set equal to the background flux. In the case of G28.20-0.05 the derived aperture radius has an angular size of $15\farcs5$ (see Figure~\ref{fig:cont-image}b), corresponding to 0.43~pc.

Figure~\ref{fig:sed_full} shows the SED of the protostar. We see that $F_\nu$ appears to peak around $70\:{\rm \mu m}$. The uncertainties due to background subtraction are seen to become significant at longer wavelengths, which reflects the fact that the protostar is surrounded by large quantities of relatively cool, dusty gas. For completeness, Figure~\ref{fig:sed_full} also shows the fluxes from ALMA, VLA and ATCA. These are seen to be enhanced with respect to the expected trend of thermal emission, indicating that they are dominated by or have significant contributions from ionized gas. We also note that these mm to cm fluxes are not evaluated with the same aperture as used at shorter wavelengths and, being measured by interferometers, are subject problems of missing flux.



The MIR to FIR SED data are then used to constrain the protostellar SED models of \cite{2018ApJ...853...18Z} with the SED fitting package sedcreator (ver. 6.0.14) \citep[][]{2022arXiv220511422F}.
The best-fitting model SEDs, in the form of $\nu F_\nu$, are plotted in Figure~\ref{fig:SED_mode_range}. We see that models give a good fit to the SED data in the range from about 20 to 100~$\rm \mu m$. At longer wavelengths the data show a modest excess of flux compared to the models. We suspect that this is caused by imperfect subtraction of surrounding background emission, perhaps also associated with an overestimation of the source aperture.


Constraints in the main parameter space of $M_c$, $\Sigma_{\rm cl}$ and $m_*$ are summarized in Figure~\ref{fig:SED_summary_plot}. We see that certain parts of parameter space are clearly favored, although there are significant degeneracies, e.g., in $\Sigma_{\rm cl}$. The physical parameters of the best five fitted models and the average and dispersion of ``good'' fitting models (defined here as having $\chi^2$ values that are $<2 \chi^2_{\rm min}$) are presented in Table~\ref{tab:dataobs}.

The average of the ``good'' SED models indicate that G28.20-0.05 harbors a protostar with a current mass of $m_*\sim 43_{27}^{68} \:M_\odot$ that is forming from a core with initial mass of $M_c\sim 300_{190}^{460} \:M_\odot$ in a clump with mass surface density of $\Sigma_{\rm cl}\sim 0.8^{2.6}_{0.3} \:{\rm g\:cm}^{-2}$. We note that this estimate for $m_*$ is consistent with our earlier dynamical mass estimate if there is also a similar mass in the dusty gas present within the 0.3\arcsec\ scale region of the protostar.



\begin{deluxetable*}{lccccc}
\tabletypesize{\normalsize}
\tablecaption{Integrated flux densities from $3.6\:{\rm \mu m}$ to 6~cm  \label{table:flux_wavelength}} 
\tablewidth{\textwidth}
\tablehead{
\colhead{Facility} &\colhead{Wavelength } & \colhead{Integrated intensity} & \colhead{Aperture radius} \\
\colhead{} & \colhead{($\rm \mu m$)} & \colhead{(Jy)} & \colhead{(\arcsec)} 
}
\startdata
\spitzer    & 3.6        & 0.424$\pm 0.598$      & 15.5         \\
\spitzer    & 4.5        & 1.34$\pm 0.15$        & 15.5           \\
\spitzer    & 5.8        & 5.07$\pm 0.57$        & 15.5                \\
\sofia       & 7.7         & 9.20$\pm 0.92$     & 15.5              \\ 
\spitzer    & 8.0        &  7.74$\pm 1.65$        & 15.5              \\
\sofia       & 19.5         & 48.5$\pm 4.9$     & 15.5               \\ 
\sofia       & 31.5         & 478$\pm 48$     & 15.5               \\ 
\sofia       & 37.1         & 696$\pm 70$     & 15.5              \\ 
\sofia       & 53         & 1449$\pm 144$     & 15.5              \\ \hline
\herschel   & 70         &  1561$\pm  77$        & 15.5            \\
\herschel   & 160        & 1222$\pm 282$        & 15.5              \\
\sofia   & 214        & 746$\pm 187$    & 15.5           \\ 
\herschel   & 350        & 167$\pm 82$        & 15.5            \\
\herschel   & 500        & 26.9$\pm 33.5$        & 15.5            \\ \hline
\alma       & 1282       &0.827(2.63)$^{a}$      & 0.5(5.0)           \\
\alma       & 1294       & 0.842(2.28)$^{a}$     & 0.5(5.0)          \\
\alma       & 1301       & 0.778(2.41)$^{a}$      & 0.5(5.0)             \\
\alma       & 1362        & 0.756(2.31)$^{a}$       & 0.5(5.0)           \\
\alma       & 1367     & 0.766(2.32)$^{a}$           & 0.5(5.0)               \\
\alma       & 1372       & 0.737(2.58)$^{a}$    & 0.5(5.0)              \\
\alma       & 1374       & 0.739(2.52)$^{a}$      & 0.5(5.0)               \\
\alma       & 1382       & 0.720(2.05)$^{a}$      & 0.5(5.0)             \\
\alma       & 1385       & 0.726(2.13)$^{a}$     & 0.5(5.0)            \\ \hline
\vla        & 6900       & 0.645$\pm 0.065^{b}$     & $0.9$        \\
\vla        & 13000      & 0.548$\pm 0.055^{c}$     & $0.5$      \\
\vla        & 20000      & 0.494$\pm 0.050^{c}$      & $0.8\times0.6^{g}$            \\
\vla        & 36000      & 0.297$\pm 0.045^{d}$     & $3.6\times3.8^{g}$      \\
\atca       & 45000      & $0.326\pm0.033^{e}$        & $1.9$            \\
\vla        & 60000      & 0.150$\pm 0.015^{f}$     & 2.19         
\enddata
\tablecomments{Information on the derivation of the MIR to FIR ($\leq500\:{\rm \mu m}$) is given in the main text. The following notes relate to the mm to cm flux measurements. 
$^{a}$The first number is the flux within the 0.5\arcsec radius aperture. The second number, in parentheses, is the C+I combined flux within the $5\arcsec$ radius aperture. The flux uncertainties are assumed to be $10\%$.
$^{b}$\citet[][]{2008ApJ...681..350S}.
$^{c}$\citet[][]{2011ApJS..194...44S}, but the 1.3~cm flux has been re-derived here.
$^{d}$\citet[][]{1994ApJS...91..659K}.
$^{e}$\citet[][]{1998MNRAS.301..640W}; uncertainties were not provided, so we adopt a fiducial value of 10\%.
$^{f}$\citet[][]{2008ASPC..387..389P}.
$^{g}$The integrated flux was measured within the source size defined by the long and short axes, which are listed here.}
\end{deluxetable*}

\begin{deluxetable*}{lcccccc}
\tabletypesize{\footnotesize}
\tablecaption{Parameters of the best five models and the average with dispersion of all 379 ``good'' models for G28.2-0.05\label{tab:dataobs}} 
\tablewidth{18pt}
\tablehead{
\multicolumn{7}{c}{G28.2-0.05 (D = 5.7~kpcs)}\\
\hline
\colhead{Parameters}&\multicolumn{5}{c}{Best 5 models}& \colhead{Average model}}
\startdata
\hline
$\chi^{2}$ & 4.08& 4.39&4.48&4.57&4.68&($N_{\mathrm{model}}=379$)\\
$M_{\rm c}$ ($M_\odot$) &200& 160&480&240&400&$303^{455}_{194}$\\
$\Sigma_{\rm cl}$ (g $\rm cm^{-2}$)& 3.16& 3.16&0.316&3.16&0.316&$0.815^{2.61}_{0.255}$\\
$R_{\rm core}$ (pc) &0.060& 0.050&0.286&0.060&0.262&$0.140^{0.306}_{0.0659}$ \\
$m_{*}$ ($M_\odot$)& 16.0& 24.0&32.0&48.0&32.0&$42.6^{67.7}_{26.8}$ \\
$\theta_{\rm view}$ (deg)&29.0& 34.0&29.0&39.0&39.0&$61.2\pm17.8$\\
$A_{V}$ (mag) & 0.00& 94.1&47.7&147&20.6&$44.8\pm 40.0$\\
$M_{\rm env}$ ($M_\odot$)&171& 114&406&138&317&$189^{301}_{119}$\\
$\theta_{w,\rm esc}$ (deg)& 15.0& 23.0&22.0&33.0&25.0&$61.2\pm17.8$\\
$\dot {M}_{\rm disk}$ ($M_{\odot}$/yr)& 1.30$\times10^{-3}$& 1.40$\times10^{-3}$&3.90$\times10^{-4}$&2.10$\times10^{-3}$&3.60$\times10^{-4}$&$7.32^{15.4}_{3.47}\times10^{-4}$\\
$L_{\rm bol, iso}$ ($L_{\odot}$)& 9.70$\times10^{4}$& 1.90$\times10^{5}$&1.40$\times10^{5}$&3.60$\times10^{5}$&1.10$\times10^{5}$&$4.37^{8.75}_{2.17}\times10^{5}$\\
$L_{\rm bol}$ ($L_{\odot}$) &1.10$\times10^{5}$&3.00$\times10^{5}$&2.00$\times10^{5}$&7.50$\times10^{5}$&2.00$\times10^{5}$&$1.36^{1.93}_{0.953}\times10^{5}$\\
\enddata
\end{deluxetable*}
\section{Fragmentation and multiplicity properties of the G28.20-0.05 protostar and protocluster}\label{sec:fragmentation}


\subsection{Dendrogram analysis of the 1.3~mm continuum image}

\begin{deluxetable*}{lcccc}
\tabletypesize{\footnotesize}
\tablecaption{Fluxes, mass surface densities and masses of structures identified in ALMA~1.3~mm continuum images of G28.20-0.05. \label{tab:mass_columndensity}} 
\tablewidth{5pt}
\tablehead{
\colhead{Aperture radius/size (ALMA config.)} &\colhead{Flux} & \colhead{$\Sigma$ } & \colhead{Mass} \\
\colhead{(\arcsec)} & \colhead{(Jy)} & \colhead{(gcm$^{-2}$)} & \colhead{($M_{\odot}$)} \\
\colhead{} & \colhead{} & \colhead{(20K,100K,300K)} & \colhead{(20K,100K,300K)} 
}
\startdata
$0.5$(C+I+E) & $0.733$ & $256,  40.4 , 13.0$ &$737,  117,  37.4$  \\
$0.5$(C+I) & $0.809$ & $282, 44.6, 14.3$ &$813, 129,  41.3$ \\
$5.0$(C+I) & $2.10$ & $8.28, 1.31, 0.421$ &$211, 333,  107$\\ \hline
All dendrogram leaves (C+I) & $0.450$ & $46.2, 7.32,  2.35$ & $452,  71.6,  23.0$\\
\hline
\enddata
\end{deluxetable*}

We characterise the fragmentation properties of G28.20-0.05 and its surroundings by applying the dendrogram algorithm \citep[][]{2008ApJ...679.1338R}.
We carry this out on images before primary beam correction, i.e., so that it has a uniform noise map.
Following \citet{2018ApJ...853..160C}, \citet{2018ApJ...862..105L} and \citet{2021ApJ...916...45O}, the fiducial dendrogram parameters that we use are minvalue = $4\sigma$ (the minimum intensity considered in the analysis); mindelta $=1\sigma$ (the minimum spacing between isocontours); minpix $=0.5$ beam area (the minimum number of pixels contained within a structure). 

\begin{figure*}
    \centering
        \begin{minipage}{0.55\textwidth}
        \includegraphics[width=\textwidth]{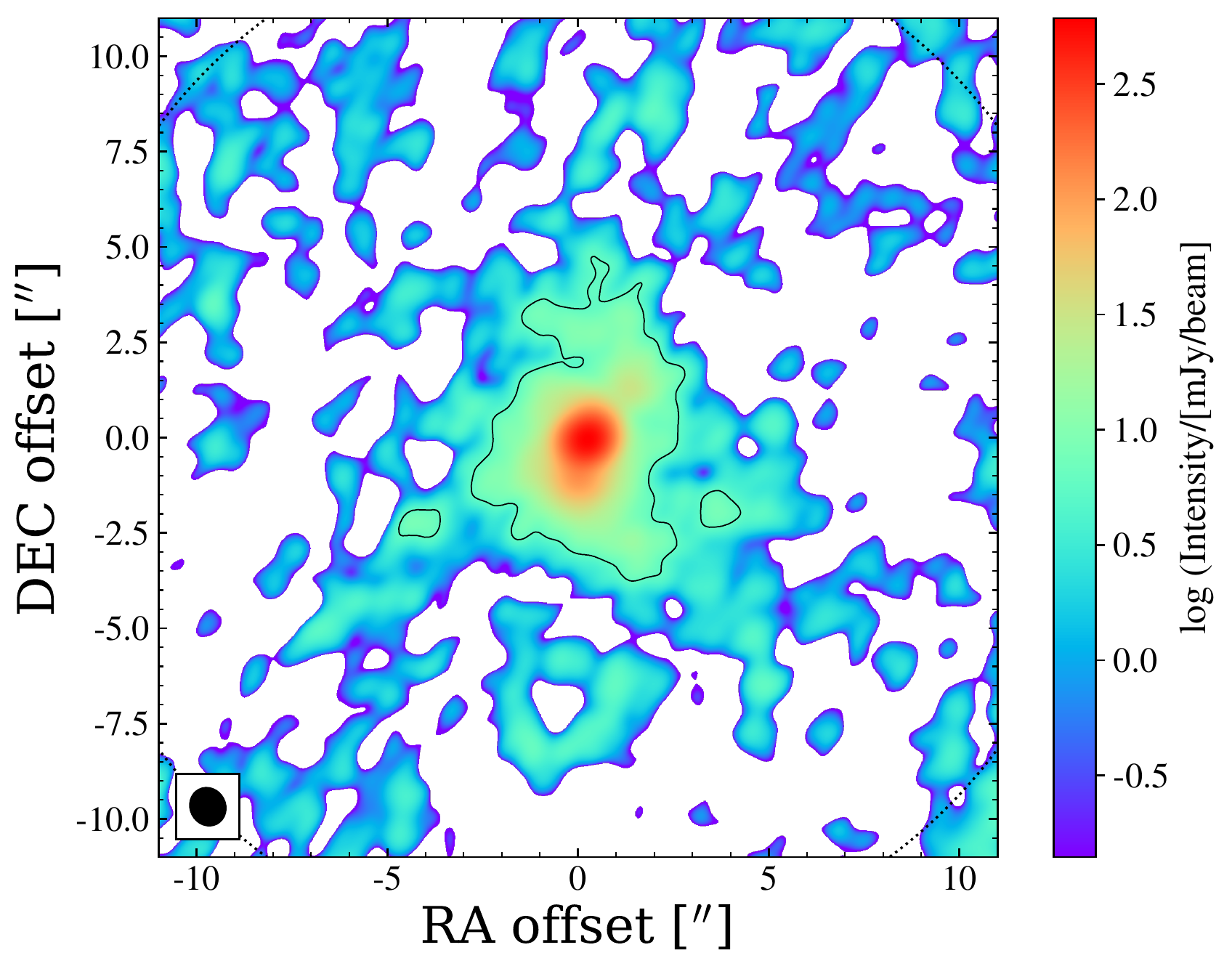}
        \end{minipage}
    
        \begin{minipage}{0.55\textwidth}
        \includegraphics[width=\textwidth]{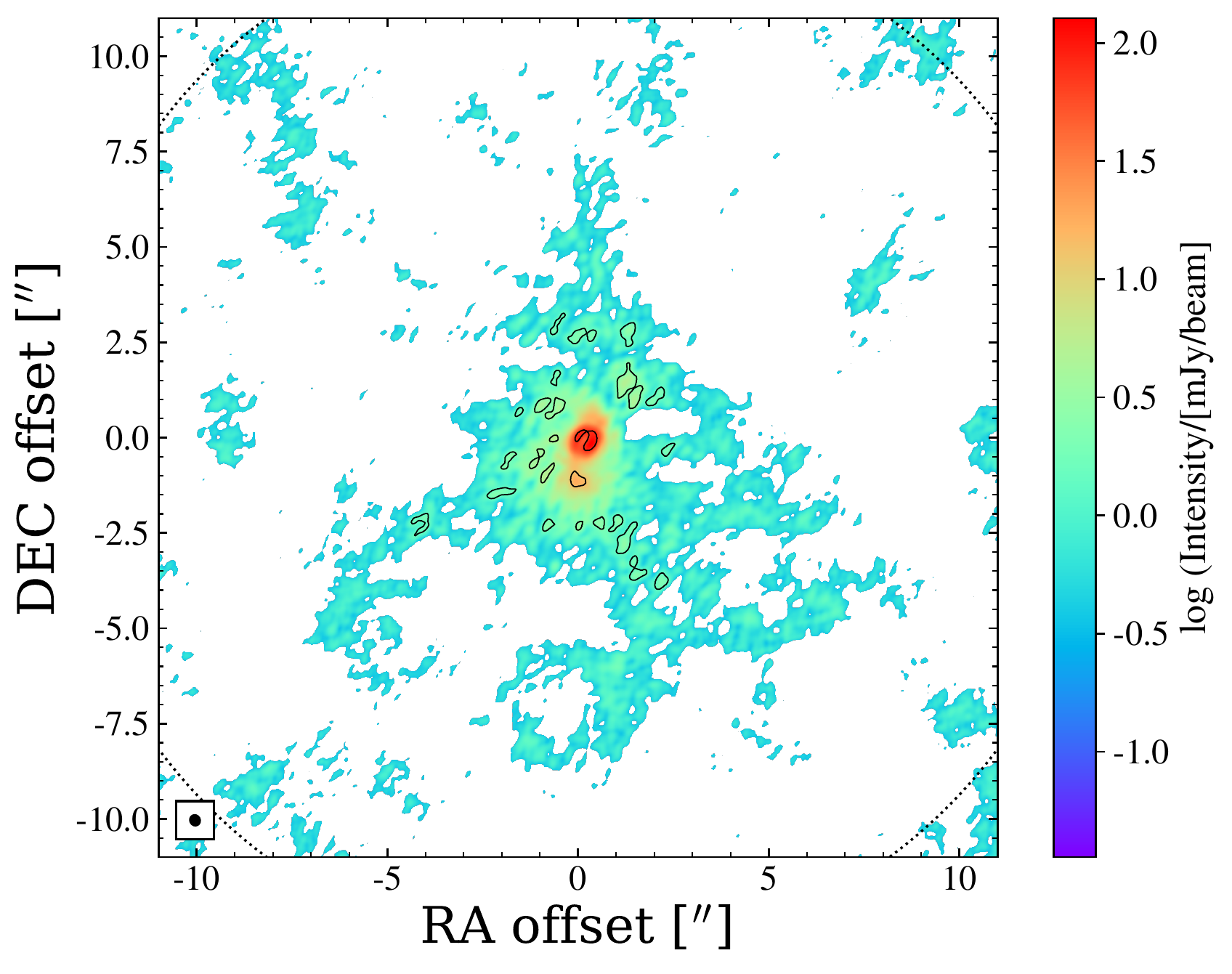}
        \end{minipage}

        \begin{minipage}{0.55\textwidth}
        \includegraphics[width=\textwidth]{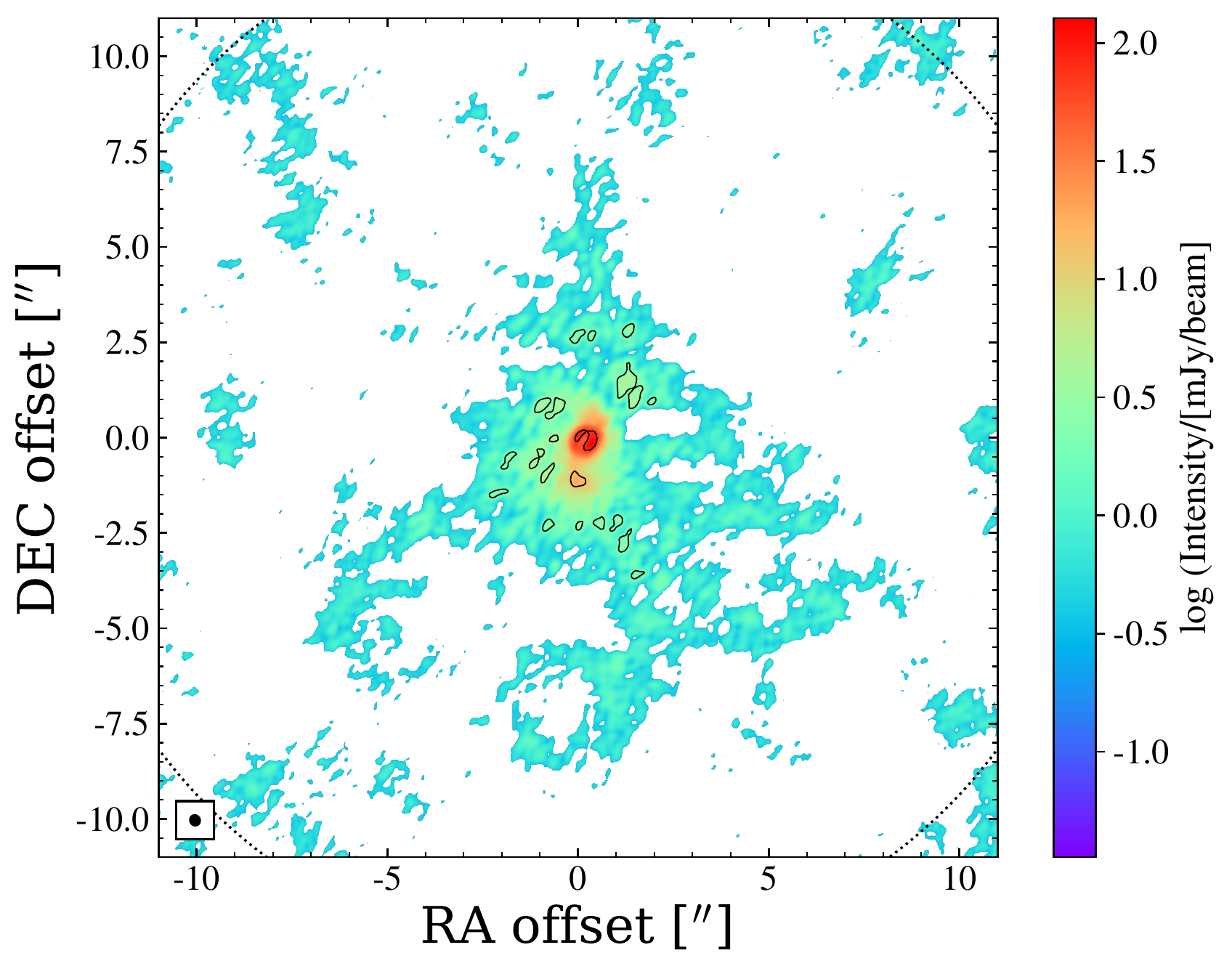}
        \end{minipage}
     \caption{Summary figure of dendrogram-identified structures (leaves) of C-only (top), C + I with minvalue $=4\sigma$ (middle), and C + I with minvalue $=5\sigma$ (bottom). The boundaries of the leaves are shown by solid black lines. 
     The corresponding rms noise levels are 1.29 $~\rm mJy\:beam^{-1}$ for C-only and 0.358$~\rm mJy\:beam^{-1}$ for C+I. The color stretch of each panel extends down to $0.1\sigma$. The dashed circle designates the ALMA primary beam.
     }
\label{fig:dendrogram_analysis}
\end{figure*}

In Figure~\ref{fig:dendrogram_analysis} we present the dendrogram identified structures respectively in the C only and C+I configurations. In the C only image, dendrogram finds the main central core, but then only two additional smaller cores (with masses 6.67~$M_\odot$, 7.66~$M_\odot$ assuming a fiducial dust temperature of 20~K). Furthermore these two additional cores are quite close to the main core, i.e., within about 4\arcsec, and only separated from the main core's boundary by less than one beam FWHM. Thus, there is an absence of dendrogram-detected sources beyond 4\arcsec, i.e., beyond $\sim 0.1~$pc (or about 23,000~au).
The dendrogram analysis of the C+I image yields a larger number of fragments (or ``cores''), but these all overlap with the central region within about 4\arcsec. We note that, assuming a temperature of 20~K, the mass sensitivity of the dendrogram analysis for the C image is 1.30~$M_\odot$ and the C+I image is 0.387~$M_\odot$. For 100~K, which we consider more realistic in the closer vicinity of a massive protostar, these mass sensitivities would decrease by a factor of 6. Thus the main conclusion to be drawn is that there is a lack of compact mm emission sources beyond about 0.1~pc from the massive protostar. Other protostars in the vicinity would be expected to appear as such compact sources. Thus G28.20-0.05 appears to be forming in near complete isolation.

Closer examination of the mm continuum images does reveal a relatively extended ring of emission about 8\arcsec\ to the south of the main source. This corresponds to a source already noted by \citet{2011ApJS..194...44S} based on VLA radio $7~$mm and $2~$cm data. Such a source likely corresponds to a small HII region around an already formed relatively massive star, e.g., a B star. However, it is not prominent in 
ALMA~1.3~mm continuum, indicating it does not have a large amount of warm dust around it.

On the scales within a radius of 5\arcsec, i.e., $\sim 0.1$~pc, the detected mm continuum emission corresponds to a mass of about $300\:M_\odot$ (assuming 100~K) (see Table~\ref{tab:mass_columndensity}). The dendrogram identified structures are within this region and correspond to about 20\% of this mass. As discussed below, there is no strong evidence that any of these structures are internally heated protostellar companions, rather than being transient density fluctuations that are generic features of massive turbulent cores.

\subsection{Multiplicity in the close vicinity of G28.20-0.05}

Here we examine the multiplicity of G28.20-0.05 on scales within a few thousand~au, i.e., toward the ring-like structure. From the dendrogram results, two continuum sources have been identified in the ring, i.e., the main source on the SW side and a secondary source on the NE side. However, as discussed above, only the main source shows a very concentrated morphology in the mm continuum. Also, this is the only source around which strong velocity gradients are seen in H30$\alpha$.

\begin{figure}
\centering
    \includegraphics[width=\columnwidth]{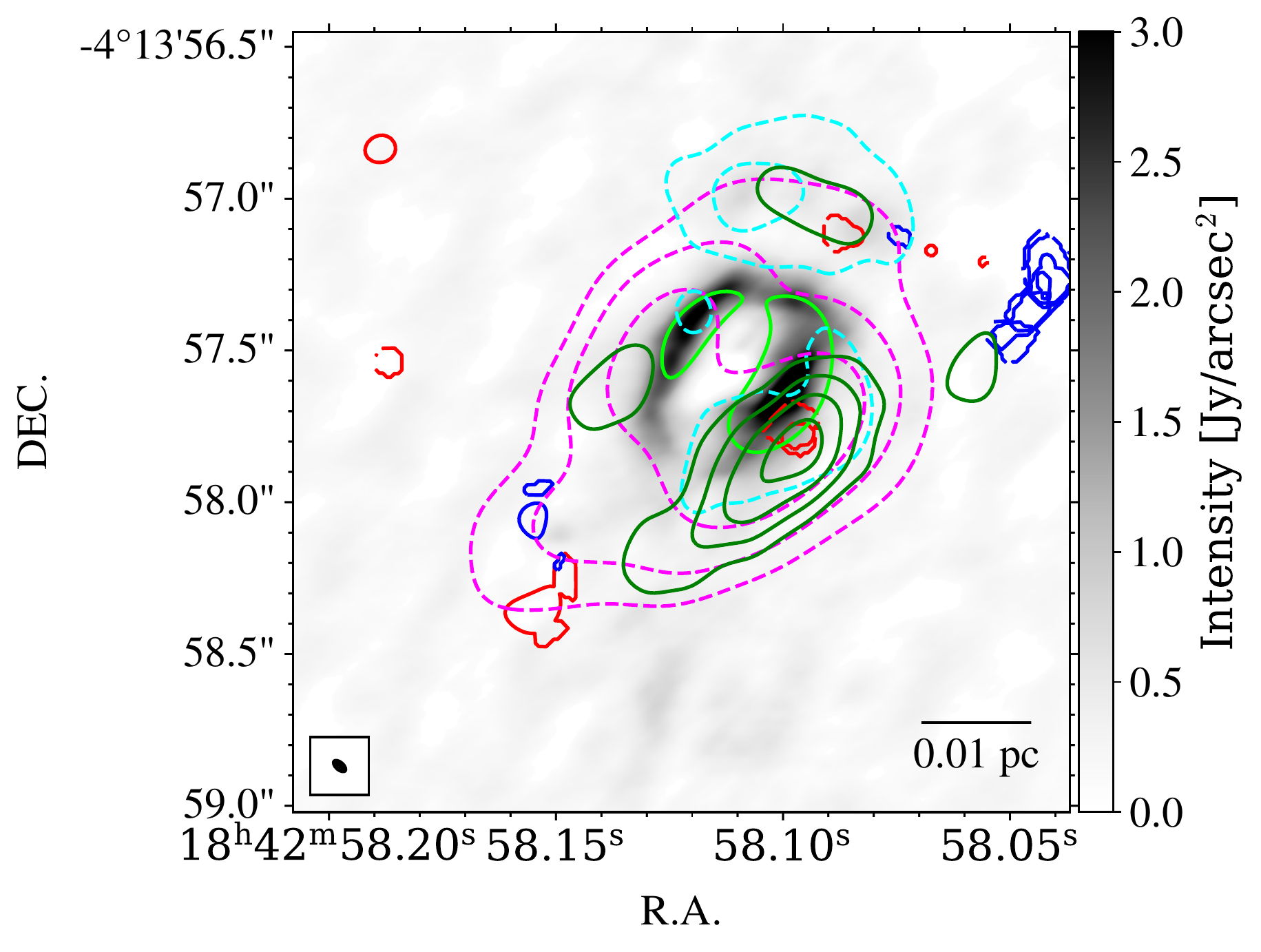}
    \caption{C+I+E continuum image (gray) overlaid with the C+I continuum dendrogram structures (light green). 
    Also overlaid are the C+I SiO(5-4) (red/blue) and C+I integrated intensity maps of hot core molecular lines C$_2$H$_5$CN($\rm{27_{1,27} - 26_{1,26}}$, E$_{\rm up} = 157.73~$K) (dark green), C$_2$H$_3$CN($\rm{25_{0,25} - 24_{0,24}}$, E$_{\rm up} = 145.54~$K) (cyan), and NH$_2$CHO($\rm{10_{1,9} - 9_{1,8}}$, E$_{\rm up} = 60.81~$K) (magenta). Three concentrations were identified for C$_2$H$_5$CN and C$_2$H$_3$CN, with both coinciding at the main continuum peak. The other two concentrations are located separately in the neighborhood of the secondary continuum structure and in the northern part outside the ring-like structure. On the other hand, NH$_2$CHO traces more the wider region around the continuum ring. 
    The intensities of the contours for C$_2$H$_5$CN are [0.16, 0.24, 0.32, 0.40]~$\rm Jy\:beam^{-1}\:{\rm km\:s}^{-1}$; for C$_2$H$_3$CN are [0.16, 0.24]~$\rm Jy\:beam^{-1}\:{\rm km\:s}^{-1}$; and for NH$_2$CHO are [0.16, 0.32, 0.64]~$\rm Jy\:beam^{-1}\:{\rm km\:s}^{-1}$.} 
    \label{fig:continuum_with_hotcore_lines}
\end{figure}

From the perspective of the hot core lines, the main mm continuum peak also shows the strongest concentration of the highest excitation species in its vicinity. However, there are more distributed hot core emission lines seen.
Figure~\ref{fig:continuum_with_hotcore_lines} presents a zoom-in of the C+I+E continuum image overlaid with hot core molecular lines C$_2$H$_5$CN($\rm{27_{1,27} - 26_{1,26}}$, E$_{\rm up} = 157.73~$K), C$_2$H$_3$CN($\rm{25_{0,25} - 24_{0,24}}$, E$_{\rm up} = 145.54~$K), and NH$_2$CHO($\rm{10_{1,9} - 9_{1,8}}$, E$_{\rm up} = 60.81~$K), as well as the SiO(5-4) outflow tracer. In addition to the main continuum peak, where we expect a massive protostar to be present, we also notice extended emission and some modest concentrations toward the NE continuum structure in the ring and in a northern region beyond the ring. Such concentrations could indicate the presence of one or two companion protostars that are forming along with the main massive protostar. However, they could equally well simply be modest overdensities in the turbulent, clump protostellar envelope of the main source, i.e., without any internal protostellar heating source. Further inspection of the moment 1 maps of H$30\alpha$ does not identify strong velocity gradients toward these locations. We have also not found any other strong velocity gradients in other molecular lines in these regions. There are some SiO(5-4) emission features in the northern region, however these are relatively weak and it is not clear that they trace the presence of a secondary protostellar outflow. 

In summary, there is no strong evidence for any companion protostar to the source located at the main mm continuum peak. There are hints of one or two surrounding concentrations in some hot core lines that could indicate the presence of protostellar companions, but could equally well be transient overdensities in the infall envelope to the primary protostar. More sensitive observations are needed to determine if there are any protostellar companions to the main source. We note that a full presentation of all detected hot core species, including their kinematics and implications for astrochemical models, will be presented in a forthcoming companion paper (Gorai et al., in prep.).



\section{Discussion and Conclusions}\label{sec:discussion}

We have presented 1.3~mm continuum and line ALMA observations of the massive protostar G28.20-0.05 ($d=5.7\:$kpc) using three array configurations to achieve sensitivity to a high spatial dynamic range from a MRS of 11\arcsec\ down to the smallest beam of $\sim 0.04\arcsec$, corresponding to about $200\:$au. Analysis of these data, along with ancillary MIR to FIR data, indicate the presence of a massive protostar with a current protostellar mass of $\sim 40\:M_\odot$. The protostar is launching powerful outflows, both in the form of a rotating ionized disk wind and as larger-scale molecular flows. Thus it appears to be still undergoing active accretion. At the same time, there is clear evidence that it is starting to produce ionizing feedback within its protostellar core, i.e., by ionizing its disk wind, but also by ionizing some surrounding, denser gas structures, as evidenced by the presence of a cm to mm free-free emitting ``ring''. The nature of this structure is still somewhat uncertain, but could involve the ionized surfaces of dense molecular accretion structures, e.g., filaments or streamers, that are generic features within a massive turbulent core \citep[e.g.,][]{2002Natur.416...59M,2003ApJ...585..850M,2013ApJ...766...97M}. Our cm to mm spectral index analysis suggests the presence of dust in and around this ring. However, higher frequency ALMA observations are needed for a more definitive characterization. Emission lines from dense and warm molecular gas are also seen in this region. These enable a dynamical mass estimate of $\sim 80\:M_\odot$ within about 2,000~au scales.


A crucial aspect in massive star formation theory is whether massive stars can form in relatively isolated environments. Competitive (clump-fed) accretion models \citep[e.g.,][]{2001MNRAS.324..573B,2010ApJ...709...27W,2020ApJ...900...82P} {\it require} the presence of a surrounding massive protocluster for a massive star to form. On the other hand, core accretion models \citep[e.g.,][]{2003ApJ...585..850M} can be valid in both isolated and relatively crowded environments.
From our analysis of the larger scale 1.3~mm continuum image, we argue that the protostar is forming in an isolated environment, i.e., with no compact and strong 1.3~mm continuum sources identified beyond a $4^{\prime\prime}$ radius, corresponding to $\sim 0.1$~pc, and extending out over the ALMA FOV to about 0.4~pc in radius. This apparent dearth of protostellar companions in a protocluster around the massive protostar is a strong constraint on massive star formation theories.

Even within the scale of about 0.1~pc, there is no strong evidence for protostellar companions. This scale matches the expected size of a massive turbulent core, e.g., $R_c = 0.074 (M_c/100\:M_\odot)^{1/2} (\Sigma_{\rm cl}/{\rm g\:cm}^{-2})^{-1/2}\:{\rm pc} \rightarrow 0.14\:{\rm pc}$ (with the last evaluation for $M_c=300\:M_\odot$ and $\Sigma_{\rm cl}=0.8\:{\rm g\:cm}^{-2}$, as inferred from our MIR to FIR SED modeling and consistent with the 1.3~mm continuum emission if assuming temperatures of $\sim 100\:$K). Such a massive turbulent core will contain overdense substructures and may have modest levels of fragmentation, especially inner disk fragmentation, leading to a few protostellar companions. However, with moderate $B-$field strengths ($\sim$mG) present, several simulations have shown that fragmentation may be completely prevented \citep[e.g.,][]{2012MNRAS.422..347S,2013ApJ...766...97M}. Such a scenario appears to be highly relevant to G28.20-0.05 and thus motivates future work to estimate the magnetic field strengths in the region.

\section*{Acknowledgements}
We thank an anonymous referee for comments that helped improve the paper. We thank M. Sewilo for providing access to VLA data of G28.20-0.05. C.-Y.L. acknowledges support from an ESO Ph.D. student fellowship. J.C.T. acknowledges ERC project MSTAR, VR grant 2017-04522. R.F. acknowledges funding from the European Union’s Horizon 2020 research and innovation programme under the Marie Sklodowska-Curie grant agreement No 101032092. G.C. and P.G acknowledge support from Chalmers Initiative on Cosmic Origins (CICO) postdoctoral fellowships.  K.E.I.T. acknowledges support by JSPS KAKENHI Grant Numbers JP19K14760, JP19H05080, JP21H00058, JP21H01145. We acknowledge support from the Nordic ALMA Regional Centre (ARC) node based at Onsala Space Observatory. The Nordic ARC node is funded through Swedish Research Council grant No 2017-00648. This paper makes use of the following \alma data: ADS/JAO.ALMA\#2015.1.01454.S, ADS/JAO.ALMA\#2016.1.00125.S. \alma is a partnership of ESO (representing its member states), NSF (USA) and  NINS (Japan), together with NRC (Canada), MOST and ASIAA (Taiwan), and KASI (Republic of Korea), in cooperation with the Republic of Chile. The Joint \alma  Observatory is operated by ESO, AUI/NRAO and NAOJ. 
\software{Matplotlib \citep{Hunter:2007}, APLpy \citep{2012ascl.soft08017R}, Astropy \citep{astropy:2013, astropy:2018}, astrodendro \citep{2019ascl.soft07016R}, Numpy \citep{harris2020array}, Photutils \citep{larry_bradley_2020_4044744}, and Spectral-Cube \citep{2019zndo...2573901G}.}
\bibliography{References}
\bibliographystyle{aasjournal}


\clearpage
\end{document}